\documentclass[superscriptaddress,twocolumn,prb,floatfix,longbibliography]{revtex4-2}

\usepackage[T2A,T1]{fontenc}
\usepackage[greek,english]{babel}
\usepackage{amsmath}
\usepackage{amssymb}
\usepackage{graphicx}
\usepackage{natbib}
\usepackage[dvipsnames]{xcolor}
\usepackage[utf8]{inputenc}
\usepackage[colorlinks=true,linkcolor=blue,citecolor=red,urlcolor=blue]{hyperref} 
\usepackage{soul}
\usepackage{bm}
\newcommand{\vect}[1]{\boldsymbol{\mathbf{#1}}}

\newcommand{\gpi}{\textrm{\greektext p}}

\begin{document}%
\setcitestyle{super}

\title{Waveguide quantum electrodynamics at the onset of spin-spin correlations}

\author{Sebastián Roca-Jerat}
\affiliation{Instituto de Nanociencia y Materiales de Aragón (INMA), CSIC-Universidad de Zaragoza, Pedro Cerbuna 12, Zaragoza 50009, Spain}

\author{Marcos Rubín-Osanz}
\affiliation{Instituto de Nanociencia y Materiales de Aragón (INMA), CSIC-Universidad de Zaragoza, Pedro Cerbuna 12, Zaragoza 50009, Spain}

\author{Mark D. Jenkins}
\affiliation{Instituto de Nanociencia y Materiales de Aragón (INMA), CSIC-Universidad de Zaragoza, Pedro Cerbuna 12, Zaragoza 50009, Spain}

\author{Agustín Camón}
\affiliation{Instituto de Nanociencia y Materiales de Aragón (INMA), CSIC-Universidad de Zaragoza, Pedro Cerbuna 12, Zaragoza 50009, Spain}

\author{Pablo J. Alonso}
\affiliation{Instituto de Nanociencia y Materiales de Aragón (INMA), CSIC-Universidad de Zaragoza, Pedro Cerbuna 12, Zaragoza 50009, Spain}

\author{David Zueco}
\affiliation{Instituto de Nanociencia y Materiales de Aragón (INMA), CSIC-Universidad de Zaragoza, Pedro Cerbuna 12, Zaragoza 50009, Spain}
\email{dzueco@unizar.es}

\author{Fernando Luis}
\affiliation{Instituto de Nanociencia y Materiales de Aragón (INMA), CSIC-Universidad de Zaragoza, Pedro Cerbuna 12, Zaragoza 50009, Spain}
\email{fluis@unizar.es}

\begin{abstract}
Waveguide quantum electrodynamics studies interactions of matter with photons travelling via a 
transmission guide and how these can be exploited to control quantum emitters and to establish 
quantum correlations between them. Here, we explore the competition between such light-mediated 
interactions with intrinsic matter-matter interactions. For this, we 
couple a superconducting line to a magnetic material made of organic free 
radical molecules. We find that molecules belonging 
to one of the two crystal sublattices form one-dimensional spin chains. Temperature 
then controls spin correlations along these chains in a continuous and 
monotonic way. In the paramagnetic region ($T > 0.7$ K), the microwave 
transmission evidences a collective coupling of quasi-identical spins 
to the propagating photons, with coupling strengths that reach values close to 
the dissipation rates. As $T$ decreases, the growth of spin 
correlations, combined with the anisotropy in the spin-spin exchange constants, 
tend to suppress the collective spin-photon coupling. In this regime, the spin visibility in 
transmission also reflects a gradual change in the nature of the dominant spin excitations, from 
single-spin flips to bosonic magnons. 
\end{abstract}
\maketitle
%
%
\section*{Introduction}\label{sec:introduction}

Waveguide Quantum Electrodynamics (wQED) studies how matter 
interacts with photons propagating along a transmission guide. 
\cite{lodahl2015interfacing, roy2017colloquium, sheremet2023waveguide} Recent 
experimental realizations of wQED span the optical to microwave 
regimes, incorporating real or artificial 
atoms and various material supports for the waveguides. \cite{Astafiev2010,vanLoo2013, Javadi2015, araujo2016, Hood2016, corzo2016, sorensen2016, Turschmann2017, Solano2017, Goban2017, Corzo2019, pennetta2022,Glicenstein2022, Tiranov2023, vylegzhanin2023} 
These experiments reveal collective effects in the emission of quantum 
emitters and help harnessing photon-induced correlations with potential 
applications for quantum technologies. 

So far, in wQED the focus has been mainly on otherwise quasi-independent 
emitters, where all 
interactions are mediated by the photons. Besides, the excitation spectra of 
systems showing long-range order, e.g. layered magnetic 
materials,\cite{MacNeill2019} have also been measured 
using microwave transmission lines. In the latter case, spin excitations are 
collective states, but correlations are dominated by interactions that are 
inherent to the material. Exploring situations in which 
the two effects compete remains virtually uncharted. Yet, there are 
compelling reasons to consider them. Firstly, because the close packing of 
emitters unavoidably generates interactions among them. 
\cite{SDjenkins2016, corman2017, petrosyan2021} And secondly, 
because their different signs and ranges can 
introduce ways to tune the order and thus create novel states. \cite{Roman-Roche2021,GarciaVidal2021,Schlawin2022} 

\begin{figure*}[t]
\begin{centering}
\includegraphics[width=0.7\textwidth, angle=-0]{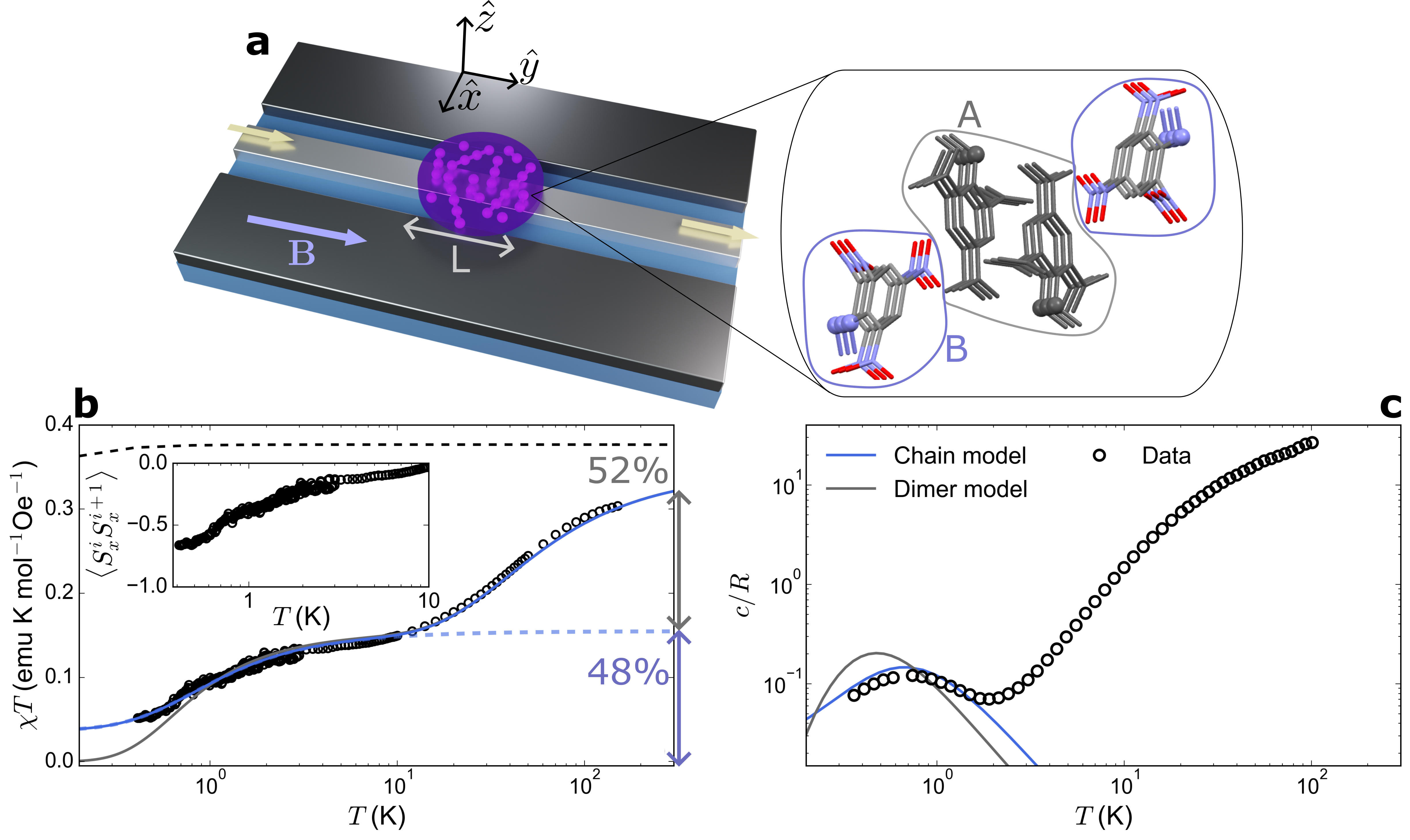}
\caption{{\bf Experimental setup and characterization of DPPH as a model 1D spin system}. (a) Sketch of the setup used in transmission experiments and image of the two molecular subspecies, A and B, present in the 
DPPH crystal structure. \cite{Zilic2010} (b) Product of the 
magnetic susceptibility $\chi$ times temperature, measured on a 
powdered DPPH sample from $0.3$ K up to $100$ K and for $B = 0.1$ 
T. The blue (grey) solid line is 
a fit with a model that considers the formation of 
antiferromagnetic dimers within sublattice A and then of spin 
chains (dimers) within sublattice B, as $T$ decreases. The inset shows the correlation function between nearest 
neighbour spins estimated as $\langle S^{x}_{i} S^{x}_{i+1} \rangle \simeq -\vert 1 - \chi T/C_{\rm B} \vert$, \cite{Blote1975,DeJongh1974} where $C_{\rm B} = 0.1536$ emu K/mol Oe is the Curie constant of the B-type DPPH molecules. (c) 
Specific heat of DPPH measured from $0.3$ K to $100$ K at $B=0$. 
The blue (grey) solid line are results calculated with the chain \cite{Blote1975} (dimer \cite{Zilic2010}) model for the magnetic contribution of spins in sub-lattice B.}
\label{fig:dpphcarac}
\end{centering}
\end{figure*}
 
In this work, we study the competition between light-mediated and 
intrinsic spin-spin interactions in a magnetic material coupled to 
microwave photons propagating via a superconducting transmission 
line (Fig. \ref{fig:dpphcarac}a). We focus on a 
particular model system, an organic free radical, in which spin-
spin correlations grow along $1D$ chains \cite{Mergenthaler2017} 
and can, therefore, be gradually controlled by either changing 
temperature or magnetic field. The broadband nature of the 
waveguide helps overcoming the limitations associated with  
narrow band in cavity QED, 
\cite{Voesch2015,Mergenthaler2017, lenz2020, Zollitsch2023} 
albeit at the price of a weaker spin-photon coupling. In 
particular, it allows studying within a single experiment the 
coupling to spin excitations at different energy scales. Here, we 
exploit this characteristic to study how the photon transmission 
reflects changes in the nature and properties of such excitations. The 
results show the possibility of attaining a very high collective spin-
photon coupling in the paramagnetic regime, at sufficiently high $T$, 
which then competes with intrinsic spin-spin correlations as $T$ 
decreases. When the latter take over, the microwave transmission bears 
evidence for a change in the statistics governing the elemental spin 
excitations, from single spin flips to bosonic magnons. 

\section*{Results and discussion}
\subsection*{DPPH as a 1D spin system}\label{subsec:sample_characterization}
All experiments described below used the 
$2,2-$diphenyl$-1-$picrylhydrazyl (hereafter DPPH) organic free 
radical \cite{Williams1966,Zilic2010} in powder form, as purchased 
from Sigma Aldrich (see methods). X-ray diffraction 
experiments, shown in Supplementary Fig. 1, 
suggest that this material exhibits the DPPH-III crystal structure 
\cite{Zilic2010} (Fig. \ref{fig:dpphcarac}a) that contains 
two inequivalent DPPH sites, referred to as A and B sublattices. 
Each DPPH radical hosts a spin $S=1/2$ with a nearly isotropic 
$g_S \simeq 2.004$ factor. At either high temperatures or in a 
sufficiently diluted form, it behaves as an ensemble of quasi-identical paramagnetic moments, which has made this material a widely used 
standard in 
Electron Paramagnetic Resonance (EPR).\cite{Yordanov1996} On the other hand, 
in dense materials and for sufficiently low temperatures, interactions between 
neighbour radicals become relevant. \cite{OhyaNishiguchi1979,Fujito1981,Zilic2010,Mergenthaler2017}

The magnetic response of DPPH has been studied via a combination 
of magnetic susceptibility and specific heat experiments 
performed in a broad temperature range $0.3$ K $\leq T \leq 100$ K. The results are outlined in 
Fig. \ref{fig:dpphcarac} and shown in detail in Supplementary Note 1 and Supplementary Figs. 4-7. 
The main conclusions are: \emph{i)} a fraction, approximately $15$ \%, of DPPH 
molecules are oxidized, thus have $S = 0$ ; 
\emph{ii)} molecules belonging to the A crystal sublattice form 
antiferromagnetic (AF) dimers, also with an $S=0$ ground state, at 
relatively high temperatures $\lesssim 50$ K, and \emph{iii)} DPPH 
spins in sublattice B form $1D$ chains along which AF 
correlations grow continuously as temperature decreases (see the 
inset of Fig. \ref{fig:dpphcarac}b). Below $10$ K, only B-type 
molecules can therefore couple to microwave photons. 

\subsection*{Broadband microwave transmission in the paramagnetic regime}\label{subsec:paramag_coupling}
Microwave transmission experiments were performed with on-chip 
superconducting coplanar waveguides using two different 
$^3$He-$^4$He dilution refrigerators (see methods and Supplementary Fig. 2 
for details). Magnetic fields up to $0.5$ T were oriented along the $y$ axis 
of the device. A $1$ mm wide polycrystalline DPPH pellet containing 
$\sim 10^{17}$ molecules, half of them belonging to 
sublattice B, was fixed on the $400$ $\mu$m wide central line by 
means of apiezon N grease. The complex transmission 
$S_{21}=|S_{21}| {\rm e}^{{\rm i} \varphi_{21}}$ and reflection 
$S_{11}=|S_{11}| {\rm e}^{{\rm i} \varphi_{11}}$ were measured in the 
frequency range $10$ MHz $\leq \omega/2 \gpi \leq 14$ GHz. Since the 
sample size is smaller 
than the shortest photon wavelengths ($\lambda \geq 4.4$ mm) we can disregard 
the dependence of the electromagnetic modes along the $y$ axis of the device 
(cf Fig. \ref{fig:dpphcarac}a). 
In order to extract the changes in transmission and reflection that arise 
from the coupling to spins and to compensate for the decay of 
$S_{21}$ with increasing $\omega$, both quantities were normalized 
using data measured at two different magnetic fields,
\cite{Clauss2013,Gimeno2023} as shown in Supplementary Note 2 and Supplementary Figs. 8-11.

\begin{figure}[h]
\begin{centering}
 \includegraphics[width=\columnwidth, angle=-0]{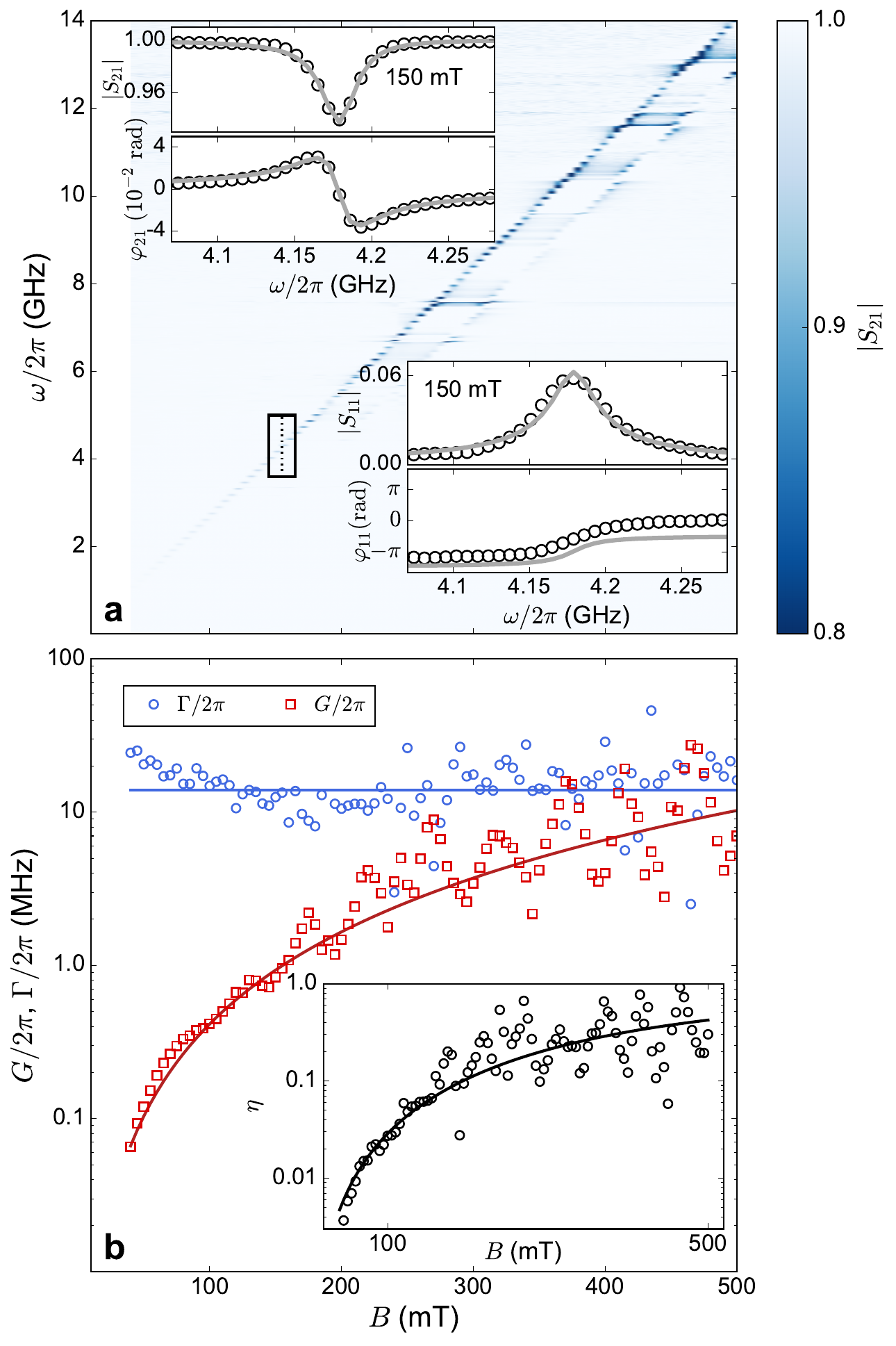}
\caption{{\bf Microwave transmission in the paramagnetic regime}. (a) Two-
dimensional plot of the normalized 
microwave transmission $\vert S_{21} \vert$ of a 
superconducting wave guide coupled to a DPPH polycrystalline 
pellet, measured at $T = 2$ K. The insets show the shape 
of both the absorption (reflection) amplitude $|S_{21}|$ ($|S_{11}|$) and 
its phase $\varphi_{21}$ ($\varphi_{11}$) with their corresponding fits 
based on 
Eq. (\ref{eq:transmission-line}), from which we 
determine the collective spin-photon coupling $G$ and the decay 
rate $\Gamma$. (b) Values for $G$ and $\Gamma$ obtained from these 
fits as a function of magnetic field. The lines are least-square 
fits based on Eq. (\ref{eq:G}), for $G$, and on a constant 
function $\Gamma / 2 \gpi \simeq 14$ MHz. The inset shows the maximum 
transmission visibility $\eta$ as a 
function of magnetic field (open dots) and the prediction derived 
from Eq. (\ref{eq:transmission-line}) (solid line).}
\label{fig:super}
\end{centering}
\end{figure}

We first examine the transmission measured at $T=2$ K. Normalized 
$|S_{21}|$ data are shown as a function of frequency and magnetic field in 
Fig.~\ref{fig:super}a, which also 
includes examples of complex transmission and reflection normalized data 
taken at $150$ mT. Transmission minima are 
observed for magnetic fields and frequencies fulfilling the condition 
$\omega = \Omega \equiv g_S\mu_{\mathrm{B}}B/\hbar$, indicating 
the resonant absorption of photons by paramagnetic spins. Note that the 
normalization of the data induces an additional line of $|S_{21}|$ maxima 
to the right of the resonance curve in Fig.~\ref{fig:super}a, as shown in 
Supplementary Fig. 8. Besides, the data show also contributions from 
spurious modes, caused by unavoidable imperfections at the chip boundaries 
or contact bonds, which manifest themselves as horizontal 
lines in Fig. ~\ref{fig:super}a. The resonances are accompanied 
by maxima in reflection. The fact that $|S_{11}| \simeq |S_{21}| - 1$ (see 
Fig. \ref{fig:super} and Supplementary Fig. 11) shows that the spin-photon 
interaction process preserves the photon coherence that characterizes 
wQED. 

These results are analyzed through the light-matter Hamiltonian 
\({\cal H} = {\cal H}_{\mathrm{M}} + {\cal H}_{\mathrm{WG}} + {\cal H}_{\mathrm{I}}\), which 
encompasses the spin system \({\cal H}_{\mathrm{M}}\), the 
waveguide \({\cal H}_{\mathrm{WG}}\), and their interaction 
\({\cal H}_{\mathrm{I}}\).  \cite{Dung2002, Dzsotjan2010, Lalumiere2013} 
In the paramagnetic regime, spins are considered as non-interacting and 
${\cal H}_{\mathrm{M}} \approx -g_{S} \mu_{\mathrm{B}} \sum_j {\bf B}{\bf S_j}$. 
The changes in $S_{21}$ and $S_{11}$ associated with the spin resonance 
can be fitted with the following expression, derived from ${\cal H}$ 
using input-output theory as described in the methods (see also Refs. 
\citenum{Fan2010,Fang2014,Burillo2016,corzo2016}),

\begin{eqnarray}
S_{21} &=& 1 - \frac{G} {G + \Gamma + {\rm i} (\Omega-\omega)} \nonumber \\
S_{11} &=& -\frac{G} {G + \Gamma + {\rm i} (\Omega-\omega)}
\label{eq:transmission-line}
\end{eqnarray} 

\noindent to obtain the collective spin-photon coupling $G$ and the decay 
rate $\Gamma$, which parametrizes the losses. Illustrative fits are 
shown as insets in Fig. \ref{fig:super}a, while the field-dependent 
parameters extracted at $T=2$ K are shown in Fig. \ref{fig:super}b.

We find a nearly constant $\Gamma / 2 \gpi \approx 14$ MHz. This 
value is larger than the spin dephasing rate 
$\gamma_{\phi}/2 \gpi = 1/2 \gpi T_2 \cong 4.8$ MHz 
($T_2 = 33$ ns) obtained from time-resolved electron 
paramagnetic resonance experiments (Supplementary Fig. 7), thus 
showing that the resonance width has contributions from 
inhomogeneous broadening. The spin-photon coupling $G$ 
increases with $B$, reaching remarkably high values $G/ 2 \gpi \simeq 12$ 
MHz for $B \simeq 0.5$ T. Oscillations in $G$ likely originate from 
deviations of the transmission line spectral density from its ideal 
behavior, caused by its spurious modes. The maximum change in 
$\vert S_{21} \vert$, denoted 
as $\eta$ (inset of Fig. \ref{fig:super}b), characterizes the collective 
spin resonance visibility, i.e. the efficiency of the spins in affecting 
photon propagation. \cite{sheremet2023waveguide} At the 
highest fields and frequencies reached in these experiments, it becomes 
higher than $1/2$, which agrees with $\eta = \frac{G}{G + \Gamma}$ 
for $\omega = \Omega$ (Eq. (\ref{eq:transmission-line})) and 
with the fact that the collective spin-photon couplings are 
then comparable to the single spin decay rates.

For $N$ quasi-independent spin emitters, $G$ is given by 
(see Ref.~\citenum{sheremet2023waveguide} and Supplementary Note 3)
\begin{equation}
G = 2 \gpi \lambda_\Omega^2 \; N_{\rm eff}  
\label{eq:G}
\end{equation}
where $\lambda_\Omega^2 = \alpha \Omega$ is the spin-photon 
coupling spectral density for a one-dimensional waveguide and 
$N_{\rm eff} = N \tanh (\hbar \Omega / 2k_{\rm B} T )$ is the effective 
spin number at the given temperature, proportional to the spin 
polarization along $\vect{B}$. 
Figure \ref{fig:super}b and Supplementary Fig. 12 show that 
Eq. (\ref{eq:G}) accounts well for the enhancement of $G$ with 
increasing $N_{\rm eff}$ that is observed above $1.2$ K, with 
$\alpha$ as the sole free parameter.

\begin{figure*}[t!]
\begin{centering}
\includegraphics[width=\textwidth, angle=-0]{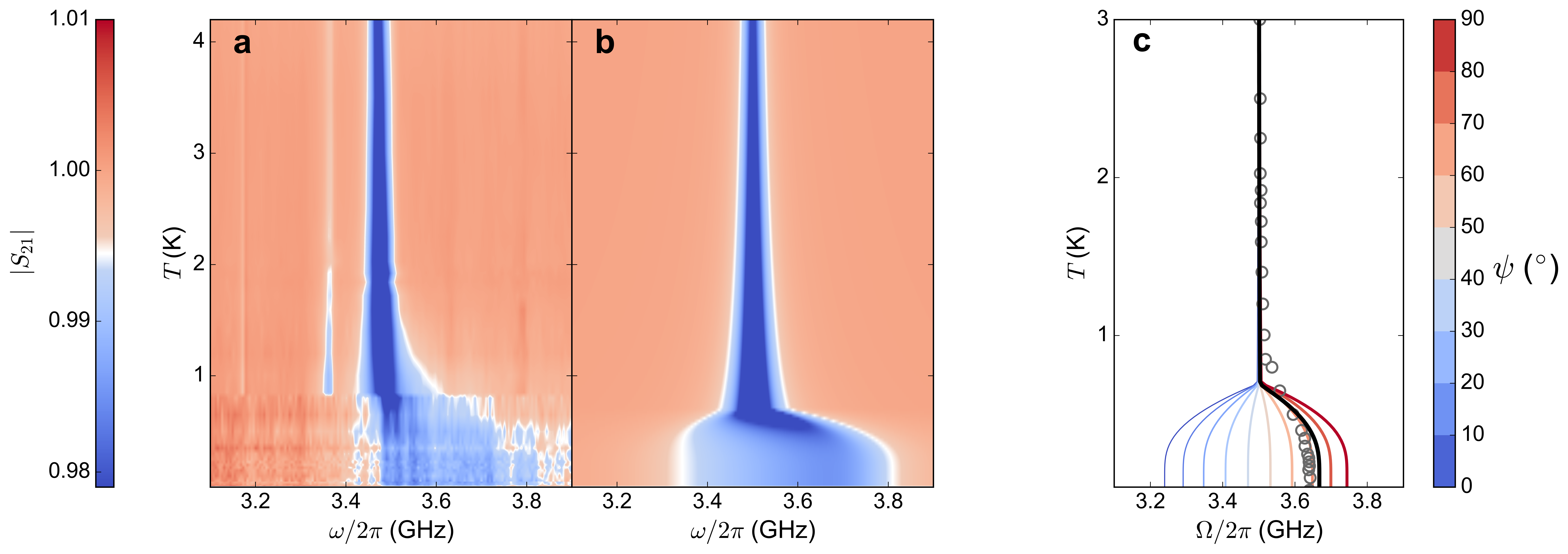}
\caption{{\bf Microwave transmission at the onset of spin-spin 
interactions}. Experimental (a) and simulated (b) 
changes in transmission associated with the coupling to DPPH 
spins measured at $B = 125$ mT as a function of frequency and 
temperature. The weak thin line near $3.38$ GHz in (a) arises from a 
spurious mode in the transmission line (horizontal lines in 
Fig. \ref{fig:super}a) that has not been fully eliminated by 
the data normalization. (c) Dependence of the 
spin resonance frequency $\Omega$ 
on the angle $\psi$ that the magnetic field forms with the anisotropy axis of 
the spin-spin interactions, derived from Eq. (\ref{eq:HMF}). The line 
thickness is proportional to the 
contribution to the spin resonance, which increases for molecules forming 
larger angles and therefore, for larger frequencies. The central $\Omega$ 
obtained from the experiments in (a) and from the simulations in (b) is 
given by the grey open symbols and the solid black line, respectively.}
\label{fig:suppresion}
\end{centering}
\end{figure*}

Detecting this remarkable enhancement in spin-photon coupling and its 
visibility has been made possible by the broadband nature of these 
transmission experiments. It reflects the 
increase in the photon-induced decay rate from superradiant collective 
states. In the single-photon limit and at zero temperature, this 
serves as a signature of the formation of Dicke states. In our 
case, these are thermal collective states that emit in a 
superradiant manner, $G \propto N_{\text{eff}}$, while the 
emission to other channels is determined by 
the single-spin decay rates $\Gamma$. The experimental data 
confirm that, at this temperature, DPPH molecules mainly act as 
non-interacting $S=1/2$ emitters and that it is 
possible to reach a regime in which collective spin-photon 
couplings become dominant over the individual spin decay rates.

\subsection*{Microwave transmission at the onset of 1D spin correlations: 
interactions competition}\label{subsec:interaction_competition}
\begin{figure}[t]
\begin{centering}
\includegraphics[width=\columnwidth, angle=-0]{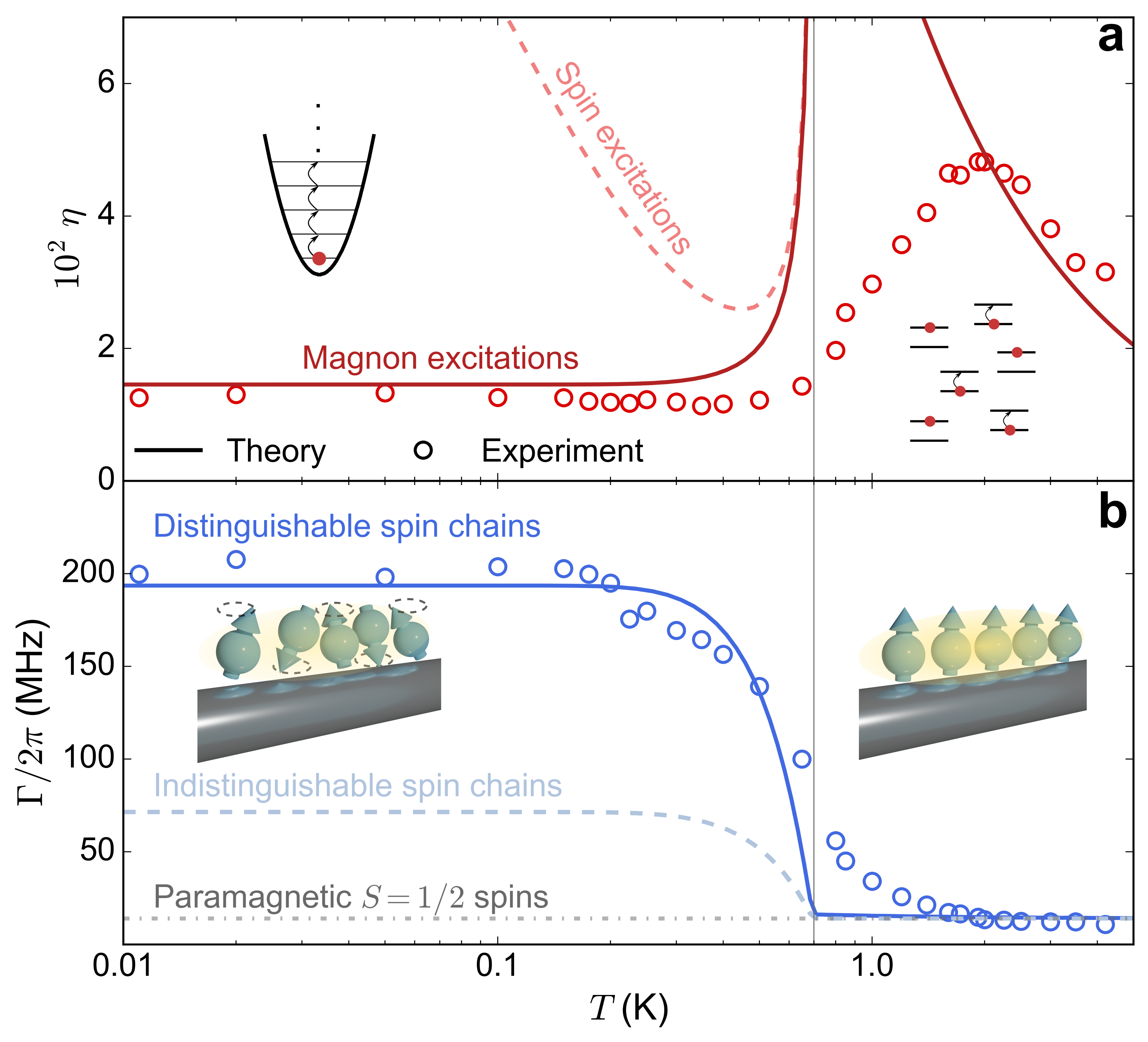}
\caption{{\bf Spin transmission visibility and elementary spin excitations}.
(a) Transmission visibility 
of the spin resonance obtained from the experimental results shown in 
Fig. \ref{fig:suppresion}a (red points) and from theoretical 
simulations based on describing the spin excitations with either a 
classical mean-field model (dashed light-red line) or as bosonic magnons (solid red line). (b) Experimental 
(open blue symbols) and theoretical (lines) spin 
resonance width. The different lines progressively add different
contributions to the line width: the dashed-pointed grey line represents the 
paramagnetic inhomogeneous broadening, the dashed light-blue line includes  
the additional damping arising from the emergence of 
antiferromagnetic spin-spin correlations
and, finally, the solid darker blue line, obtained from the results in  Fig.~\ref{fig:suppresion}b, adds also the broadening caused by the 
spin-spin interaction anisotropy that makes chains with different 
orientations interact differently with the microwave field. The inset 
illustrates the nature of the  excitations that couple to propagating 
photons: single spin excitations in the paramagnetic regime and 
bosonic spin waves in the low-$T$ region.}
\label{fig:spinwaves}
\end{centering}
\end{figure}

Lowering temperature allows exploring how 
AF correlations along spin chains \cite{Bonner1964,DeJongh1974,Blote1975} 
modify the interaction with microwave photons. 
Figure \ref{fig:suppresion} shows transmission data measured at a fixed 
$B = 125$ mT as a function of $\omega$ and (decreasing) $T$. 
The spin resonance first gets enhanced on 
cooling below $2$ K, as a result of 
the higher spin polarization, but then, for $T \lesssim 0.8$ K, its 
visibility decreases (Fig. \ref{fig:spinwaves}a and Supplementary Fig. 10).
Moreover, the effective spin-photon coupling $G$ also decreases and its 
dependence with temperature and magnetic field deviate from what is 
expected for a paramagnetic system (Supplementary Fig. 13). The data 
therefore point to a crossover to a different regime in the spin-photon 
interaction. The access to a wide frequency window allows observing that 
this is accompanied by a significant broadening, by a factor ten between 
$1.5$ K to $10$ mK and, as Fig. \ref{fig:suppresion}c shows, 
by a $7$ \% upwards shift of the resonance center frequency. This 
qualitative behavior persists for magnetic fields up to $300$ mT (see 
Supplementary Figs. 13-15), although its magnitude, as gauged by the 
resonance broadening and by the drop in its visibility, seems to get 
somewhat reduced by increasing $B$.  

Experiments performed for different input 
microwave powers, shown in Supplementary Note 5 and Supplementary Fig. 16, lead to the same results. 
Therefore, the broadening is not associated with any spin 
saturation effect. It is also much larger than the line width 
arising from the weak anisotropy, less 
than $0.1$ \%, between the principal $g_S$-factors of DPPH. 
\cite{Zilic2010} A possible origin is the 
combined effect of growing spin correlations and of a weak 
anisotropy in the spin-spin exchange interactions, which is not unusual 
with free-radical molecules.\cite{Yamauchi1971} As a result, 
each DPPH crystallite in the powder would acquire a 
different $\Omega(\psi)$, depending on its orientation $\psi$ 
with respect to ${\bf B}$. The ensuing additional inhomogeneous 
broadening makes DPPH molecules located in different powder grains 
`distinguishable' and therefore tends to suppress the collective emission 
from superradiant states. By contrast, increasing $B$ brings the spins closer 
to the paramagnetic state, as shown by field-dependent heat capacity 
experiments (Supplementary Fig. 5), and would then reduce such 
effects, as it is indeed observed (see Supplementary note 4 and Supplementary 
Fig. 15). 


In order to provide a firmer basis to this interpretation, we 
model the light-matter Hamiltonian incorporating intrinsic
interactions into the spin Hamiltonian 
${\cal H}_{\rm M}$. We treat the spin chains in DPPH by a mean field (MF) 
approximation (see Ref. \citenum{Nagamiya1955}, Supplementary Note 6 
and Supplementary Figs. 17-20 for details):

\begin{align}
  {\mathcal H}_{\rm M} = & - g_{S} \mu_{\rm B}  {\bf B} ({\bf M_1} + {\bf M_2})
  +  {\bf M_1} \hat J {\bf M_2} 
  \label{eq:HMF}
\end{align}

\noindent where ${\bf M_{1}}$ and ${\bf M_{2}}$ are the 
macro-spin vectors of the two magnetic sublattices within each 
chain and the exchange interaction tensor 
$\hat J = {\rm diag} \left( J, J(1+\epsilon\sin\psi), 
J(1+\epsilon\cos\psi)\right)$ 
introduces an uniaxial anisotropy in the spin-spin interactions.
The exchange constant $J$ has been determined 
comparing the Weiss temperature $\theta = -0.7$ K, which 
determines the paramagnetic 
susceptibility measured between $1$ K and $10$ K 
(Fig. \ref{fig:dpphcarac}b and 
Supplementary Note 1), with the MF prediction 
$\theta = -J / k_{\rm{B}}$.
Then, $\Omega$ and its distribution can be obtained by 
solving the Landau-Lifshitz-Gilbert equation in the linear regime and considering that the 
powder sample has a random distribution of anisotropy axes. 
\cite{Keffer1952} The comparison with experimental data allows to determine both $|\epsilon|$, 
from the resonance broadening, and its sign, from the observed shift in the resonance center 
frequency (grey dots in Fig.~\ref{fig:suppresion}c). In particular, a negative $\epsilon$ leads 
to higher frequencies for angles closer to $\gpi/2$, which then dominate the average in a random 
distribution. We then find $\epsilon = -0.086$, which accounts well for the experimental results 
in Fig.~\ref{fig:suppresion}. Note that the solid black curve in Fig.~\ref{fig:suppresion}c 
emerges naturally from the resonance curves shown in Fig.~\ref{fig:suppresion}b. Therefore, it is not an additional fit. 

Again, the transmission can be computed using input-output theory,\cite{Gardiner1985} 
as shown by Eqs. (5-12) in the Methods. However, it is 
essential to consider two limitations of the MF theory. First, MF 
overestimates the sharpness of changes occurring when $T$ approaches 
$\vert \theta \vert \equiv T_{\rm N}$, as it 
predicts a transition towards an AF phase that 
does not occur in 1$D$ systems. Additionally, the MF low-energy 
excitations involve tilting the relative orientations of $\vect{M_1}$ 
and $\vect{M_2}$. In reality, bosonic spin-wave excitations occur. 
Our theory incorporates these modes by assuming that, below 
$T_{\rm N}$, excitations are magnons with  $\Omega (\psi)$ calculated 
by MF. Consequently, the classical Hamiltonian 
(\ref{eq:HMF}) is replaced by the spin-wave one 
${\cal H}_{\rm M} = \hbar \Omega (\psi) (b^{\dagger}b + 1/2)$, 
where $b$ and $b^\dagger$ are magnon annihilation and creation 
operators that fulfill the condition $[b, b^\dagger]=1$. Note that, 
since the excitations are bosonic, the system then behaves 
exactly as an harmonic oscillator and its coupling to the 
waveguide loses its dependence with temperature. Then, $S_{21}$ can 
be calculated as follows (see Methods):
\begin{equation}
\label{t-sw}
    S_{21} = \frac{1}{1 + \int d \psi \sin \psi \frac{ 2\gpi\lambda_{\Omega}^2 N_{\mathrm{eff}} (T_{\rm N})  }{\Gamma + {\rm i} \left( \Omega(\psi)-\omega \right)} } \ .
\end{equation}
\noindent This formula gives the transmission shown in 
Fig.~\ref{fig:suppresion}b, which, apart from the discussed sharpness, 
accounts well for the experimental results in the whole temperature region. 

It is worth taking a closer look at the temperature 
dependence of the visibility shown in Fig.~\ref{fig:spinwaves}a. The dashed 
red curve shows $\eta$ calculated with the classical MF Hamiltonian, 
considering the usual thermal dependence of the coupling between the single 
spin excitations and the photons. In this case, $\eta$ increases as 
$T$ decreases. The solid red line follows instead from freezing this thermal 
dependence, accounting for the spin-wave character of the 
excitations. Clearly, only the latter agrees with the experimental 
data. The nearly constant $\eta$ observed below $700$ mK therefore 
provides direct evidence for a change in the statistics of the 
elementary spin excitations that couple to the microwave 
photons. 

This conclusion is further supported by the line width $\Gamma$ of the 
transmission resonances, whose temperature dependence is shown in 
Fig.~\ref{fig:spinwaves}b. It has been both observed and 
proven\cite{Kamra2018, Suto2019, Wang2024} that the line width in 
antiferromagnetic resonance is broader than its ferromagnetic
or paramagnetic counterparts. The origin of this effect is linked to the 
intrinsic nature of magnon excitations in antiferromagnetic chains: the 
magnon resonance frequency $\Omega(\psi)$ acquires an imaginary component 
that adds to the paramagnetic $\Gamma$ in Eq.~\eqref{t-sw}. This broadening
emerges naturally from MF calculations (see Eqs. (24-25) in Supplementary Note 6) without 
further fitting and it is represented by the dashed blue curve in 
Fig.~\ref{fig:spinwaves}b. However, as can be seen in this plot, the sources 
of broadening present in isotropic antiferromagnetic chains are not 
sufficient to account for the experimentally observed $\Gamma$. Only when we 
include in the model an anisotropic exchange (i.e. we take $\epsilon \neq 0$) 
with randomly oriented anisotropy axes does it account for $\Gamma$ 
measured at very low temperatures. The theoretical $\Gamma$ {\em vs} $T$ 
curve labeled as "distinguishable spin chains", which includes all these 
effects, is directly extracted from the theoretical transmission curves in 
Fig.~\ref{fig:suppresion}b. Thus, the increase in $\Gamma$ has two 
sources: the onset of antiferromagnetic correlations and the anisotropic 
nature of this interaction in randomly oriented chains.

%
%
\section*{Conclusions}\label{sec:discussion}
In summary, we have observed a crossover from quasi identical 
spins interacting collectively with microwave photons to a 
regime governed by $1D$ magnetic correlations. In the former, 
paramagnetic region, spin-photon couplings comparable to the 
decay rates have been achieved, which lead to a coupling 
efficiency $> 0.5$. When $T$ decreases 
and spin-spin interactions gain prominence, they tend to 
suppress the  collective emission from superradiant modes  
because  different molecules are no longer equivalent to each 
other. Transmission data allow detecting, and estimating, the 
anisotropy in the spin-spin interaction 
along each chain. Besides, we find that its temperature 
dependence reflects the change in the spin excitation statistics. As it
is illustrated in Fig. \ref{fig:spinwaves}, when the material is paramagnetic 
these are single spin flips, while antiferromagnetic interactions 
stabilize bosonic magnon excitations. These results illustrate a regime 
in waveguide Quantum Electrodynamics that arises from the delicate interplay 
between light-mediated and intrinsic matter interactions. Besides, they 
provide a method to characterize in detail the strength and nature of spin-
spin interactions in low-dimensional systems.

\section*{Methods}\label{sec:methods}

\subsection*{Sample preparation and structural characterization}
\label{sec:dpph}

DPPH ($2,2-$diphenyl$-1-$picrylhydrazyl) is an organic free radical molecule 
hosting a spin $S = 1/2$ with a nearly isotropic $g_S = 2.004$ factor, very 
close to that of a free electron. \cite{Williams1966,Zilic2010} DPPH samples 
used throughout this work were purchased, in powder form, 
from Sigma Aldrich (reference D9132). Even though solvent-free DPPH has been 
known for a long time,\cite{Williams1966} a full structural determination was 
not performed until quite recently.\cite{Zilic2010} The results of powder X-
ray diffraction experiments performed on our sample 
are shown in Supplementary Fig. 1. They suggest that it corresponds to the 
DPPH-III structure,\cite{Zilic2010} shown in Fig. \ref{fig:dpphcarac}a, which 
contains two inequivalent DPPH sites, referred to as A and B sublattices.

\subsection*{Magnetic measurements}
\label{ssec:mag_properties}

The dc magnetic susceptibility and magnetic isotherms were measured above 
$2$ K in a Magnetic Properties Measurement System (MPMS) by Quantum Design, a 
commercial magnetometer based on a dc-SQUID, operated by the Servicio de 
Apoyo a la Investigaci\'{o}n (SAI) of the University of 
Zaragoza. In addition, magnetic measurements were also performed, for 
$0.3$ K $\leq T \leq 10$ K, with a homemade micro-Hall 
magnetometer \cite{Kent1994} mounted in the insert of a Physical Properties 
Measurement System (PPMS), also by Quantum Design and operated by the SAI. 
The magnetometer (Supplementary Fig. 3) consists of two semiconducting 
layers of GaAs and Al$_{1-\alpha}$Ga$_{\alpha}$As, shaped in the 
form of a double cross, with a two-dimensional electron gas 
confined at the interface between the two materials. A current $I_{\rm ac}$ 
is introduced through the 
main arm and a magnetic field ${\bf B}$ is applied parallel to it. The DPPH 
sample was placed on the 
edge of one of the two Hall crosses of the device and fixed/thermalized with 
Apiezon N grease. Its 
magnetization ${\bf M}$ generates an additional magnetic field ${\bf B}_{M}$, 
perpendicular to the 
plane of the device, which induces a Hall voltage $V_{\rm Hall}$ proportional 
to $|{\bf M}|$. In 
order to reduce noise and detect this signal, the current was modulated at a 
frequency of $107$ Hz, 
with an $I_{\rm ac} = 10$ $\mu$A amplitude, and the reference voltage 
$V_{\rm ref}$ in the bare cross 
was subtracted from $V_{\rm Hall}$ by means of a differential lock-in 
amplifier.

\subsection*{Heat capacity measurements}
\label{ssec:heatCap}

Heat capacity experiments were also carried out in a PPMS. The calorimeter 
consists of a sapphire disc on which the sample is placed. This holder 
integrates a heater and a thermometer, and it is connected to a thermal 
reservoir by thin gold wires. Sample and holder are mutually thermalized 
with apiezon N grease. The constant-pressure heat capacity is measured using 
a relaxation technique.\cite{Stewart1983} A heat power is applied for a short 
time and then removed. The time constant of the exponential increase and 
decrease in temperature during this process is proportional to the heat 
capacity of the combined sample-calorimeter system. The contributions from 
the empty calorimeter and the grease are known from previous calibration 
experiments, and then subtracted from the measurement in order to get the 
sample specific heat.

\subsection*{Electron paramagnetic resonance}
\label{sup:magRes}

Electron Paramagnetic Resonance (EPR) experiments were performed in a 
commercial Elexsys E-580 by 
Bruker operating in the X-band ($\sim 9.8$ GHz). The spectrometer consists of 
a resonant cavity, in 
which the paramagnetic sample is placed, located at the center of a $0-1.3$ T 
electromagnet. Temperatures 
between $7$ K and $300$ K can be accessed by means of a Helium flow cryostat. 
In continuous 
wave EPR (cw-EPR) experiments, the cavity is continuously irradiated while 
the DC magnetic field is 
swept slowly. The absorption signal is obtained with a field-modulation 
detection. Time-domain EPR 
experiments measure the signal emitted by the spin system after sending a 
series of microwave pulses 
to the cavity. When homogeneous broadening is dominant, as it is the case 
with DPPH, the phase memory 
time $T_{\rm{m}}$ of the spin system is obtained as the characteristic time 
constant of the free induction decay, which is the signal that the spin 
system induces in the cavity after a microwave 
pulse generates a coherence between the two spin states of a given transition.

\subsection*{On-chip microwave transmission and reflection measurements}
\label{sec:transmission}
Microwave transmission experiments were performed with on-chip 
superconducting coplanar waveguides 
fabricated by optical lithography on a $100$ nm Nb thin film deposited onto 
crystalline sapphire wafers. A $100$ $\mu$m thick DPPH pellet was fixed on 
the $400$ $\mu$m wide central 
line by means of apiezon N grease, matching roughly the width of the 
central line plus the separation to both ground planes, see 
Supplementary Fig. 2a. Based on this, we estimate a sample
area and volume of the order of $1$ mm$^2$ and $0.1$ mm$^3$, 
respectively. DPPH-III has a molecular density of $1.06 \times 10^{27}$ 
molecules/m$^3$, with two independent DPPH molecules (A and B) per unit 
cell.\cite{Kiers1976} Only type B DPPH molecules couple to the waveguide 
at low temperatures, thus the number of spins that effectively 
interact with the propagating photons is $N \sim 5 \times 10^{16}$. The 
superconducting chips were integrated into two different $^{3}$He-$^{4}$He 
dilution cryostats. The first one, sketched in 
Supplementary Fig. 2, gives access to temperatures between $7$ and $800$ mK. 
A cold finger (Supplementary Fig. 2b) places the chip at the center of an 
axial superconducting magnet (Supplementary Fig. 2c), which applies magnetic 
fields up to $1$ T along the $\hat{y}$ axis of the 
laboratory reference frame, which is parallel to the waveguide. The complex 
transmission $S_{21} = \vert S_{21} \vert {\rm e}^{{\rm i} \varphi_{21}}$ 
through the device was measured in the frequency range 
$10$ MHz $\leq \omega/2 \gpi \leq 14$ GHz with a vector 
network analyzer. The input signal was attenuated by $-50$ dB before reaching 
the chip, while the transmitted signal was amplified by $30$ dB at $4$ K with 
a low-noise cryogenic amplifier. This setup is tailored to measure only the 
transmission, with the reflection $S_{11}$ doubly attenuated instead of 
amplified. For this reason, microwave transmission 
and reflection experiments were also performed, for 
$130$ mK $\leq T \leq 4.2$ K, in a different setup with less attenuation in 
the lines inside the cryostat and without an output cryogenic amplifier. This 
$^{3}$He-$^{4}$He cryostat is equipped with a $9$T$/1$T$/1$T superconducting 
vector magnet. 

\subsection*{Waveguide QED theory}
\label{sec:methods_qedtheory}
The $B_{\rm rms}$ field lies in the $xz$ plane, that is, perpendicular to the 
applied dc magnetic field $B$. Therefore, the photons from the guide will 
essentially induce spin-flips in the DPPH sample. Thus, the Hamiltonian for
our calculations will be 
$\mathcal{H} = {\mathcal H}_{\rm WG} + {\mathcal H}_{\rm I} + {\mathcal H}_{\rm M}$, where
\begin{equation}
    \label{eq:hamwg}
    {\mathcal H}_{\rm WG} = \int 
{\rm d} \omega 
\, 
\omega
\, 
(  r_\omega^\dagger r_\omega +  l_\omega^\dagger l_\omega )\ ,
\end{equation}
with $r_\omega^\dagger$ ( $l_\omega^\dagger$ ) bosonic operators that create 
right (left) moving photons, and
\begin{equation}
    \label{eq:hamI}
    {\mathcal H}_{\rm I} = \sum_{j, n} \int_{\Omega_-}^{\Omega_+} d {\omega} \;  
 \lambda_\omega ({\bf r_{j, n}})
 \; 
 \sigma^x_{j,n}
 \,  
 X(\omega)\ .
\end{equation}

In Eq.~\eqref{eq:hamI}, $X(\omega) = (r_\omega + l_\omega) + {\rm h.c.}$ and 
$\sigma^x_{j,n} = \sigma^+_{j,n}+\sigma^-_{j,n}$
with $\sigma^\pm_{j,n}$ the Pauli spin-raising and lowering 
operators for the $j$-th DPPH molecule from the $n$-th chain. The coupling 
between spins and photons is determined by the spectral function 
$\lambda_\omega ({\rm r_{j, n}})$,
which essentially depends on the magnetic field generated by the waveguide at
the position ${\rm r_{j, n}}$ of the molecule. Given that the sample size is 
much smaller than the wavelength of the photons, we neglect distance effects 
and assume $\lambda_\omega ({\rm r_{j, n}})\simeq \lambda_\omega$.

For computing the transmission, $S_{21}$, and reflection, $S_{11}$, we use 
input-output theory \cite{Gardiner1985}. We need to compute the 
input/output right/left fields such that 
$S_{21}=\frac{\langle r_{\rm out}\rangle}{\langle r_{\rm in}\rangle}$ and 
$S_{11}=\frac{\langle l_{\rm out}\rangle}{\langle r_{\rm in}\rangle}$. First, 
let's focus on the single spin scenario. For a coherent driving on our system 
$\mathcal{H}_{drive}=\sqrt{2\gpi}\lambda_\Omega\alpha_{\rm in}e^{-{\rm i}\omega t}\sigma_x$, with
$\Omega = g_S\mu_B B/\hbar$, we can express
\begin{equation}
    \label{eq:meth_rout}
    \langle r_{\rm out} \rangle(t) = \langle r_{\rm in} \rangle(t) - 2\gpi\lambda_\omega^2\chi_{\sigma_x}(\omega)\langle r_{\rm in} \rangle(t)\ ,
\end{equation}
\begin{equation}
    \label{eq:meth_lout}
    \langle l_{\rm out} \rangle(t) = \langle l_{\rm in} \rangle(t) - 2\gpi\lambda_\omega^2\chi_{\sigma_x}(\omega)\langle r_{\rm in} \rangle(t)\ .
\end{equation}
The spectral density $\lambda_\omega$ is considered to be a smooth function 
so that for frequencies $\omega$ close to the frequency of the spin system 
$\lambda_\omega\approx\lambda_\Omega$. In order to obtain $\chi_{\sigma_x}$ 
we compute the time dependence of $\langle\sigma_\pm\rangle$ from the quantum 
master equation, which leads to
\begin{equation}
    \label{eq:meth_sigmapm}
    \frac{d}{dt}\langle\sigma_\pm\rangle = \pm {\rm i}\Omega\langle\sigma_\pm\rangle - \Gamma\langle\sigma_\pm\rangle \pm {\rm i}\sqrt{2\gpi}\lambda_\Omega\langle\sigma_z\rangle_\beta\alpha_{\rm in}e^{-{\rm i}\omega t}\ ,
\end{equation}
where $\langle\sigma_z\rangle_\beta=\tanh(\hbar\Omega/2k_{\rm B}T)$ is the 
thermal population of the spin and 
$\Gamma = \gamma_\phi + [2\bar n(\Omega, T)+1]\cdot 2\gpi\lambda_\Omega^2$ 
includes the contributions of dissipation 
channels that are either intrinsic to the spin sample ($\gamma_\phi$) or 
related to its coupling to the transmission line through the bosonic 
occupation number for thermal photons ($\bar n$). Typically, the single spin-
photon coupling $2\gpi\lambda_\Omega^2 \ll \gamma_\phi^{-1}$, thus 
$\Gamma \approx\gamma_\phi$. Solving Eq.~\eqref{eq:meth_sigmapm} in the 
stationary limit and applying the previous relations we get
\begin{equation}
    \label{eq:meth_s21}
    S_{21}^{(1)}(\omega) = 1 - \frac{2\gpi\lambda_\Omega^2\langle\sigma_z\rangle_\beta}{{\rm i}(\Omega - \omega)+\Gamma}\ ,
\end{equation}
\begin{equation}
    \label{eq:meth_s11}
    S_{11}^{(1)}(\omega) = - \frac{2\gpi\lambda_\Omega^2\langle\sigma_z\rangle_\beta}{{\rm i}(\Omega - \omega)+\Gamma}\ .
\end{equation}
The superscript indicates that these relations account for a system with a 
single spin. In order to extend these relations to the case of an spin 
ensemble we use the transfer matrix formalism, where the output field of one 
spin becomes the input field for the next one. Defining 
$\theta_j(\omega) = S_{11}^{(1)}(\omega)/S_{21}^{(1)}(\omega)$ and using 
that $S_{21}=1+S_{11}$, the transmission and reflection read
\begin{equation}
    S_{21}(\omega) = \frac{1}{1-\sum_j\theta_j}\ ;\hspace{2 mm} S_{11}(\omega) = \frac{\sum_j\theta_j}{1 - \sum_j\theta_j}\ .
\end{equation}
In the paramagnetic regime, $\Omega_j\approx\Omega$ and defining 
$G=2 \gpi \lambda_\Omega^2N\langle\sigma_z\rangle_\beta$ we get the 
relations given in Eq.~\eqref{eq:transmission-line}. When the AFM 
correlations grow below $T_N$, 
the resonant frequency $\Omega_j$ of each DPPH chain depends on its 
orientation $\psi$ with respect to the magnetic field. Then, we replace the 
sum for an integral over solid angle (since our sample is a powder pellet, 
all orientations are present), which leads to Eq.~\eqref{t-sw}. Moreover, 
since spin excitations are then of a bosonic nature, the temperature 
dependence is lost and $\langle\sigma_z\rangle(T) \approx \langle\sigma_z\rangle(T_{\rm N})$ for $T\leq T_{\rm N}$.

\section*{Data availability} 
All experimental and simulation data that support the findings of 
this study will be deposited in the EU Open Research Community within the Zenodo open repository with accession code 10.5281/zenodo.15854400. 

\section*{References}
\bibliography{superra}

\begin{thebibliography}{18}%
\makeatletter
\providecommand \@ifxundefined [1]{%
 \@ifx{#1\undefined}
}%
\providecommand \@ifnum [1]{%
 \ifnum #1\expandafter \@firstoftwo
 \else \expandafter \@secondoftwo
 \fi
}%
\providecommand \@ifx [1]{%
 \ifx #1\expandafter \@firstoftwo
 \else \expandafter \@secondoftwo
 \fi
}%
\providecommand \natexlab [1]{#1}%
\providecommand \enquote  [1]{``#1''}%
\providecommand \bibnamefont  [1]{#1}%
\providecommand \bibfnamefont [1]{#1}%
\providecommand \citenamefont [1]{#1}%
\providecommand \href@noop [0]{\@secondoftwo}%
\providecommand \href [0]{\begingroup \@sanitize@url \@href}%
\providecommand \@href[1]{\@@startlink{#1}\@@href}%
\providecommand \@@href[1]{\endgroup#1\@@endlink}%
\providecommand \@sanitize@url [0]{\catcode `\\12\catcode `\$12\catcode
  `\&12\catcode `\#12\catcode `\^12\catcode `\_12\catcode `\%12\relax}%
\providecommand \@@startlink[1]{}%
\providecommand \@@endlink[0]{}%
\providecommand \url  [0]{\begingroup\@sanitize@url \@url }%
\providecommand \@url [1]{\endgroup\@href {#1}{\urlprefix }}%
\providecommand \urlprefix  [0]{URL }%
\providecommand \Eprint [0]{\href }%
\providecommand \doibase [0]{http://dx.doi.org/}%
\providecommand \selectlanguage [0]{\@gobble}%
\providecommand \bibinfo  [0]{\@secondoftwo}%
\providecommand \bibfield  [0]{\@secondoftwo}%
\providecommand \translation [1]{[#1]}%
\providecommand \BibitemOpen [0]{}%
\providecommand \bibitemStop [0]{}%
\providecommand \bibitemNoStop [0]{.\EOS\space}%
\providecommand \EOS [0]{\spacefactor3000\relax}%
\providecommand \BibitemShut  [1]{\csname bibitem#1\endcsname}%
\let\auto@bib@innerbib\@empty
\bibitem [{\citenamefont {{\v{Z}}ili{\'{c}}}\ \emph {et~al.}(2010)\citenamefont
  {{\v{Z}}ili{\'{c}}}, \citenamefont {Paji{\'{c}}}, \citenamefont
  {Juri{\'{c}}}, \citenamefont {Mol{\v{c}}anov}, \citenamefont {Rakvin},
  \citenamefont {Planini{\'{c}}},\ and\ \citenamefont {Zadro}}]{Zilic2010}%
  \BibitemOpen
  \bibfield  {author} {\bibinfo {author} {\bibfnamefont {D.}~\bibnamefont
  {{\v{Z}}ili{\'{c}}}}, \bibinfo {author} {\bibfnamefont {D.}~\bibnamefont
  {Paji{\'{c}}}}, \bibinfo {author} {\bibfnamefont {M.}~\bibnamefont
  {Juri{\'{c}}}}, \bibinfo {author} {\bibfnamefont {K.}~\bibnamefont
  {Mol{\v{c}}anov}}, \bibinfo {author} {\bibfnamefont {B.}~\bibnamefont
  {Rakvin}}, \bibinfo {author} {\bibfnamefont {P.}~\bibnamefont
  {Planini{\'{c}}}}, \ and\ \bibinfo {author} {\bibfnamefont {K.}~\bibnamefont
  {Zadro}},\ }\href {\doibase 10.1016/j.jmr.2010.08.005} {\bibfield  {journal}
  {\bibinfo  {journal} {J. Magn. Reson.}\ }\textbf {\bibinfo {volume} {207}},\
  \bibinfo {pages} {34} (\bibinfo {year} {2010})}\BibitemShut {NoStop}%
\bibitem [{\citenamefont {Ohya-Nishiguchi}(1979)}]{OhyaNishiguchi1979}%
  \BibitemOpen
  \bibfield  {author} {\bibinfo {author} {\bibfnamefont {H.}~\bibnamefont
  {Ohya-Nishiguchi}},\ }\href {\doibase 10.1246/bcsj.52.3480} {\bibfield
  {journal} {\bibinfo  {journal} {Bull. Chem. Soc. Jpn.}\ }\textbf {\bibinfo
  {volume} {52}},\ \bibinfo {pages} {3480} (\bibinfo {year}
  {1979})}\BibitemShut {NoStop}%
\bibitem [{\citenamefont {Fujito}(1981)}]{Fujito1981}%
  \BibitemOpen
  \bibfield  {author} {\bibinfo {author} {\bibfnamefont {T.}~\bibnamefont
  {Fujito}},\ }\href {\doibase 10.1246/bcsj.54.3110} {\bibfield  {journal}
  {\bibinfo  {journal} {Bull. Chem. Soc. Jpn.}\ }\textbf {\bibinfo {volume}
  {54}},\ \bibinfo {pages} {3110} (\bibinfo {year} {1981})}\BibitemShut
  {NoStop}%
\bibitem [{\citenamefont {Blöte}(1975)}]{Blote1975}%
  \BibitemOpen
  \bibfield  {author} {\bibinfo {author} {\bibfnamefont {H.}~\bibnamefont
  {Blöte}},\ }\href {\doibase https://doi.org/10.1016/0378-4363(75)90001-7}
  {\bibfield  {journal} {\bibinfo  {journal} {Physica B+C}\ }\textbf {\bibinfo
  {volume} {79}},\ \bibinfo {pages} {427} (\bibinfo {year} {1975})}\BibitemShut
  {NoStop}%
\bibitem [{\citenamefont {Vleck}(1948)}]{VanVleck1948}%
  \BibitemOpen
  \bibfield  {author} {\bibinfo {author} {\bibfnamefont {J.~H.~V.}\
  \bibnamefont {Vleck}},\ }\href {\doibase 10.1103/physrev.74.1168} {\bibfield
  {journal} {\bibinfo  {journal} {Phys. Rev.}\ }\textbf {\bibinfo {volume}
  {74}},\ \bibinfo {pages} {1168} (\bibinfo {year} {1948})}\BibitemShut
  {NoStop}%
\bibitem [{\citenamefont {Anderson}\ and\ \citenamefont
  {Weiss}(1953)}]{Anderson1953}%
  \BibitemOpen
  \bibfield  {author} {\bibinfo {author} {\bibfnamefont {P.~W.}\ \bibnamefont
  {Anderson}}\ and\ \bibinfo {author} {\bibfnamefont {P.~R.}\ \bibnamefont
  {Weiss}},\ }\href {\doibase 10.1103/RevModPhys.25.269} {\bibfield  {journal}
  {\bibinfo  {journal} {Rev. Mod. Phys.}\ }\textbf {\bibinfo {volume} {25}},\
  \bibinfo {pages} {269} (\bibinfo {year} {1953})}\BibitemShut {NoStop}%
\bibitem [{\citenamefont {Yordanov}(1996)}]{Yordanov1996}%
  \BibitemOpen
  \bibfield  {author} {\bibinfo {author} {\bibfnamefont {N.~D.}\ \bibnamefont
  {Yordanov}},\ }\href {\doibase 10.1007/BF03163117} {\bibfield  {journal}
  {\bibinfo  {journal} {Appl. Magn. Reson.}\ }\textbf {\bibinfo {volume}
  {10}},\ \bibinfo {pages} {339} (\bibinfo {year} {1996})}\BibitemShut
  {NoStop}%
\bibitem [{\citenamefont {Clauss}\ \emph {et~al.}(2013)\citenamefont {Clauss},
  \citenamefont {Bothner}, \citenamefont {Koelle}, \citenamefont {Kleiner},
  \citenamefont {Bogani}, \citenamefont {Scheffler},\ and\ \citenamefont
  {Dressel}}]{Clauss2013}%
  \BibitemOpen
  \bibfield  {author} {\bibinfo {author} {\bibfnamefont {C.}~\bibnamefont
  {Clauss}}, \bibinfo {author} {\bibfnamefont {D.}~\bibnamefont {Bothner}},
  \bibinfo {author} {\bibfnamefont {D.}~\bibnamefont {Koelle}}, \bibinfo
  {author} {\bibfnamefont {R.}~\bibnamefont {Kleiner}}, \bibinfo {author}
  {\bibfnamefont {L.}~\bibnamefont {Bogani}}, \bibinfo {author} {\bibfnamefont
  {M.}~\bibnamefont {Scheffler}}, \ and\ \bibinfo {author} {\bibfnamefont
  {M.}~\bibnamefont {Dressel}},\ }\href {\doibase 10.1063/1.4802956} {\bibfield
   {journal} {\bibinfo  {journal} {Appl. Phys. Lett.}\ }\textbf {\bibinfo
  {volume} {102}},\ \bibinfo {pages} {162601} (\bibinfo {year}
  {2013})}\BibitemShut {NoStop}%
\bibitem [{\citenamefont {Gimeno}\ \emph {et~al.}(2023)\citenamefont {Gimeno},
  \citenamefont {Rollano}, \citenamefont {Zueco}, \citenamefont {Duan},
  \citenamefont {de~Ory}, \citenamefont {Gomez}, \citenamefont {Gaita-Ari\~no},
  \citenamefont {S\'anchez-Azqueta}, \citenamefont {Astner}, \citenamefont
  {Granados}, \citenamefont {Hill}, \citenamefont {Majer}, \citenamefont
  {Coronado},\ and\ \citenamefont {Luis}}]{Gimeno2023}%
  \BibitemOpen
  \bibfield  {author} {\bibinfo {author} {\bibfnamefont {I.}~\bibnamefont
  {Gimeno}}, \bibinfo {author} {\bibfnamefont {V.}~\bibnamefont {Rollano}},
  \bibinfo {author} {\bibfnamefont {D.}~\bibnamefont {Zueco}}, \bibinfo
  {author} {\bibfnamefont {Y.}~\bibnamefont {Duan}}, \bibinfo {author}
  {\bibfnamefont {M.~C.}\ \bibnamefont {de~Ory}}, \bibinfo {author}
  {\bibfnamefont {A.}~\bibnamefont {Gomez}}, \bibinfo {author} {\bibfnamefont
  {A.}~\bibnamefont {Gaita-Ari\~no}}, \bibinfo {author} {\bibfnamefont
  {C.}~\bibnamefont {S\'anchez-Azqueta}}, \bibinfo {author} {\bibfnamefont
  {T.}~\bibnamefont {Astner}}, \bibinfo {author} {\bibfnamefont
  {D.}~\bibnamefont {Granados}}, \bibinfo {author} {\bibfnamefont
  {S.}~\bibnamefont {Hill}}, \bibinfo {author} {\bibfnamefont {J.}~\bibnamefont
  {Majer}}, \bibinfo {author} {\bibfnamefont {E.}~\bibnamefont {Coronado}}, \
  and\ \bibinfo {author} {\bibfnamefont {F.}~\bibnamefont {Luis}},\ }\href
  {\doibase 10.1103/PhysRevApplied.20.044070} {\bibfield  {journal} {\bibinfo
  {journal} {Phys. Rev. Appl.}\ }\textbf {\bibinfo {volume} {20}},\ \bibinfo
  {pages} {044070} (\bibinfo {year} {2023})}\BibitemShut {NoStop}%
\bibitem [{\citenamefont {Gimeno}\ \emph {et~al.}(2020)\citenamefont {Gimeno},
  \citenamefont {Kersten}, \citenamefont {Pallar{\'{e}}s}, \citenamefont
  {Hermosilla}, \citenamefont {Mart{\'{\i}}nez-P{\'{e}}rez}, \citenamefont
  {Jenkins}, \citenamefont {Angerer}, \citenamefont {S{\'{a}}nchez-Azqueta},
  \citenamefont {Zueco}, \citenamefont {Majer}, \citenamefont {Lostao},\ and\
  \citenamefont {Luis}}]{Gimeno2020}%
  \BibitemOpen
  \bibfield  {author} {\bibinfo {author} {\bibfnamefont {I.}~\bibnamefont
  {Gimeno}}, \bibinfo {author} {\bibfnamefont {W.}~\bibnamefont {Kersten}},
  \bibinfo {author} {\bibfnamefont {M.~C.}\ \bibnamefont {Pallar{\'{e}}s}},
  \bibinfo {author} {\bibfnamefont {P.}~\bibnamefont {Hermosilla}}, \bibinfo
  {author} {\bibfnamefont {M.~J.}\ \bibnamefont {Mart{\'{\i}}nez-P{\'{e}}rez}},
  \bibinfo {author} {\bibfnamefont {M.~D.}\ \bibnamefont {Jenkins}}, \bibinfo
  {author} {\bibfnamefont {A.}~\bibnamefont {Angerer}}, \bibinfo {author}
  {\bibfnamefont {C.}~\bibnamefont {S{\'{a}}nchez-Azqueta}}, \bibinfo {author}
  {\bibfnamefont {D.}~\bibnamefont {Zueco}}, \bibinfo {author} {\bibfnamefont
  {J.}~\bibnamefont {Majer}}, \bibinfo {author} {\bibfnamefont
  {A.}~\bibnamefont {Lostao}}, \ and\ \bibinfo {author} {\bibfnamefont
  {F.}~\bibnamefont {Luis}},\ }\href {\doibase 10.1021/acsnano.0c03167}
  {\bibfield  {journal} {\bibinfo  {journal} {{ACS} Nano}\ }\textbf {\bibinfo
  {volume} {14}},\ \bibinfo {pages} {8707} (\bibinfo {year}
  {2020})}\BibitemShut {NoStop}%
\bibitem [{\citenamefont {Sheremet}\ \emph {et~al.}(2023)\citenamefont
  {Sheremet}, \citenamefont {Petrov}, \citenamefont {Iorsh}, \citenamefont
  {Poshakinskiy},\ and\ \citenamefont {Poddubny}}]{sheremet2023waveguide}%
  \BibitemOpen
  \bibfield  {author} {\bibinfo {author} {\bibfnamefont {A.~S.}\ \bibnamefont
  {Sheremet}}, \bibinfo {author} {\bibfnamefont {M.~I.}\ \bibnamefont
  {Petrov}}, \bibinfo {author} {\bibfnamefont {I.~V.}\ \bibnamefont {Iorsh}},
  \bibinfo {author} {\bibfnamefont {A.~V.}\ \bibnamefont {Poshakinskiy}}, \
  and\ \bibinfo {author} {\bibfnamefont {A.~N.}\ \bibnamefont {Poddubny}},\
  }\href {\doibase 10.1103/RevModPhys.95.015002} {\bibfield  {journal}
  {\bibinfo  {journal} {Rev. Mod. Phys.}\ }\textbf {\bibinfo {volume} {95}},\
  \bibinfo {pages} {015002} (\bibinfo {year} {2023})}\BibitemShut {NoStop}%
\bibitem [{\citenamefont {H\"ummer}\ \emph {et~al.}(2013)\citenamefont
  {H\"ummer}, \citenamefont {Garc\'{\i}a-Vidal}, \citenamefont
  {Mart\'{\i}n-Moreno},\ and\ \citenamefont {Zueco}}]{Hummer2013}%
  \BibitemOpen
  \bibfield  {author} {\bibinfo {author} {\bibfnamefont {T.}~\bibnamefont
  {H\"ummer}}, \bibinfo {author} {\bibfnamefont {F.~J.}\ \bibnamefont
  {Garc\'{\i}a-Vidal}}, \bibinfo {author} {\bibfnamefont {L.}~\bibnamefont
  {Mart\'{\i}n-Moreno}}, \ and\ \bibinfo {author} {\bibfnamefont
  {D.}~\bibnamefont {Zueco}},\ }\href {\doibase 10.1103/PhysRevB.87.115419}
  {\bibfield  {journal} {\bibinfo  {journal} {Phys. Rev. B}\ }\textbf {\bibinfo
  {volume} {87}},\ \bibinfo {pages} {115419} (\bibinfo {year}
  {2013})}\BibitemShut {NoStop}%
\bibitem [{\citenamefont {Ghirri}\ \emph {et~al.}(2016)\citenamefont {Ghirri},
  \citenamefont {Bonizzoni}, \citenamefont {Troiani}, \citenamefont {Buccheri},
  \citenamefont {Beverina}, \citenamefont {Cassinese},\ and\ \citenamefont
  {Affronte}}]{Ghirri2016}%
  \BibitemOpen
  \bibfield  {author} {\bibinfo {author} {\bibfnamefont {A.}~\bibnamefont
  {Ghirri}}, \bibinfo {author} {\bibfnamefont {C.}~\bibnamefont {Bonizzoni}},
  \bibinfo {author} {\bibfnamefont {F.}~\bibnamefont {Troiani}}, \bibinfo
  {author} {\bibfnamefont {N.}~\bibnamefont {Buccheri}}, \bibinfo {author}
  {\bibfnamefont {L.}~\bibnamefont {Beverina}}, \bibinfo {author}
  {\bibfnamefont {A.}~\bibnamefont {Cassinese}}, \ and\ \bibinfo {author}
  {\bibfnamefont {M.}~\bibnamefont {Affronte}},\ }\href {\doibase
  10.1103/PhysRevA.93.063855} {\bibfield  {journal} {\bibinfo  {journal} {Phys.
  Rev. A}\ }\textbf {\bibinfo {volume} {93}},\ \bibinfo {pages} {063855}
  (\bibinfo {year} {2016})}\BibitemShut {NoStop}%
\bibitem [{\citenamefont {Mergenthaler}\ \emph {et~al.}(2017)\citenamefont
  {Mergenthaler}, \citenamefont {Liu}, \citenamefont {Le~Roy}, \citenamefont
  {Ares}, \citenamefont {Thompson}, \citenamefont {Bogani}, \citenamefont
  {Luis}, \citenamefont {Blundell}, \citenamefont {Lancaster}, \citenamefont
  {Ardavan}, \citenamefont {Briggs}, \citenamefont {Leek},\ and\ \citenamefont
  {Laird}}]{Mergenthaler2017}%
  \BibitemOpen
  \bibfield  {author} {\bibinfo {author} {\bibfnamefont {M.}~\bibnamefont
  {Mergenthaler}}, \bibinfo {author} {\bibfnamefont {J.}~\bibnamefont {Liu}},
  \bibinfo {author} {\bibfnamefont {J.~J.}\ \bibnamefont {Le~Roy}}, \bibinfo
  {author} {\bibfnamefont {N.}~\bibnamefont {Ares}}, \bibinfo {author}
  {\bibfnamefont {A.~L.}\ \bibnamefont {Thompson}}, \bibinfo {author}
  {\bibfnamefont {L.}~\bibnamefont {Bogani}}, \bibinfo {author} {\bibfnamefont
  {F.}~\bibnamefont {Luis}}, \bibinfo {author} {\bibfnamefont {S.~J.}\
  \bibnamefont {Blundell}}, \bibinfo {author} {\bibfnamefont {T.}~\bibnamefont
  {Lancaster}}, \bibinfo {author} {\bibfnamefont {A.}~\bibnamefont {Ardavan}},
  \bibinfo {author} {\bibfnamefont {G.~A.~D.}\ \bibnamefont {Briggs}}, \bibinfo
  {author} {\bibfnamefont {P.~J.}\ \bibnamefont {Leek}}, \ and\ \bibinfo
  {author} {\bibfnamefont {E.~A.}\ \bibnamefont {Laird}},\ }\href
  {http://dx.doi.org/10.1103/PhysRevLett.119.147701} {\bibfield  {journal}
  {\bibinfo  {journal} {Phys. Rev. Lett.}\ }\textbf {\bibinfo {volume} {119}},\
  \bibinfo {pages} {147701} (\bibinfo {year} {2017})}\BibitemShut {NoStop}%
\bibitem [{\citenamefont {Garcia-Palacios}\ \emph {et~al.}(2009)\citenamefont
  {Garcia-Palacios}, \citenamefont {Gong},\ and\ \citenamefont
  {Luis}}]{Garcia2009}%
  \BibitemOpen
  \bibfield  {author} {\bibinfo {author} {\bibfnamefont {J.}~\bibnamefont
  {Garcia-Palacios}}, \bibinfo {author} {\bibfnamefont {J.}~\bibnamefont
  {Gong}}, \ and\ \bibinfo {author} {\bibfnamefont {F.}~\bibnamefont {Luis}},\
  }\href {\doibase 10.1088/0953-8984/21/45/456006} {\bibfield  {journal}
  {\bibinfo  {journal} {J. Phys. Condens. Matter}\ }\textbf {\bibinfo {volume}
  {21}},\ \bibinfo {pages} {456006} (\bibinfo {year} {2009})}\BibitemShut
  {NoStop}%
\bibitem [{\citenamefont {Vonsovskii}(1966)}]{Vonsovskii}%
  \BibitemOpen
  \bibfield  {author} {\bibinfo {author} {\bibfnamefont {S.~V.}\ \bibnamefont
  {Vonsovskii}},\ }\href@noop {} {\emph {\bibinfo {title} {Ferromagnetic
  resonance}}}\ (\bibinfo  {publisher} {Pergamon Press},\ \bibinfo {year}
  {1966})\BibitemShut {NoStop}%
\bibitem [{\citenamefont {Ross}(2013)}]{Ross2013}%
  \BibitemOpen
  \bibfield  {author} {\bibinfo {author} {\bibfnamefont {M.~P.}\ \bibnamefont
  {Ross}},\ }\emph {\bibinfo {title} {Spin dynamics in an antiferromagnet}},\
  \href@noop {} {\bibinfo {type} {Diploma thesis}},\ \bibinfo  {school}
  {Technische Universität München} (\bibinfo {year} {2013})\BibitemShut
  {NoStop}%
\bibitem [{\citenamefont {Wang}\ and\ \citenamefont {Xiao}(2024)}]{Wang2024}%
  \BibitemOpen
  \bibfield  {author} {\bibinfo {author} {\bibfnamefont {Y.}~\bibnamefont
  {Wang}}\ and\ \bibinfo {author} {\bibfnamefont {J.}~\bibnamefont {Xiao}},\
  }\href {\doibase 10.1103/PhysRevB.110.134409} {\bibfield  {journal} {\bibinfo
   {journal} {Phys. Rev. B}\ }\textbf {\bibinfo {volume} {110}},\ \bibinfo
  {pages} {134409} (\bibinfo {year} {2024})}\BibitemShut {NoStop}%
\end{thebibliography}%


\begin{thebibliography}{55}%
\makeatletter
\providecommand \@ifxundefined [1]{%
 \@ifx{#1\undefined}
}%
\providecommand \@ifnum [1]{%
 \ifnum #1\expandafter \@firstoftwo
 \else \expandafter \@secondoftwo
 \fi
}%
\providecommand \@ifx [1]{%
 \ifx #1\expandafter \@firstoftwo
 \else \expandafter \@secondoftwo
 \fi
}%
\providecommand \natexlab [1]{#1}%
\providecommand \enquote  [1]{``#1''}%
\providecommand \bibnamefont  [1]{#1}%
\providecommand \bibfnamefont [1]{#1}%
\providecommand \citenamefont [1]{#1}%
\providecommand \href@noop [0]{\@secondoftwo}%
\providecommand \href [0]{\begingroup \@sanitize@url \@href}%
\providecommand \@href[1]{\@@startlink{#1}\@@href}%
\providecommand \@@href[1]{\endgroup#1\@@endlink}%
\providecommand \@sanitize@url [0]{\catcode `\\12\catcode `\$12\catcode
  `\&12\catcode `\#12\catcode `\^12\catcode `\_12\catcode `\%12\relax}%
\providecommand \@@startlink[1]{}%
\providecommand \@@endlink[0]{}%
\providecommand \url  [0]{\begingroup\@sanitize@url \@url }%
\providecommand \@url [1]{\endgroup\@href {#1}{\urlprefix }}%
\providecommand \urlprefix  [0]{URL }%
\providecommand \Eprint [0]{\href }%
\providecommand \doibase [0]{https://doi.org/}%
\providecommand \selectlanguage [0]{\@gobble}%
\providecommand \bibinfo  [0]{\@secondoftwo}%
\providecommand \bibfield  [0]{\@secondoftwo}%
\providecommand \translation [1]{[#1]}%
\providecommand \BibitemOpen [0]{}%
\providecommand \bibitemStop [0]{}%
\providecommand \bibitemNoStop [0]{.\EOS\space}%
\providecommand \EOS [0]{\spacefactor3000\relax}%
\providecommand \BibitemShut  [1]{\csname bibitem#1\endcsname}%
\let\auto@bib@innerbib\@empty
\bibitem [{\citenamefont {Lodahl}\ \emph {et~al.}(2015)\citenamefont {Lodahl},
  \citenamefont {Mahmoodian},\ and\ \citenamefont
  {Stobbe}}]{lodahl2015interfacing}%
  \BibitemOpen
  \bibfield  {author} {\bibinfo {author} {\bibfnamefont {P.}~\bibnamefont
  {Lodahl}}, \bibinfo {author} {\bibfnamefont {S.}~\bibnamefont {Mahmoodian}},\
  and\ \bibinfo {author} {\bibfnamefont {S.}~\bibnamefont {Stobbe}},\
  }\bibfield  {title} {\bibinfo {title} {Interfacing single photons and single
  quantum dots with photonic nanostructures},\ }\href
  {https://doi.org/10.1103/revmodphys.87.347} {\bibfield  {journal} {\bibinfo
  {journal} {Rev. Mod. Phys.}\ }\textbf {\bibinfo {volume} {87}},\ \bibinfo
  {pages} {347} (\bibinfo {year} {2015})}\BibitemShut {NoStop}%
\bibitem [{\citenamefont {Roy}\ \emph {et~al.}(2017)\citenamefont {Roy},
  \citenamefont {Wilson},\ and\ \citenamefont
  {Firstenberg}}]{roy2017colloquium}%
  \BibitemOpen
  \bibfield  {author} {\bibinfo {author} {\bibfnamefont {D.}~\bibnamefont
  {Roy}}, \bibinfo {author} {\bibfnamefont {C.~M.}\ \bibnamefont {Wilson}},\
  and\ \bibinfo {author} {\bibfnamefont {O.}~\bibnamefont {Firstenberg}},\
  }\bibfield  {title} {\bibinfo {title} {Colloquium: Strongly interacting
  photons in one-dimensional continuum},\ }\href
  {https://doi.org/10.1103/RevModPhys.89.021001} {\bibfield  {journal}
  {\bibinfo  {journal} {Rev. Mod. Phys.}\ }\textbf {\bibinfo {volume} {89}},\
  \bibinfo {pages} {021001} (\bibinfo {year} {2017})}\BibitemShut {NoStop}%
\bibitem [{\citenamefont {Sheremet}\ \emph {et~al.}(2023)\citenamefont
  {Sheremet}, \citenamefont {Petrov}, \citenamefont {Iorsh}, \citenamefont
  {Poshakinskiy},\ and\ \citenamefont {Poddubny}}]{sheremet2023waveguide}%
  \BibitemOpen
  \bibfield  {author} {\bibinfo {author} {\bibfnamefont {A.~S.}\ \bibnamefont
  {Sheremet}}, \bibinfo {author} {\bibfnamefont {M.~I.}\ \bibnamefont
  {Petrov}}, \bibinfo {author} {\bibfnamefont {I.~V.}\ \bibnamefont {Iorsh}},
  \bibinfo {author} {\bibfnamefont {A.~V.}\ \bibnamefont {Poshakinskiy}},\ and\
  \bibinfo {author} {\bibfnamefont {A.~N.}\ \bibnamefont {Poddubny}},\
  }\bibfield  {title} {\bibinfo {title} {Waveguide quantum electrodynamics:
  Collective radiance and photon-photon correlations},\ }\href
  {https://doi.org/10.1103/RevModPhys.95.015002} {\bibfield  {journal}
  {\bibinfo  {journal} {Rev. Mod. Phys.}\ }\textbf {\bibinfo {volume} {95}},\
  \bibinfo {pages} {015002} (\bibinfo {year} {2023})}\BibitemShut {NoStop}%
\bibitem [{\citenamefont {Astafiev}\ \emph {et~al.}(2010)\citenamefont
  {Astafiev}, \citenamefont {Zagoskin}, \citenamefont {Abdumalikov},
  \citenamefont {Pashkin}, \citenamefont {Yamamoto}, \citenamefont {Inomata},
  \citenamefont {Nakamura},\ and\ \citenamefont {Tsai}}]{Astafiev2010}%
  \BibitemOpen
  \bibfield  {author} {\bibinfo {author} {\bibfnamefont {O.}~\bibnamefont
  {Astafiev}}, \bibinfo {author} {\bibfnamefont {A.~M.}\ \bibnamefont
  {Zagoskin}}, \bibinfo {author} {\bibfnamefont {A.~A.}\ \bibnamefont
  {Abdumalikov}}, \bibinfo {author} {\bibfnamefont {Y.~A.}\ \bibnamefont
  {Pashkin}}, \bibinfo {author} {\bibfnamefont {T.}~\bibnamefont {Yamamoto}},
  \bibinfo {author} {\bibfnamefont {K.}~\bibnamefont {Inomata}}, \bibinfo
  {author} {\bibfnamefont {Y.}~\bibnamefont {Nakamura}},\ and\ \bibinfo
  {author} {\bibfnamefont {J.~S.}\ \bibnamefont {Tsai}},\ }\bibfield  {title}
  {\bibinfo {title} {Resonance fluorescence of a single artificial atom},\
  }\href {https://doi.org/10.1126/science.1181918} {\bibfield  {journal}
  {\bibinfo  {journal} {Science}\ }\textbf {\bibinfo {volume} {327}},\ \bibinfo
  {pages} {840} (\bibinfo {year} {2010})}\BibitemShut {NoStop}%
\bibitem [{\citenamefont {van Loo}\ \emph {et~al.}(2013)\citenamefont {van
  Loo}, \citenamefont {Fedorov}, \citenamefont {Lalumiere}, \citenamefont
  {Sanders}, \citenamefont {Blais},\ and\ \citenamefont
  {Wallraff}}]{vanLoo2013}%
  \BibitemOpen
  \bibfield  {author} {\bibinfo {author} {\bibfnamefont {A.~F.}\ \bibnamefont
  {van Loo}}, \bibinfo {author} {\bibfnamefont {A.}~\bibnamefont {Fedorov}},
  \bibinfo {author} {\bibfnamefont {K.}~\bibnamefont {Lalumiere}}, \bibinfo
  {author} {\bibfnamefont {B.~C.}\ \bibnamefont {Sanders}}, \bibinfo {author}
  {\bibfnamefont {A.}~\bibnamefont {Blais}},\ and\ \bibinfo {author}
  {\bibfnamefont {A.}~\bibnamefont {Wallraff}},\ }\bibfield  {title} {\bibinfo
  {title} {Photon-mediated interactions between distant artificial atoms},\
  }\href {https://doi.org/10.1126/science.1244324} {\bibfield  {journal}
  {\bibinfo  {journal} {Science}\ }\textbf {\bibinfo {volume} {342}},\ \bibinfo
  {pages} {1494} (\bibinfo {year} {2013})}\BibitemShut {NoStop}%
\bibitem [{\citenamefont {Javadi}\ \emph {et~al.}(2015)\citenamefont {Javadi},
  \citenamefont {S\"{o}llner}, \citenamefont {Arcari}, \citenamefont {Hansen},
  \citenamefont {Midolo}, \citenamefont {Mahmoodian}, \citenamefont
  {Kir{\v{s}}ansk{\.{e}}}, \citenamefont {Pregnolato}, \citenamefont {Lee},
  \citenamefont {Song}, \citenamefont {Stobbe},\ and\ \citenamefont
  {Lodahl}}]{Javadi2015}%
  \BibitemOpen
  \bibfield  {author} {\bibinfo {author} {\bibfnamefont {A.}~\bibnamefont
  {Javadi}}, \bibinfo {author} {\bibfnamefont {I.}~\bibnamefont {S\"{o}llner}},
  \bibinfo {author} {\bibfnamefont {M.}~\bibnamefont {Arcari}}, \bibinfo
  {author} {\bibfnamefont {S.~L.}\ \bibnamefont {Hansen}}, \bibinfo {author}
  {\bibfnamefont {L.}~\bibnamefont {Midolo}}, \bibinfo {author} {\bibfnamefont
  {S.}~\bibnamefont {Mahmoodian}}, \bibinfo {author} {\bibfnamefont
  {G.}~\bibnamefont {Kir{\v{s}}ansk{\.{e}}}}, \bibinfo {author} {\bibfnamefont
  {T.}~\bibnamefont {Pregnolato}}, \bibinfo {author} {\bibfnamefont {E.~H.}\
  \bibnamefont {Lee}}, \bibinfo {author} {\bibfnamefont {J.~D.}\ \bibnamefont
  {Song}}, \bibinfo {author} {\bibfnamefont {S.}~\bibnamefont {Stobbe}},\ and\
  \bibinfo {author} {\bibfnamefont {P.}~\bibnamefont {Lodahl}},\ }\bibfield
  {title} {\bibinfo {title} {Single-photon non-linear optics with a quantum dot
  in a waveguide},\ }\href {https://doi.org/10.1038/ncomms9655} {\bibfield
  {journal} {\bibinfo  {journal} {Nat. Commun.}\ }\textbf {\bibinfo {volume}
  {6}},\ \bibinfo {pages} {8655} (\bibinfo {year} {2015})}\BibitemShut
  {NoStop}%
\bibitem [{\citenamefont {Ara\'ujo}\ \emph {et~al.}(2016)\citenamefont
  {Ara\'ujo}, \citenamefont {Kre\ifmmode \check{s}\else
  \v{s}\fi{}i\ifmmode~\acute{c}\else \'{c}\fi{}}, \citenamefont {Kaiser},\ and\
  \citenamefont {Guerin}}]{araujo2016}%
  \BibitemOpen
  \bibfield  {author} {\bibinfo {author} {\bibfnamefont {M.~O.}\ \bibnamefont
  {Ara\'ujo}}, \bibinfo {author} {\bibfnamefont {I.}~\bibnamefont {Kre\ifmmode
  \check{s}\else \v{s}\fi{}i\ifmmode~\acute{c}\else \'{c}\fi{}}}, \bibinfo
  {author} {\bibfnamefont {R.}~\bibnamefont {Kaiser}},\ and\ \bibinfo {author}
  {\bibfnamefont {W.}~\bibnamefont {Guerin}},\ }\bibfield  {title} {\bibinfo
  {title} {Superradiance in a large and dilute cloud of cold atoms in the
  linear-optics regime},\ }\href
  {https://doi.org/10.1103/PhysRevLett.117.073002} {\bibfield  {journal}
  {\bibinfo  {journal} {Phys. Rev. Lett.}\ }\textbf {\bibinfo {volume} {117}},\
  \bibinfo {pages} {073002} (\bibinfo {year} {2016})}\BibitemShut {NoStop}%
\bibitem [{\citenamefont {Hood}\ \emph {et~al.}(2016)\citenamefont {Hood},
  \citenamefont {Goban}, \citenamefont {Asenjo-Garcia}, \citenamefont {Lu},
  \citenamefont {Yu}, \citenamefont {Chang},\ and\ \citenamefont
  {Kimble}}]{Hood2016}%
  \BibitemOpen
  \bibfield  {author} {\bibinfo {author} {\bibfnamefont {J.~D.}\ \bibnamefont
  {Hood}}, \bibinfo {author} {\bibfnamefont {A.}~\bibnamefont {Goban}},
  \bibinfo {author} {\bibfnamefont {A.}~\bibnamefont {Asenjo-Garcia}}, \bibinfo
  {author} {\bibfnamefont {M.}~\bibnamefont {Lu}}, \bibinfo {author}
  {\bibfnamefont {S.-P.}\ \bibnamefont {Yu}}, \bibinfo {author} {\bibfnamefont
  {D.~E.}\ \bibnamefont {Chang}},\ and\ \bibinfo {author} {\bibfnamefont
  {H.~J.}\ \bibnamefont {Kimble}},\ }\bibfield  {title} {\bibinfo {title}
  {Atom{\textendash}atom interactions around the band edge of a photonic
  crystal waveguide},\ }\href {https://doi.org/10.1073/pnas.1603788113}
  {\bibfield  {journal} {\bibinfo  {journal} {Proc. Natl. Acad. Sci. U.S.A.}\
  }\textbf {\bibinfo {volume} {113}},\ \bibinfo {pages} {10507} (\bibinfo
  {year} {2016})}\BibitemShut {NoStop}%
\bibitem [{\citenamefont {Corzo}\ \emph {et~al.}(2016)\citenamefont {Corzo},
  \citenamefont {Gouraud}, \citenamefont {Chandra}, \citenamefont {Goban},
  \citenamefont {Sheremet}, \citenamefont {Kupriyanov},\ and\ \citenamefont
  {Laurat}}]{corzo2016}%
  \BibitemOpen
  \bibfield  {author} {\bibinfo {author} {\bibfnamefont {N.~V.}\ \bibnamefont
  {Corzo}}, \bibinfo {author} {\bibfnamefont {B.}~\bibnamefont {Gouraud}},
  \bibinfo {author} {\bibfnamefont {A.}~\bibnamefont {Chandra}}, \bibinfo
  {author} {\bibfnamefont {A.}~\bibnamefont {Goban}}, \bibinfo {author}
  {\bibfnamefont {A.~S.}\ \bibnamefont {Sheremet}}, \bibinfo {author}
  {\bibfnamefont {D.~V.}\ \bibnamefont {Kupriyanov}},\ and\ \bibinfo {author}
  {\bibfnamefont {J.}~\bibnamefont {Laurat}},\ }\bibfield  {title} {\bibinfo
  {title} {Large bragg reflection from one-dimensional chains of trapped atoms
  near a nanoscale waveguide},\ }\href
  {https://doi.org/10.1103/PhysRevLett.117.133603} {\bibfield  {journal}
  {\bibinfo  {journal} {Phys. Rev. Lett.}\ }\textbf {\bibinfo {volume} {117}},\
  \bibinfo {pages} {133603} (\bibinfo {year} {2016})}\BibitemShut {NoStop}%
\bibitem [{\citenamefont {S\o{}rensen}\ \emph {et~al.}(2016)\citenamefont
  {S\o{}rensen}, \citenamefont {B\'eguin}, \citenamefont {Kluge}, \citenamefont
  {Iakoupov}, \citenamefont {S\o{}rensen}, \citenamefont {M\"uller},
  \citenamefont {Polzik},\ and\ \citenamefont {Appel}}]{sorensen2016}%
  \BibitemOpen
  \bibfield  {author} {\bibinfo {author} {\bibfnamefont {H.~L.}\ \bibnamefont
  {S\o{}rensen}}, \bibinfo {author} {\bibfnamefont {J.-B.}\ \bibnamefont
  {B\'eguin}}, \bibinfo {author} {\bibfnamefont {K.~W.}\ \bibnamefont {Kluge}},
  \bibinfo {author} {\bibfnamefont {I.}~\bibnamefont {Iakoupov}}, \bibinfo
  {author} {\bibfnamefont {A.~S.}\ \bibnamefont {S\o{}rensen}}, \bibinfo
  {author} {\bibfnamefont {J.~H.}\ \bibnamefont {M\"uller}}, \bibinfo {author}
  {\bibfnamefont {E.~S.}\ \bibnamefont {Polzik}},\ and\ \bibinfo {author}
  {\bibfnamefont {J.}~\bibnamefont {Appel}},\ }\bibfield  {title} {\bibinfo
  {title} {Coherent backscattering of light off one-dimensional atomic
  strings},\ }\href {https://doi.org/10.1103/PhysRevLett.117.133604} {\bibfield
   {journal} {\bibinfo  {journal} {Phys. Rev. Lett.}\ }\textbf {\bibinfo
  {volume} {117}},\ \bibinfo {pages} {133604} (\bibinfo {year}
  {2016})}\BibitemShut {NoStop}%
\bibitem [{\citenamefont {Türschmann}\ \emph {et~al.}(2017)\citenamefont
  {Türschmann}, \citenamefont {Rotenberg}, \citenamefont {Renger},
  \citenamefont {Harder}, \citenamefont {Lohse}, \citenamefont {Utikal},
  \citenamefont {Götzinger},\ and\ \citenamefont
  {Sandoghdar}}]{Turschmann2017}%
  \BibitemOpen
  \bibfield  {author} {\bibinfo {author} {\bibfnamefont {P.}~\bibnamefont
  {Türschmann}}, \bibinfo {author} {\bibfnamefont {N.}~\bibnamefont
  {Rotenberg}}, \bibinfo {author} {\bibfnamefont {J.}~\bibnamefont {Renger}},
  \bibinfo {author} {\bibfnamefont {I.}~\bibnamefont {Harder}}, \bibinfo
  {author} {\bibfnamefont {O.}~\bibnamefont {Lohse}}, \bibinfo {author}
  {\bibfnamefont {T.}~\bibnamefont {Utikal}}, \bibinfo {author} {\bibfnamefont
  {S.}~\bibnamefont {Götzinger}},\ and\ \bibinfo {author} {\bibfnamefont
  {V.}~\bibnamefont {Sandoghdar}},\ }\href {https://arxiv.org/abs/1702.05923}
  {\bibinfo {title} {On-chip linear and nonlinear control of single molecules
  coupled to a nanoguide}} (\bibinfo {year} {2017}),\ \Eprint
  {https://arxiv.org/abs/1702.05923} {arXiv:1702.05923 [quant-ph]} \BibitemShut
  {NoStop}%
\bibitem [{\citenamefont {Solano}\ \emph {et~al.}(2017)\citenamefont {Solano},
  \citenamefont {Barberis-Blostein}, \citenamefont {Fatemi}, \citenamefont
  {Orozco},\ and\ \citenamefont {Rolston}}]{Solano2017}%
  \BibitemOpen
  \bibfield  {author} {\bibinfo {author} {\bibfnamefont {P.}~\bibnamefont
  {Solano}}, \bibinfo {author} {\bibfnamefont {P.}~\bibnamefont
  {Barberis-Blostein}}, \bibinfo {author} {\bibfnamefont {F.~K.}\ \bibnamefont
  {Fatemi}}, \bibinfo {author} {\bibfnamefont {L.~A.}\ \bibnamefont {Orozco}},\
  and\ \bibinfo {author} {\bibfnamefont {S.~L.}\ \bibnamefont {Rolston}},\
  }\bibfield  {title} {\bibinfo {title} {Super-radiance reveals infinite-range
  dipole interactions through a nanofiber},\ }\href
  {https://doi.org/10.1038/s41467-017-01994-3} {\bibfield  {journal} {\bibinfo
  {journal} {Nat. Commun.}\ }\textbf {\bibinfo {volume} {8}},\ \bibinfo {pages}
  {1857} (\bibinfo {year} {2017})}\BibitemShut {NoStop}%
\bibitem [{\citenamefont {Goban}\ \emph {et~al.}(2015)\citenamefont {Goban},
  \citenamefont {Hung}, \citenamefont {Hood}, \citenamefont {Yu}, \citenamefont
  {Muniz}, \citenamefont {Painter},\ and\ \citenamefont {Kimble}}]{Goban2017}%
  \BibitemOpen
  \bibfield  {author} {\bibinfo {author} {\bibfnamefont {A.}~\bibnamefont
  {Goban}}, \bibinfo {author} {\bibfnamefont {C.-L.}\ \bibnamefont {Hung}},
  \bibinfo {author} {\bibfnamefont {J.~D.}\ \bibnamefont {Hood}}, \bibinfo
  {author} {\bibfnamefont {S.-P.}\ \bibnamefont {Yu}}, \bibinfo {author}
  {\bibfnamefont {J.~A.}\ \bibnamefont {Muniz}}, \bibinfo {author}
  {\bibfnamefont {O.}~\bibnamefont {Painter}},\ and\ \bibinfo {author}
  {\bibfnamefont {H.~J.}\ \bibnamefont {Kimble}},\ }\bibfield  {title}
  {\bibinfo {title} {Superradiance for atoms trapped along a photonic crystal
  waveguide},\ }\href {https://doi.org/10.1103/PhysRevLett.115.063601}
  {\bibfield  {journal} {\bibinfo  {journal} {Phys. Rev. Lett.}\ }\textbf
  {\bibinfo {volume} {115}},\ \bibinfo {pages} {063601} (\bibinfo {year}
  {2015})}\BibitemShut {NoStop}%
\bibitem [{\citenamefont {Corzo}\ \emph {et~al.}(2019)\citenamefont {Corzo},
  \citenamefont {Raskop}, \citenamefont {Chandra}, \citenamefont {Sheremet},
  \citenamefont {Gouraud},\ and\ \citenamefont {Laurat}}]{Corzo2019}%
  \BibitemOpen
  \bibfield  {author} {\bibinfo {author} {\bibfnamefont {N.~V.}\ \bibnamefont
  {Corzo}}, \bibinfo {author} {\bibfnamefont {J.}~\bibnamefont {Raskop}},
  \bibinfo {author} {\bibfnamefont {A.}~\bibnamefont {Chandra}}, \bibinfo
  {author} {\bibfnamefont {A.~S.}\ \bibnamefont {Sheremet}}, \bibinfo {author}
  {\bibfnamefont {B.}~\bibnamefont {Gouraud}},\ and\ \bibinfo {author}
  {\bibfnamefont {J.}~\bibnamefont {Laurat}},\ }\bibfield  {title} {\bibinfo
  {title} {Waveguide-coupled single collective excitation of atomic arrays},\
  }\href {https://doi.org/10.1038/s41586-019-0902-3} {\bibfield  {journal}
  {\bibinfo  {journal} {Nature}\ }\textbf {\bibinfo {volume} {566}},\ \bibinfo
  {pages} {359} (\bibinfo {year} {2019})}\BibitemShut {NoStop}%
\bibitem [{\citenamefont {Pennetta}\ \emph {et~al.}(2022)\citenamefont
  {Pennetta}, \citenamefont {Lechner}, \citenamefont {Blaha}, \citenamefont
  {Rauschenbeutel}, \citenamefont {Schneeweiss},\ and\ \citenamefont
  {Volz}}]{pennetta2022}%
  \BibitemOpen
  \bibfield  {author} {\bibinfo {author} {\bibfnamefont {R.}~\bibnamefont
  {Pennetta}}, \bibinfo {author} {\bibfnamefont {D.}~\bibnamefont {Lechner}},
  \bibinfo {author} {\bibfnamefont {M.}~\bibnamefont {Blaha}}, \bibinfo
  {author} {\bibfnamefont {A.}~\bibnamefont {Rauschenbeutel}}, \bibinfo
  {author} {\bibfnamefont {P.}~\bibnamefont {Schneeweiss}},\ and\ \bibinfo
  {author} {\bibfnamefont {J.}~\bibnamefont {Volz}},\ }\bibfield  {title}
  {\bibinfo {title} {Observation of coherent coupling between super- and
  subradiant states of an ensemble of cold atoms collectively coupled to a
  single propagating optical mode},\ }\href
  {https://doi.org/10.1103/PhysRevLett.128.203601} {\bibfield  {journal}
  {\bibinfo  {journal} {Phys. Rev. Lett.}\ }\textbf {\bibinfo {volume} {128}},\
  \bibinfo {pages} {203601} (\bibinfo {year} {2022})}\BibitemShut {NoStop}%
\bibitem [{\citenamefont {Glicenstein}\ \emph {et~al.}(2022)\citenamefont
  {Glicenstein}, \citenamefont {Ferioli}, \citenamefont {Browaeys},\ and\
  \citenamefont {Ferrier-Barbut}}]{Glicenstein2022}%
  \BibitemOpen
  \bibfield  {author} {\bibinfo {author} {\bibfnamefont {A.}~\bibnamefont
  {Glicenstein}}, \bibinfo {author} {\bibfnamefont {G.}~\bibnamefont
  {Ferioli}}, \bibinfo {author} {\bibfnamefont {A.}~\bibnamefont {Browaeys}},\
  and\ \bibinfo {author} {\bibfnamefont {I.}~\bibnamefont {Ferrier-Barbut}},\
  }\bibfield  {title} {\bibinfo {title} {From superradiance to subradiance:
  exploring the many-body dicke ladder},\ }\href
  {https://doi.org/10.1364/ol.451903} {\bibfield  {journal} {\bibinfo
  {journal} {Opt. Lett.}\ }\textbf {\bibinfo {volume} {47}},\ \bibinfo {pages}
  {1541} (\bibinfo {year} {2022})}\BibitemShut {NoStop}%
\bibitem [{\citenamefont {Tiranov}\ \emph {et~al.}(2023)\citenamefont
  {Tiranov}, \citenamefont {Angelopoulou}, \citenamefont {van Diepen},
  \citenamefont {Schrinski}, \citenamefont {Sandberg}, \citenamefont {Wang},
  \citenamefont {Midolo}, \citenamefont {Scholz}, \citenamefont {Wieck},
  \citenamefont {Ludwig}, \citenamefont {Sørensen},\ and\ \citenamefont
  {Lodahl}}]{Tiranov2023}%
  \BibitemOpen
  \bibfield  {author} {\bibinfo {author} {\bibfnamefont {A.}~\bibnamefont
  {Tiranov}}, \bibinfo {author} {\bibfnamefont {V.}~\bibnamefont
  {Angelopoulou}}, \bibinfo {author} {\bibfnamefont {C.~J.}\ \bibnamefont {van
  Diepen}}, \bibinfo {author} {\bibfnamefont {B.}~\bibnamefont {Schrinski}},
  \bibinfo {author} {\bibfnamefont {O.~A.~D.}\ \bibnamefont {Sandberg}},
  \bibinfo {author} {\bibfnamefont {Y.}~\bibnamefont {Wang}}, \bibinfo {author}
  {\bibfnamefont {L.}~\bibnamefont {Midolo}}, \bibinfo {author} {\bibfnamefont
  {S.}~\bibnamefont {Scholz}}, \bibinfo {author} {\bibfnamefont {A.~D.}\
  \bibnamefont {Wieck}}, \bibinfo {author} {\bibfnamefont {A.}~\bibnamefont
  {Ludwig}}, \bibinfo {author} {\bibfnamefont {A.~S.}\ \bibnamefont
  {Sørensen}},\ and\ \bibinfo {author} {\bibfnamefont {P.}~\bibnamefont
  {Lodahl}},\ }\bibfield  {title} {\bibinfo {title} {Collective super- and
  subradiant dynamics between distant optical quantum emitters},\ }\href
  {https://doi.org/10.1126/science.ade9324} {\bibfield  {journal} {\bibinfo
  {journal} {Science}\ }\textbf {\bibinfo {volume} {379}},\ \bibinfo {pages}
  {389} (\bibinfo {year} {2023})}\BibitemShut {NoStop}%
\bibitem [{\citenamefont {Vylegzhanin}\ \emph {et~al.}(2023)\citenamefont
  {Vylegzhanin}, \citenamefont {Brown}, \citenamefont {Raj}, \citenamefont
  {Kornovan}, \citenamefont {Everett}, \citenamefont {Brion}, \citenamefont
  {Robert},\ and\ \citenamefont {Chormaic}}]{vylegzhanin2023}%
  \BibitemOpen
  \bibfield  {author} {\bibinfo {author} {\bibfnamefont {A.}~\bibnamefont
  {Vylegzhanin}}, \bibinfo {author} {\bibfnamefont {D.~J.}\ \bibnamefont
  {Brown}}, \bibinfo {author} {\bibfnamefont {A.}~\bibnamefont {Raj}}, \bibinfo
  {author} {\bibfnamefont {D.~F.}\ \bibnamefont {Kornovan}}, \bibinfo {author}
  {\bibfnamefont {J.~L.}\ \bibnamefont {Everett}}, \bibinfo {author}
  {\bibfnamefont {E.}~\bibnamefont {Brion}}, \bibinfo {author} {\bibfnamefont
  {J.}~\bibnamefont {Robert}},\ and\ \bibinfo {author} {\bibfnamefont {S.~N.}\
  \bibnamefont {Chormaic}},\ }\href {https://arxiv.org/abs/2305.05186}
  {\bibinfo {title} {Excitation of $^{87}$rb rydberg atoms to ns and nd states
  (n$\leq$68) via an optical nanofiber}} (\bibinfo {year} {2023}),\ \Eprint
  {https://arxiv.org/abs/2305.05186} {arXiv:2305.05186 [physics.atom-ph]}
  \BibitemShut {NoStop}%
\bibitem [{\citenamefont {MacNeill}\ \emph {et~al.}(2019)\citenamefont
  {MacNeill}, \citenamefont {Hou}, \citenamefont {Klein}, \citenamefont
  {Zhang}, \citenamefont {Jarillo-Herrero},\ and\ \citenamefont
  {Liu}}]{MacNeill2019}%
  \BibitemOpen
  \bibfield  {author} {\bibinfo {author} {\bibfnamefont {D.}~\bibnamefont
  {MacNeill}}, \bibinfo {author} {\bibfnamefont {J.~T.}\ \bibnamefont {Hou}},
  \bibinfo {author} {\bibfnamefont {D.~R.}\ \bibnamefont {Klein}}, \bibinfo
  {author} {\bibfnamefont {P.}~\bibnamefont {Zhang}}, \bibinfo {author}
  {\bibfnamefont {P.}~\bibnamefont {Jarillo-Herrero}},\ and\ \bibinfo {author}
  {\bibfnamefont {L.}~\bibnamefont {Liu}},\ }\bibfield  {title} {\bibinfo
  {title} {Gigahertz frequency antiferromagnetic resonance and strong
  magnon-magnon coupling in the layered crystal ${\mathrm{crcl}}_{3}$},\ }\href
  {https://doi.org/10.1103/PhysRevLett.123.047204} {\bibfield  {journal}
  {\bibinfo  {journal} {Phys. Rev. Lett.}\ }\textbf {\bibinfo {volume} {123}},\
  \bibinfo {pages} {047204} (\bibinfo {year} {2019})}\BibitemShut {NoStop}%
\bibitem [{\citenamefont {Jenkins}\ \emph {et~al.}(2016)\citenamefont
  {Jenkins}, \citenamefont {Ruostekoski}, \citenamefont {Javanainen},
  \citenamefont {Jennewein}, \citenamefont {Bourgain}, \citenamefont
  {Pellegrino}, \citenamefont {Sortais},\ and\ \citenamefont
  {Browaeys}}]{SDjenkins2016}%
  \BibitemOpen
  \bibfield  {author} {\bibinfo {author} {\bibfnamefont {S.~D.}\ \bibnamefont
  {Jenkins}}, \bibinfo {author} {\bibfnamefont {J.}~\bibnamefont
  {Ruostekoski}}, \bibinfo {author} {\bibfnamefont {J.}~\bibnamefont
  {Javanainen}}, \bibinfo {author} {\bibfnamefont {S.}~\bibnamefont
  {Jennewein}}, \bibinfo {author} {\bibfnamefont {R.}~\bibnamefont {Bourgain}},
  \bibinfo {author} {\bibfnamefont {J.}~\bibnamefont {Pellegrino}}, \bibinfo
  {author} {\bibfnamefont {Y.~R.~P.}\ \bibnamefont {Sortais}},\ and\ \bibinfo
  {author} {\bibfnamefont {A.}~\bibnamefont {Browaeys}},\ }\bibfield  {title}
  {\bibinfo {title} {Collective resonance fluorescence in small and dense atom
  clouds: Comparison between theory and experiment},\ }\href
  {https://doi.org/10.1103/PhysRevA.94.023842} {\bibfield  {journal} {\bibinfo
  {journal} {Phys. Rev. A}\ }\textbf {\bibinfo {volume} {94}},\ \bibinfo
  {pages} {023842} (\bibinfo {year} {2016})}\BibitemShut {NoStop}%
\bibitem [{\citenamefont {Corman}\ \emph {et~al.}(2017)\citenamefont {Corman},
  \citenamefont {Ville}, \citenamefont {Saint-Jalm}, \citenamefont
  {Aidelsburger}, \citenamefont {Bienaim\'e}, \citenamefont {Nascimb\`ene},
  \citenamefont {Dalibard},\ and\ \citenamefont {Beugnon}}]{corman2017}%
  \BibitemOpen
  \bibfield  {author} {\bibinfo {author} {\bibfnamefont {L.}~\bibnamefont
  {Corman}}, \bibinfo {author} {\bibfnamefont {J.~L.}\ \bibnamefont {Ville}},
  \bibinfo {author} {\bibfnamefont {R.}~\bibnamefont {Saint-Jalm}}, \bibinfo
  {author} {\bibfnamefont {M.}~\bibnamefont {Aidelsburger}}, \bibinfo {author}
  {\bibfnamefont {T.}~\bibnamefont {Bienaim\'e}}, \bibinfo {author}
  {\bibfnamefont {S.}~\bibnamefont {Nascimb\`ene}}, \bibinfo {author}
  {\bibfnamefont {J.}~\bibnamefont {Dalibard}},\ and\ \bibinfo {author}
  {\bibfnamefont {J.}~\bibnamefont {Beugnon}},\ }\bibfield  {title} {\bibinfo
  {title} {Transmission of near-resonant light through a dense slab of cold
  atoms},\ }\href {https://doi.org/10.1103/PhysRevA.96.053629} {\bibfield
  {journal} {\bibinfo  {journal} {Phys. Rev. A}\ }\textbf {\bibinfo {volume}
  {96}},\ \bibinfo {pages} {053629} (\bibinfo {year} {2017})}\BibitemShut
  {NoStop}%
\bibitem [{\citenamefont {Petrosyan}\ and\ \citenamefont
  {M\o{}lmer}(2021)}]{petrosyan2021}%
  \BibitemOpen
  \bibfield  {author} {\bibinfo {author} {\bibfnamefont {D.}~\bibnamefont
  {Petrosyan}}\ and\ \bibinfo {author} {\bibfnamefont {K.}~\bibnamefont
  {M\o{}lmer}},\ }\bibfield  {title} {\bibinfo {title} {Collective emission of
  photons from dense, dipole-dipole interacting atomic ensembles},\ }\href
  {https://doi.org/10.1103/PhysRevA.103.023703} {\bibfield  {journal} {\bibinfo
   {journal} {Phys. Rev. A}\ }\textbf {\bibinfo {volume} {103}},\ \bibinfo
  {pages} {023703} (\bibinfo {year} {2021})}\BibitemShut {NoStop}%
\bibitem [{\citenamefont {Rom\'an-Roche}\ \emph {et~al.}(2021)\citenamefont
  {Rom\'an-Roche}, \citenamefont {Luis},\ and\ \citenamefont
  {Zueco}}]{Roman-Roche2021}%
  \BibitemOpen
  \bibfield  {author} {\bibinfo {author} {\bibfnamefont {J.}~\bibnamefont
  {Rom\'an-Roche}}, \bibinfo {author} {\bibfnamefont {F.}~\bibnamefont
  {Luis}},\ and\ \bibinfo {author} {\bibfnamefont {D.}~\bibnamefont {Zueco}},\
  }\bibfield  {title} {\bibinfo {title} {Photon condensation and enhanced
  magnetism in cavity qed},\ }\href
  {https://doi.org/10.1103/PhysRevLett.127.167201} {\bibfield  {journal}
  {\bibinfo  {journal} {Phys. Rev. Lett.}\ }\textbf {\bibinfo {volume} {127}},\
  \bibinfo {pages} {167201} (\bibinfo {year} {2021})}\BibitemShut {NoStop}%
\bibitem [{\citenamefont {Garcia-Vidal}\ \emph {et~al.}(2021)\citenamefont
  {Garcia-Vidal}, \citenamefont {Ciuti},\ and\ \citenamefont
  {Ebbesen}}]{GarciaVidal2021}%
  \BibitemOpen
  \bibfield  {author} {\bibinfo {author} {\bibfnamefont {F.~J.}\ \bibnamefont
  {Garcia-Vidal}}, \bibinfo {author} {\bibfnamefont {C.}~\bibnamefont
  {Ciuti}},\ and\ \bibinfo {author} {\bibfnamefont {T.~W.}\ \bibnamefont
  {Ebbesen}},\ }\bibfield  {title} {\bibinfo {title} {Manipulating matter by
  strong coupling to vacuum fields},\ }\href
  {https://doi.org/10.1126/science.abd0336} {\bibfield  {journal} {\bibinfo
  {journal} {Science}\ }\textbf {\bibinfo {volume} {373}},\ \bibinfo {pages}
  {eabd0336} (\bibinfo {year} {2021})}\BibitemShut {NoStop}%
\bibitem [{\citenamefont {Schlawin}\ \emph {et~al.}(2022)\citenamefont
  {Schlawin}, \citenamefont {Kennes},\ and\ \citenamefont
  {Sentef}}]{Schlawin2022}%
  \BibitemOpen
  \bibfield  {author} {\bibinfo {author} {\bibfnamefont {F.}~\bibnamefont
  {Schlawin}}, \bibinfo {author} {\bibfnamefont {D.~M.}\ \bibnamefont
  {Kennes}},\ and\ \bibinfo {author} {\bibfnamefont {M.~A.}\ \bibnamefont
  {Sentef}},\ }\bibfield  {title} {\bibinfo {title} {Cavity quantum
  materials},\ }\href {http://dx.doi.org/10.1063/5.0083825} {\bibfield
  {journal} {\bibinfo  {journal} {Appl. Phys. Rev.}\ }\textbf {\bibinfo
  {volume} {9}},\ \bibinfo {pages} {011312} (\bibinfo {year}
  {2022})}\BibitemShut {NoStop}%
\bibitem [{\citenamefont {{\v{Z}}ili{\'{c}}}\ \emph {et~al.}(2010)\citenamefont
  {{\v{Z}}ili{\'{c}}}, \citenamefont {Paji{\'{c}}}, \citenamefont
  {Juri{\'{c}}}, \citenamefont {Mol{\v{c}}anov}, \citenamefont {Rakvin},
  \citenamefont {Planini{\'{c}}},\ and\ \citenamefont {Zadro}}]{Zilic2010}%
  \BibitemOpen
  \bibfield  {author} {\bibinfo {author} {\bibfnamefont {D.}~\bibnamefont
  {{\v{Z}}ili{\'{c}}}}, \bibinfo {author} {\bibfnamefont {D.}~\bibnamefont
  {Paji{\'{c}}}}, \bibinfo {author} {\bibfnamefont {M.}~\bibnamefont
  {Juri{\'{c}}}}, \bibinfo {author} {\bibfnamefont {K.}~\bibnamefont
  {Mol{\v{c}}anov}}, \bibinfo {author} {\bibfnamefont {B.}~\bibnamefont
  {Rakvin}}, \bibinfo {author} {\bibfnamefont {P.}~\bibnamefont
  {Planini{\'{c}}}},\ and\ \bibinfo {author} {\bibfnamefont {K.}~\bibnamefont
  {Zadro}},\ }\bibfield  {title} {\bibinfo {title} {Single crystals of {DPPH}
  grown from diethyl ether and carbon disulfide solutions {\textendash} crystal
  structures, {IR}, {EPR} and magnetization studies},\ }\href
  {https://doi.org/10.1016/j.jmr.2010.08.005} {\bibfield  {journal} {\bibinfo
  {journal} {J. Magn. Reson.}\ }\textbf {\bibinfo {volume} {207}},\ \bibinfo
  {pages} {34} (\bibinfo {year} {2010})}\BibitemShut {NoStop}%
\bibitem [{\citenamefont {Blöte}(1975)}]{Blote1975}%
  \BibitemOpen
  \bibfield  {author} {\bibinfo {author} {\bibfnamefont {H.}~\bibnamefont
  {Blöte}},\ }\bibfield  {title} {\bibinfo {title} {The specific heat of
  magnetic linear chains},\ }\href
  {https://doi.org/https://doi.org/10.1016/0378-4363(75)90001-7} {\bibfield
  {journal} {\bibinfo  {journal} {Physica B+C}\ }\textbf {\bibinfo {volume}
  {79}},\ \bibinfo {pages} {427} (\bibinfo {year} {1975})}\BibitemShut
  {NoStop}%
\bibitem [{\citenamefont {de~Jongh}\ and\ \citenamefont
  {Miedema}(1974)}]{DeJongh1974}%
  \BibitemOpen
  \bibfield  {author} {\bibinfo {author} {\bibfnamefont {L.}~\bibnamefont
  {de~Jongh}}\ and\ \bibinfo {author} {\bibfnamefont {A.}~\bibnamefont
  {Miedema}},\ }\bibfield  {title} {\bibinfo {title} {Experiments on simple
  magnetic model systems},\ }\href {https://doi.org/10.1080/00018739700101558}
  {\bibfield  {journal} {\bibinfo  {journal} {Adv. Phys.}\ }\textbf {\bibinfo
  {volume} {23}},\ \bibinfo {pages} {1} (\bibinfo {year} {1974})}\BibitemShut
  {NoStop}%
\bibitem [{\citenamefont {Mergenthaler}\ \emph {et~al.}(2017)\citenamefont
  {Mergenthaler}, \citenamefont {Liu}, \citenamefont {Le~Roy}, \citenamefont
  {Ares}, \citenamefont {Thompson}, \citenamefont {Bogani}, \citenamefont
  {Luis}, \citenamefont {Blundell}, \citenamefont {Lancaster}, \citenamefont
  {Ardavan}, \citenamefont {Briggs}, \citenamefont {Leek},\ and\ \citenamefont
  {Laird}}]{Mergenthaler2017}%
  \BibitemOpen
  \bibfield  {author} {\bibinfo {author} {\bibfnamefont {M.}~\bibnamefont
  {Mergenthaler}}, \bibinfo {author} {\bibfnamefont {J.}~\bibnamefont {Liu}},
  \bibinfo {author} {\bibfnamefont {J.~J.}\ \bibnamefont {Le~Roy}}, \bibinfo
  {author} {\bibfnamefont {N.}~\bibnamefont {Ares}}, \bibinfo {author}
  {\bibfnamefont {A.~L.}\ \bibnamefont {Thompson}}, \bibinfo {author}
  {\bibfnamefont {L.}~\bibnamefont {Bogani}}, \bibinfo {author} {\bibfnamefont
  {F.}~\bibnamefont {Luis}}, \bibinfo {author} {\bibfnamefont {S.~J.}\
  \bibnamefont {Blundell}}, \bibinfo {author} {\bibfnamefont {T.}~\bibnamefont
  {Lancaster}}, \bibinfo {author} {\bibfnamefont {A.}~\bibnamefont {Ardavan}},
  \bibinfo {author} {\bibfnamefont {G.~A.~D.}\ \bibnamefont {Briggs}}, \bibinfo
  {author} {\bibfnamefont {P.~J.}\ \bibnamefont {Leek}},\ and\ \bibinfo
  {author} {\bibfnamefont {E.~A.}\ \bibnamefont {Laird}},\ }\bibfield  {title}
  {\bibinfo {title} {Strong coupling of microwave photons to antiferromagnetic
  fluctuations in an organic magnet},\ }\href
  {http://dx.doi.org/10.1103/PhysRevLett.119.147701} {\bibfield  {journal}
  {\bibinfo  {journal} {Phys. Rev. Lett.}\ }\textbf {\bibinfo {volume} {119}},\
  \bibinfo {pages} {147701} (\bibinfo {year} {2017})}\BibitemShut {NoStop}%
\bibitem [{\citenamefont {Voesch}\ \emph {et~al.}(2015)\citenamefont {Voesch},
  \citenamefont {Thiemann}, \citenamefont {Bothner}, \citenamefont {Dressel},\
  and\ \citenamefont {Scheffler}}]{Voesch2015}%
  \BibitemOpen
  \bibfield  {author} {\bibinfo {author} {\bibfnamefont {W.}~\bibnamefont
  {Voesch}}, \bibinfo {author} {\bibfnamefont {M.}~\bibnamefont {Thiemann}},
  \bibinfo {author} {\bibfnamefont {D.}~\bibnamefont {Bothner}}, \bibinfo
  {author} {\bibfnamefont {M.}~\bibnamefont {Dressel}},\ and\ \bibinfo {author}
  {\bibfnamefont {M.}~\bibnamefont {Scheffler}},\ }\bibfield  {title} {\bibinfo
  {title} {On-chip esr measurements of dpph at mk temperatures},\ }\href
  {https://doi.org/10.1016/j.phpro.2015.12.063} {\bibfield  {journal} {\bibinfo
   {journal} {Phys. Procedia}\ }\textbf {\bibinfo {volume} {75}},\ \bibinfo
  {pages} {503} (\bibinfo {year} {2015})}\BibitemShut {NoStop}%
\bibitem [{\citenamefont {Lenz}\ \emph {et~al.}(2020)\citenamefont {Lenz},
  \citenamefont {Hunger},\ and\ \citenamefont {van Slageren}}]{lenz2020}%
  \BibitemOpen
  \bibfield  {author} {\bibinfo {author} {\bibfnamefont {S.}~\bibnamefont
  {Lenz}}, \bibinfo {author} {\bibfnamefont {D.}~\bibnamefont {Hunger}},\ and\
  \bibinfo {author} {\bibfnamefont {J.}~\bibnamefont {van Slageren}},\
  }\bibfield  {title} {\bibinfo {title} {Strong coupling between resonators and
  spin ensembles in the presence of exchange couplings},\ }\href
  {https://doi.org/10.1039/d0cc04841k} {\bibfield  {journal} {\bibinfo
  {journal} {Chem. Commun.}\ }\textbf {\bibinfo {volume} {56}},\ \bibinfo
  {pages} {12837} (\bibinfo {year} {2020})}\BibitemShut {NoStop}%
\bibitem [{\citenamefont {Zollitsch}\ \emph {et~al.}(2023)\citenamefont
  {Zollitsch}, \citenamefont {Khan}, \citenamefont {Nam}, \citenamefont
  {Verzhbitskiy}, \citenamefont {Sagkovits}, \citenamefont {O’Sullivan},
  \citenamefont {Kennedy}, \citenamefont {Strungaru}, \citenamefont {Santos},
  \citenamefont {Morton}, \citenamefont {Eda},\ and\ \citenamefont
  {Kurebayashi}}]{Zollitsch2023}%
  \BibitemOpen
  \bibfield  {author} {\bibinfo {author} {\bibfnamefont {C.~W.}\ \bibnamefont
  {Zollitsch}}, \bibinfo {author} {\bibfnamefont {S.}~\bibnamefont {Khan}},
  \bibinfo {author} {\bibfnamefont {V.~T.~T.}\ \bibnamefont {Nam}}, \bibinfo
  {author} {\bibfnamefont {I.~A.}\ \bibnamefont {Verzhbitskiy}}, \bibinfo
  {author} {\bibfnamefont {D.}~\bibnamefont {Sagkovits}}, \bibinfo {author}
  {\bibfnamefont {J.}~\bibnamefont {O’Sullivan}}, \bibinfo {author}
  {\bibfnamefont {O.~W.}\ \bibnamefont {Kennedy}}, \bibinfo {author}
  {\bibfnamefont {M.}~\bibnamefont {Strungaru}}, \bibinfo {author}
  {\bibfnamefont {E.~J.~G.}\ \bibnamefont {Santos}}, \bibinfo {author}
  {\bibfnamefont {J.~J.~L.}\ \bibnamefont {Morton}}, \bibinfo {author}
  {\bibfnamefont {G.}~\bibnamefont {Eda}},\ and\ \bibinfo {author}
  {\bibfnamefont {H.}~\bibnamefont {Kurebayashi}},\ }\bibfield  {title}
  {\bibinfo {title} {Probing spin dynamics of ultra-thin van der waals magnets
  via photon-magnon coupling},\ }\href
  {http://dx.doi.org/10.1038/s41467-023-38322-x} {\bibfield  {journal}
  {\bibinfo  {journal} {Nat. Commun.}\ }\textbf {\bibinfo {volume} {14}},\
  \bibinfo {pages} {2619} (\bibinfo {year} {2023})}\BibitemShut {NoStop}%
\bibitem [{\citenamefont {Williams}(1966)}]{Williams1966}%
  \BibitemOpen
  \bibfield  {author} {\bibinfo {author} {\bibfnamefont {D.~E.}\ \bibnamefont
  {Williams}},\ }\bibfield  {title} {\bibinfo {title} {Structure of 2,
  2-diphenyl-1-picrylhydrazyl free radical1},\ }\href
  {https://doi.org/10.1021/ja00975a064} {\bibfield  {journal} {\bibinfo
  {journal} {J. Am. Chem. Soc.}\ }\textbf {\bibinfo {volume} {88}},\ \bibinfo
  {pages} {5665} (\bibinfo {year} {1966})}\BibitemShut {NoStop}%
\bibitem [{\citenamefont {Yordanov}(1996)}]{Yordanov1996}%
  \BibitemOpen
  \bibfield  {author} {\bibinfo {author} {\bibfnamefont {N.~D.}\ \bibnamefont
  {Yordanov}},\ }\bibfield  {title} {\bibinfo {title} {Is our knowledge about
  the chemical and physical properties of dpph enough to consider it as a
  primary standard for quantitative epr spectrometry},\ }\href
  {https://doi.org/10.1007/BF03163117} {\bibfield  {journal} {\bibinfo
  {journal} {Appl. Magn. Reson.}\ }\textbf {\bibinfo {volume} {10}},\ \bibinfo
  {pages} {339} (\bibinfo {year} {1996})}\BibitemShut {NoStop}%
\bibitem [{\citenamefont {Ohya-Nishiguchi}(1979)}]{OhyaNishiguchi1979}%
  \BibitemOpen
  \bibfield  {author} {\bibinfo {author} {\bibfnamefont {H.}~\bibnamefont
  {Ohya-Nishiguchi}},\ }\bibfield  {title} {\bibinfo {title} {On the magnetic
  susceptibility of interacting spin-pair systems},\ }\href
  {https://doi.org/10.1246/bcsj.52.3480} {\bibfield  {journal} {\bibinfo
  {journal} {Bull. Chem. Soc. Jpn.}\ }\textbf {\bibinfo {volume} {52}},\
  \bibinfo {pages} {3480} (\bibinfo {year} {1979})}\BibitemShut {NoStop}%
\bibitem [{\citenamefont {Fujito}(1981)}]{Fujito1981}%
  \BibitemOpen
  \bibfield  {author} {\bibinfo {author} {\bibfnamefont {T.}~\bibnamefont
  {Fujito}},\ }\bibfield  {title} {\bibinfo {title} {Magnetic interaction in
  solvent-free {DPPH} and {DPPH}{\textendash}solvent complexes},\ }\href
  {https://doi.org/10.1246/bcsj.54.3110} {\bibfield  {journal} {\bibinfo
  {journal} {Bull. Chem. Soc. Jpn.}\ }\textbf {\bibinfo {volume} {54}},\
  \bibinfo {pages} {3110} (\bibinfo {year} {1981})}\BibitemShut {NoStop}%
\bibitem [{\citenamefont {Clauss}\ \emph {et~al.}(2013)\citenamefont {Clauss},
  \citenamefont {Bothner}, \citenamefont {Koelle}, \citenamefont {Kleiner},
  \citenamefont {Bogani}, \citenamefont {Scheffler},\ and\ \citenamefont
  {Dressel}}]{Clauss2013}%
  \BibitemOpen
  \bibfield  {author} {\bibinfo {author} {\bibfnamefont {C.}~\bibnamefont
  {Clauss}}, \bibinfo {author} {\bibfnamefont {D.}~\bibnamefont {Bothner}},
  \bibinfo {author} {\bibfnamefont {D.}~\bibnamefont {Koelle}}, \bibinfo
  {author} {\bibfnamefont {R.}~\bibnamefont {Kleiner}}, \bibinfo {author}
  {\bibfnamefont {L.}~\bibnamefont {Bogani}}, \bibinfo {author} {\bibfnamefont
  {M.}~\bibnamefont {Scheffler}},\ and\ \bibinfo {author} {\bibfnamefont
  {M.}~\bibnamefont {Dressel}},\ }\bibfield  {title} {\bibinfo {title}
  {Broadband electron spin resonance from 500{\hspace{0.167em}}{MHz} to
  40{\hspace{0.167em}}{GHz} using superconducting coplanar waveguides},\ }\href
  {https://doi.org/10.1063/1.4802956} {\bibfield  {journal} {\bibinfo
  {journal} {Appl. Phys. Lett.}\ }\textbf {\bibinfo {volume} {102}},\ \bibinfo
  {pages} {162601} (\bibinfo {year} {2013})}\BibitemShut {NoStop}%
\bibitem [{\citenamefont {Gimeno}\ \emph {et~al.}(2023)\citenamefont {Gimeno},
  \citenamefont {Rollano}, \citenamefont {Zueco}, \citenamefont {Duan},
  \citenamefont {de~Ory}, \citenamefont {Gomez}, \citenamefont {Gaita-Ari\~no},
  \citenamefont {S\'anchez-Azqueta}, \citenamefont {Astner}, \citenamefont
  {Granados}, \citenamefont {Hill}, \citenamefont {Majer}, \citenamefont
  {Coronado},\ and\ \citenamefont {Luis}}]{Gimeno2023}%
  \BibitemOpen
  \bibfield  {author} {\bibinfo {author} {\bibfnamefont {I.}~\bibnamefont
  {Gimeno}}, \bibinfo {author} {\bibfnamefont {V.}~\bibnamefont {Rollano}},
  \bibinfo {author} {\bibfnamefont {D.}~\bibnamefont {Zueco}}, \bibinfo
  {author} {\bibfnamefont {Y.}~\bibnamefont {Duan}}, \bibinfo {author}
  {\bibfnamefont {M.~C.}\ \bibnamefont {de~Ory}}, \bibinfo {author}
  {\bibfnamefont {A.}~\bibnamefont {Gomez}}, \bibinfo {author} {\bibfnamefont
  {A.}~\bibnamefont {Gaita-Ari\~no}}, \bibinfo {author} {\bibfnamefont
  {C.}~\bibnamefont {S\'anchez-Azqueta}}, \bibinfo {author} {\bibfnamefont
  {T.}~\bibnamefont {Astner}}, \bibinfo {author} {\bibfnamefont
  {D.}~\bibnamefont {Granados}}, \bibinfo {author} {\bibfnamefont
  {S.}~\bibnamefont {Hill}}, \bibinfo {author} {\bibfnamefont {J.}~\bibnamefont
  {Majer}}, \bibinfo {author} {\bibfnamefont {E.}~\bibnamefont {Coronado}},\
  and\ \bibinfo {author} {\bibfnamefont {F.}~\bibnamefont {Luis}},\ }\bibfield
  {title} {\bibinfo {title} {Optimal coupling of ${\mathrm{ho}\mathrm{w}}_{10}$
  molecular magnets to superconducting circuits near spin clock transitions},\
  }\href {https://doi.org/10.1103/PhysRevApplied.20.044070} {\bibfield
  {journal} {\bibinfo  {journal} {Phys. Rev. Appl.}\ }\textbf {\bibinfo
  {volume} {20}},\ \bibinfo {pages} {044070} (\bibinfo {year}
  {2023})}\BibitemShut {NoStop}%
\bibitem [{\citenamefont {Dung}\ \emph {et~al.}(2002)\citenamefont {Dung},
  \citenamefont {Kn\"oll},\ and\ \citenamefont {Welsch}}]{Dung2002}%
  \BibitemOpen
  \bibfield  {author} {\bibinfo {author} {\bibfnamefont {H.~T.}\ \bibnamefont
  {Dung}}, \bibinfo {author} {\bibfnamefont {L.}~\bibnamefont {Kn\"oll}},\ and\
  \bibinfo {author} {\bibfnamefont {D.-G.}\ \bibnamefont {Welsch}},\ }\bibfield
   {title} {\bibinfo {title} {Resonant dipole-dipole interaction in the
  presence of dispersing and absorbing surroundings},\ }\href
  {https://doi.org/10.1103/PhysRevA.66.063810} {\bibfield  {journal} {\bibinfo
  {journal} {Phys. Rev. A}\ }\textbf {\bibinfo {volume} {66}},\ \bibinfo
  {pages} {063810} (\bibinfo {year} {2002})}\BibitemShut {NoStop}%
\bibitem [{\citenamefont {Dzsotjan}\ \emph {et~al.}(2010)\citenamefont
  {Dzsotjan}, \citenamefont {S\o{}rensen},\ and\ \citenamefont
  {Fleischhauer}}]{Dzsotjan2010}%
  \BibitemOpen
  \bibfield  {author} {\bibinfo {author} {\bibfnamefont {D.}~\bibnamefont
  {Dzsotjan}}, \bibinfo {author} {\bibfnamefont {A.~S.}\ \bibnamefont
  {S\o{}rensen}},\ and\ \bibinfo {author} {\bibfnamefont {M.}~\bibnamefont
  {Fleischhauer}},\ }\bibfield  {title} {\bibinfo {title} {Quantum emitters
  coupled to surface plasmons of a nanowire: A green's function approach},\
  }\href {https://doi.org/10.1103/PhysRevB.82.075427} {\bibfield  {journal}
  {\bibinfo  {journal} {Phys. Rev. B}\ }\textbf {\bibinfo {volume} {82}},\
  \bibinfo {pages} {075427} (\bibinfo {year} {2010})}\BibitemShut {NoStop}%
\bibitem [{\citenamefont {Lalumi\`ere}\ \emph {et~al.}(2013)\citenamefont
  {Lalumi\`ere}, \citenamefont {Sanders}, \citenamefont {van Loo},
  \citenamefont {Fedorov}, \citenamefont {Wallraff},\ and\ \citenamefont
  {Blais}}]{Lalumiere2013}%
  \BibitemOpen
  \bibfield  {author} {\bibinfo {author} {\bibfnamefont {K.}~\bibnamefont
  {Lalumi\`ere}}, \bibinfo {author} {\bibfnamefont {B.~C.}\ \bibnamefont
  {Sanders}}, \bibinfo {author} {\bibfnamefont {A.~F.}\ \bibnamefont {van
  Loo}}, \bibinfo {author} {\bibfnamefont {A.}~\bibnamefont {Fedorov}},
  \bibinfo {author} {\bibfnamefont {A.}~\bibnamefont {Wallraff}},\ and\
  \bibinfo {author} {\bibfnamefont {A.}~\bibnamefont {Blais}},\ }\bibfield
  {title} {\bibinfo {title} {Input-output theory for waveguide qed with an
  ensemble of inhomogeneous atoms},\ }\href
  {https://doi.org/10.1103/PhysRevA.88.043806} {\bibfield  {journal} {\bibinfo
  {journal} {Phys. Rev. A}\ }\textbf {\bibinfo {volume} {88}},\ \bibinfo
  {pages} {043806} (\bibinfo {year} {2013})}\BibitemShut {NoStop}%
\bibitem [{\citenamefont {Fan}\ \emph {et~al.}(2010)\citenamefont {Fan},
  \citenamefont {Kocaba\ifmmode~\mbox{\c{s}}\else \c{s}\fi{}},\ and\
  \citenamefont {Shen}}]{Fan2010}%
  \BibitemOpen
  \bibfield  {author} {\bibinfo {author} {\bibfnamefont {S.}~\bibnamefont
  {Fan}}, \bibinfo {author} {\bibfnamefont {S.~E.}\ \bibnamefont
  {Kocaba\ifmmode~\mbox{\c{s}}\else \c{s}\fi{}}},\ and\ \bibinfo {author}
  {\bibfnamefont {J.-T.}\ \bibnamefont {Shen}},\ }\bibfield  {title} {\bibinfo
  {title} {Input-output formalism for few-photon transport in one-dimensional
  nanophotonic waveguides coupled to a qubit},\ }\href
  {https://doi.org/10.1103/PhysRevA.82.063821} {\bibfield  {journal} {\bibinfo
  {journal} {Phys. Rev. A}\ }\textbf {\bibinfo {volume} {82}},\ \bibinfo
  {pages} {063821} (\bibinfo {year} {2010})}\BibitemShut {NoStop}%
\bibitem [{\citenamefont {Fang}\ \emph {et~al.}(2014)\citenamefont {Fang},
  \citenamefont {Zheng},\ and\ \citenamefont {Baranger}}]{Fang2014}%
  \BibitemOpen
  \bibfield  {author} {\bibinfo {author} {\bibfnamefont {Y.-L.~L.}\
  \bibnamefont {Fang}}, \bibinfo {author} {\bibfnamefont {H.}~\bibnamefont
  {Zheng}},\ and\ \bibinfo {author} {\bibfnamefont {H.~U.}\ \bibnamefont
  {Baranger}},\ }\bibfield  {title} {\bibinfo {title} {One-dimensional
  waveguide coupled to multiple qubits: photon-photon correlations},\ }\href
  {http://dx.doi.org/10.1140/epjqt3} {\bibfield  {journal} {\bibinfo  {journal}
  {EPJ Quantum Technol.}\ }\textbf {\bibinfo {volume} {1}},\ \bibinfo {pages}
  {3} (\bibinfo {year} {2014})}\BibitemShut {NoStop}%
\bibitem [{\citenamefont {S\'{a}nchez-Burillo}\ \emph
  {et~al.}(2016)\citenamefont {S\'{a}nchez-Burillo}, \citenamefont
  {Mart\'{\i}n-Moreno}, \citenamefont {Garc\'{\i}a-Ripoll},\ and\ \citenamefont
  {Zueco}}]{Burillo2016}%
  \BibitemOpen
  \bibfield  {author} {\bibinfo {author} {\bibfnamefont {E.}~\bibnamefont
  {S\'{a}nchez-Burillo}}, \bibinfo {author} {\bibfnamefont {L.}~\bibnamefont
  {Mart\'{\i}n-Moreno}}, \bibinfo {author} {\bibfnamefont {J.~J.}\ \bibnamefont
  {Garc\'{\i}a-Ripoll}},\ and\ \bibinfo {author} {\bibfnamefont
  {D.}~\bibnamefont {Zueco}},\ }\bibfield  {title} {\bibinfo {title} {{Full
  two-photon down-conversion of a single photon}},\ }\href
  {https://doi.org/10.1103/PhysRevA.94.053814} {\bibfield  {journal} {\bibinfo
  {journal} {Phys. Rev. A}\ }\textbf {\bibinfo {volume} {94}},\ \bibinfo
  {pages} {053814} (\bibinfo {year} {2016})}\BibitemShut {NoStop}%
\bibitem [{\citenamefont {Bonner}\ and\ \citenamefont
  {Fisher}(1964)}]{Bonner1964}%
  \BibitemOpen
  \bibfield  {author} {\bibinfo {author} {\bibfnamefont {J.~C.}\ \bibnamefont
  {Bonner}}\ and\ \bibinfo {author} {\bibfnamefont {M.~E.}\ \bibnamefont
  {Fisher}},\ }\bibfield  {title} {\bibinfo {title} {Linear magnetic chains
  with anisotropic coupling},\ }\href
  {https://doi.org/10.1103/PhysRev.135.A640} {\bibfield  {journal} {\bibinfo
  {journal} {Phys. Rev.}\ }\textbf {\bibinfo {volume} {135}},\ \bibinfo {pages}
  {A640} (\bibinfo {year} {1964})}\BibitemShut {NoStop}%
\bibitem [{\citenamefont {Yamauchi}(2006)}]{Yamauchi1971}%
  \BibitemOpen
  \bibfield  {author} {\bibinfo {author} {\bibfnamefont {J.}~\bibnamefont
  {Yamauchi}},\ }\bibfield  {title} {\bibinfo {title} {{Linear
  Antiferromagnetic Interaction in Organic Free Radicals}},\ }\href
  {https://doi.org/10.1246/bcsj.44.2301} {\bibfield  {journal} {\bibinfo
  {journal} {Bull. Chem. Soc. Jpn.}\ }\textbf {\bibinfo {volume} {44}},\
  \bibinfo {pages} {2301} (\bibinfo {year} {2006})}\BibitemShut {NoStop}%
\bibitem [{\citenamefont {Nagamiya}\ \emph {et~al.}(1955)\citenamefont
  {Nagamiya}, \citenamefont {Yosida},\ and\ \citenamefont
  {Kubo}}]{Nagamiya1955}%
  \BibitemOpen
  \bibfield  {author} {\bibinfo {author} {\bibfnamefont {T.}~\bibnamefont
  {Nagamiya}}, \bibinfo {author} {\bibfnamefont {K.}~\bibnamefont {Yosida}},\
  and\ \bibinfo {author} {\bibfnamefont {R.}~\bibnamefont {Kubo}},\ }\bibfield
  {title} {\bibinfo {title} {Antiferromagnetism},\ }\href
  {https://doi.org/10.1080/00018735500101154} {\bibfield  {journal} {\bibinfo
  {journal} {Adv. Phys.}\ }\textbf {\bibinfo {volume} {4}},\ \bibinfo {pages}
  {1} (\bibinfo {year} {1955})}\BibitemShut {NoStop}%
\bibitem [{\citenamefont {Keffer}\ and\ \citenamefont
  {Kittel}(1952)}]{Keffer1952}%
  \BibitemOpen
  \bibfield  {author} {\bibinfo {author} {\bibfnamefont {F.}~\bibnamefont
  {Keffer}}\ and\ \bibinfo {author} {\bibfnamefont {C.}~\bibnamefont
  {Kittel}},\ }\bibfield  {title} {\bibinfo {title} {Theory of
  antiferromagnetic resonance},\ }\href
  {https://doi.org/10.1103/physrev.85.329} {\bibfield  {journal} {\bibinfo
  {journal} {Phys. Rev.}\ }\textbf {\bibinfo {volume} {85}},\ \bibinfo {pages}
  {329} (\bibinfo {year} {1952})}\BibitemShut {NoStop}%
\bibitem [{\citenamefont {Gardiner}\ and\ \citenamefont
  {Collett}(1985)}]{Gardiner1985}%
  \BibitemOpen
  \bibfield  {author} {\bibinfo {author} {\bibfnamefont {C.~W.}\ \bibnamefont
  {Gardiner}}\ and\ \bibinfo {author} {\bibfnamefont {M.~J.}\ \bibnamefont
  {Collett}},\ }\bibfield  {title} {\bibinfo {title} {Input and output in
  damped quantum systems: Quantum stochastic differential equations and the
  master equation},\ }\href {https://doi.org/10.1103/PhysRevA.31.3761}
  {\bibfield  {journal} {\bibinfo  {journal} {Phys. Rev. A}\ }\textbf {\bibinfo
  {volume} {31}},\ \bibinfo {pages} {3761} (\bibinfo {year}
  {1985})}\BibitemShut {NoStop}%
\bibitem [{\citenamefont {Kamra}\ \emph {et~al.}(2018)\citenamefont {Kamra},
  \citenamefont {Troncoso}, \citenamefont {Belzig},\ and\ \citenamefont
  {Brataas}}]{Kamra2018}%
  \BibitemOpen
  \bibfield  {author} {\bibinfo {author} {\bibfnamefont {A.}~\bibnamefont
  {Kamra}}, \bibinfo {author} {\bibfnamefont {R.~E.}\ \bibnamefont {Troncoso}},
  \bibinfo {author} {\bibfnamefont {W.}~\bibnamefont {Belzig}},\ and\ \bibinfo
  {author} {\bibfnamefont {A.}~\bibnamefont {Brataas}},\ }\bibfield  {title}
  {\bibinfo {title} {Gilbert damping phenomenology for two-sublattice
  magnets},\ }\href {https://doi.org/10.1103/PhysRevB.98.184402} {\bibfield
  {journal} {\bibinfo  {journal} {Phys. Rev. B}\ }\textbf {\bibinfo {volume}
  {98}},\ \bibinfo {pages} {184402} (\bibinfo {year} {2018})}\BibitemShut
  {NoStop}%
\bibitem [{\citenamefont {Suto}\ \emph {et~al.}(2019)\citenamefont {Suto},
  \citenamefont {Kanao}, \citenamefont {Nagasawa}, \citenamefont {Mizushima},
  \citenamefont {Sato}, \citenamefont {Kikuchi},\ and\ \citenamefont
  {Okamoto}}]{Suto2019}%
  \BibitemOpen
  \bibfield  {author} {\bibinfo {author} {\bibfnamefont {H.}~\bibnamefont
  {Suto}}, \bibinfo {author} {\bibfnamefont {T.}~\bibnamefont {Kanao}},
  \bibinfo {author} {\bibfnamefont {T.}~\bibnamefont {Nagasawa}}, \bibinfo
  {author} {\bibfnamefont {K.}~\bibnamefont {Mizushima}}, \bibinfo {author}
  {\bibfnamefont {R.}~\bibnamefont {Sato}}, \bibinfo {author} {\bibfnamefont
  {N.}~\bibnamefont {Kikuchi}},\ and\ \bibinfo {author} {\bibfnamefont
  {S.}~\bibnamefont {Okamoto}},\ }\bibfield  {title} {\bibinfo {title}
  {Microwave-magnetic-field-induced magnetization excitation and assisted
  switching of antiferromagnetically coupled magnetic bilayer with
  perpendicular magnetization},\ }\bibfield  {journal} {\bibinfo  {journal}
  {Journal of Applied Physics}\ }\textbf {\bibinfo {volume} {125}},\ \href
  {https://doi.org/10.1063/1.5089799} {10.1063/1.5089799} (\bibinfo {year}
  {2019})\BibitemShut {NoStop}%
\bibitem [{\citenamefont {Wang}\ and\ \citenamefont {Xiao}(2024)}]{Wang2024}%
  \BibitemOpen
  \bibfield  {author} {\bibinfo {author} {\bibfnamefont {Y.}~\bibnamefont
  {Wang}}\ and\ \bibinfo {author} {\bibfnamefont {J.}~\bibnamefont {Xiao}},\
  }\bibfield  {title} {\bibinfo {title} {Mechanism for broadened linewidth in
  antiferromagnetic resonance},\ }\href
  {https://doi.org/10.1103/PhysRevB.110.134409} {\bibfield  {journal} {\bibinfo
   {journal} {Phys. Rev. B}\ }\textbf {\bibinfo {volume} {110}},\ \bibinfo
  {pages} {134409} (\bibinfo {year} {2024})}\BibitemShut {NoStop}%
\bibitem [{\citenamefont {Kent}\ \emph {et~al.}(1994)\citenamefont {Kent},
  \citenamefont {von Moln\'{a}r}, \citenamefont {Gider},\ and\ \citenamefont
  {Awschalom}}]{Kent1994}%
  \BibitemOpen
  \bibfield  {author} {\bibinfo {author} {\bibfnamefont {A.~D.}\ \bibnamefont
  {Kent}}, \bibinfo {author} {\bibfnamefont {S.}~\bibnamefont {von
  Moln\'{a}r}}, \bibinfo {author} {\bibfnamefont {S.}~\bibnamefont {Gider}},\
  and\ \bibinfo {author} {\bibfnamefont {D.~D.}\ \bibnamefont {Awschalom}},\
  }\bibfield  {title} {\bibinfo {title} {{Properties and measurement of
  scanning tunneling microscope fabricated ferromagnetic particle arrays}},\
  }\href {https://doi.org/10.1063/1.358160} {\bibfield  {journal} {\bibinfo
  {journal} {J. Appl. Phys.}\ }\textbf {\bibinfo {volume} {76}},\ \bibinfo
  {pages} {6656} (\bibinfo {year} {1994})}\BibitemShut {NoStop}%
\bibitem [{\citenamefont {Stewart}(1983)}]{Stewart1983}%
  \BibitemOpen
  \bibfield  {author} {\bibinfo {author} {\bibfnamefont {G.~R.}\ \bibnamefont
  {Stewart}},\ }\bibfield  {title} {\bibinfo {title} {{Measurement of
  low-temperature specific heat}},\ }\href {https://doi.org/10.1063/1.1137207}
  {\bibfield  {journal} {\bibinfo  {journal} {Rev. Sci. Instrum.}\ }\textbf
  {\bibinfo {volume} {54}},\ \bibinfo {pages} {1} (\bibinfo {year}
  {1983})}\BibitemShut {NoStop}%
\bibitem [{\citenamefont {Kiers}\ \emph {et~al.}(1976)\citenamefont {Kiers},
  \citenamefont {de~Boer}, \citenamefont {Olthof},\ and\ \citenamefont
  {Spek}}]{Kiers1976}%
  \BibitemOpen
  \bibfield  {author} {\bibinfo {author} {\bibfnamefont {C.~T.}\ \bibnamefont
  {Kiers}}, \bibinfo {author} {\bibfnamefont {J.~L.}\ \bibnamefont {de~Boer}},
  \bibinfo {author} {\bibfnamefont {R.}~\bibnamefont {Olthof}},\ and\ \bibinfo
  {author} {\bibfnamefont {A.~L.}\ \bibnamefont {Spek}},\ }\bibfield  {title}
  {\bibinfo {title} {The crystal structure of a 2,2-diphenyl-1-pierylhydrazyl
  (dpph) modification},\ }\href {https://doi.org/10.1107/S0567740876007632}
  {\bibfield  {journal} {\bibinfo  {journal} {Acta Cryst. B}\ }\textbf
  {\bibinfo {volume} {32}},\ \bibinfo {pages} {2297} (\bibinfo {year}
  {1976})}\BibitemShut {NoStop}%
\end{thebibliography}%

\begin{acknowledgments}
This work has received support from grants CEX2023-001286-S, 
TED2021-131447B-C21 and
PID2022-140923NB-C21 funded by MCIN/AEI/10.13039/501100011033,
ERDF `A way of making Europe' and ESF `Investing in your future', 
from the Gobierno de Arag\'on grant E09-23R-Q-MAD, from the European Union 
Horizon 2020 research and innovation programme through FET-
OPEN grant FATMOLS-No862893, and from the Spanish Ministry for Digital 
Transformation and Civil Service and NextGenerationEU through the Quantum 
Spain project (Digital Spain 2026 Agenda). It also 
forms part of the Advanced Materials and Quantum Communication programmes 
with funding from European Union NextGenerationEU (PRTR-C17.I1), MCIN, 
Gobierno de Arag\'on, and CSIC (PTI001).
\end{acknowledgments}

\section*{Author contributions}

SR-J and DZ developed the theoretical models 
to describe the magnetic response of DPPH and the microwave experiments. 
MR-O, MDJ and FL carried 
out and analyzed the microwave measurements. AC and FL performed and analyzed 
the magnetic and heat capacity characterization. PJA carried out and analyzed 
the EPR experiments. All authors contributed to writing the manuscript.  

\section*{Competing interests}
The authors declare no competing interests.

\section*{Additional information}
{\bf Supplementary information} The online version contains supplementary 
material available at https://doi.org/xx.xxxx/syyyyy-yyy-yyyyy-y

{\bf Correspondence} and requests for materials should be addressed to David 
Zueco (dzueco@unizar.es) and Fernando Luis (fluis@unizar.es).

\end{document}


\title{Waveguide quantum electrodynamics at the onset of spin-spin correlations. Supplementary Information}

\author{Sebastián Roca-Jerat}
\affiliation{Instituto de Nanociencia y Materiales de Aragón (INMA), CSIC-Universidad de Zaragoza, Pedro Cerbuna 12, Zaragoza 50009, Spain}

\author{Marcos Rubín-Osanz}
\affiliation{Instituto de Nanociencia y Materiales de Aragón (INMA), CSIC-Universidad de Zaragoza, Pedro Cerbuna 12, Zaragoza 50009, Spain}

\author{Mark D. Jenkins}
\affiliation{Instituto de Nanociencia y Materiales de Aragón (INMA), CSIC-Universidad de Zaragoza, Pedro Cerbuna 12, Zaragoza 50009, Spain}

\author{Agustín Camón}
\affiliation{Instituto de Nanociencia y Materiales de Aragón (INMA), CSIC-Universidad de Zaragoza, Pedro Cerbuna 12, Zaragoza 50009, Spain}

\author{Pablo J. Alonso}
\affiliation{Instituto de Nanociencia y Materiales de Aragón (INMA), CSIC-Universidad de Zaragoza, Pedro Cerbuna 12, Zaragoza 50009, Spain}

\author{David Zueco}
\affiliation{Instituto de Nanociencia y Materiales de Aragón (INMA), CSIC-Universidad de Zaragoza, Pedro Cerbuna 12, Zaragoza 50009, Spain}
\email{dzueco@unizar.es}

\author{Fernando Luis}
\affiliation{Instituto de Nanociencia y Materiales de Aragón (INMA), CSIC-Universidad de Zaragoza, Pedro Cerbuna 12, Zaragoza 50009, Spain}
\email{fluis@unizar.es}

\maketitle
\tableofcontents
\newpage

\addcontentsline{toc}{subsection}{Supplementary Figure 1: Determination of the crystal structure of the DPPH samples with X-ray diffraction experiments.}
\begin{figure}[h!]
\begin{centering}
 \includegraphics[width=1\columnwidth, angle=-0]{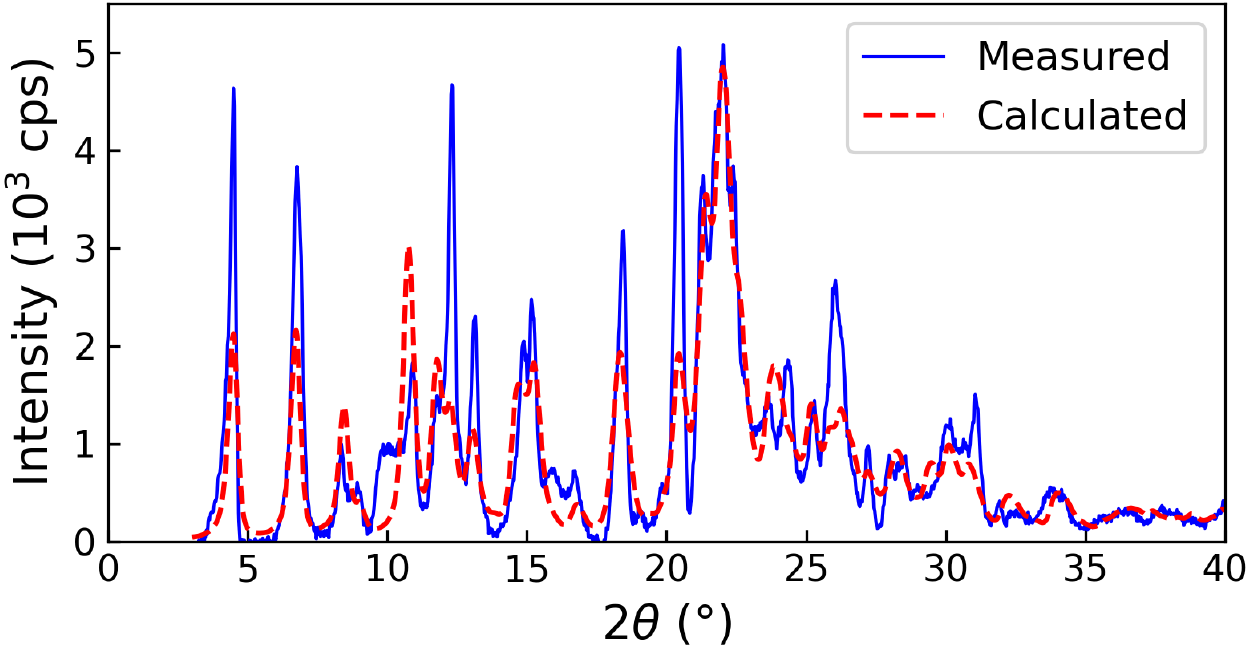}
\caption{\textbf{Determination of the crystal structure of the DPPH samples with X-ray diffraction experiments.} Powder diffraction experiment performed on a DPPH sample (blue solid line) and the calculated pattern for the DPPH-III crystal structure (red dashed line).\cite{Zilic2010}}
\label{fig:Xray_and_structure}
\end{centering}
\end{figure}

\newpage

\addcontentsline{toc}{subsection}{Supplementary Figure 2: Micro-Hall magnetometry.}
\begin{figure}[h!]
\begin{centering}
\includegraphics[width=0.7\columnwidth, angle=-0]{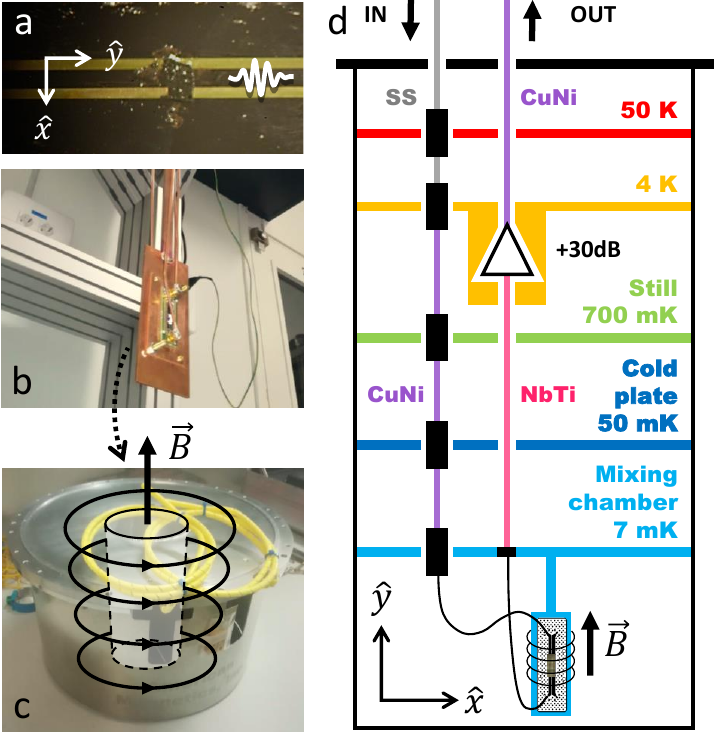}
\caption{\textbf{Experimental setup for on-chip transmission experiments.} 
(a) Microscopy image of a DPPH powder sample on top of the waveguide. (b) 
Cold finger hosting the superconducting chip. (c) 
Superconducting magnet in which the cold finger is inserted, generating a 
magnetic field parallel to the waveguide. (d) Scheme of the 
$^3$He-$^4$He dilution refrigerator with a base temperature $ \simeq 7$ mK, 
and of the input and output microwave lines used for the microwave 
transmission measurements.}
\label{fig:transmission_setup}
\end{centering}
\end{figure}

\newpage

\addcontentsline{toc}{subsection}{Supplementary Figure 3: Experimental setup for on-chip transmission experiments.}
\begin{figure}[h!]
\begin{centering}
\includegraphics[width=0.7\columnwidth, angle=-0]{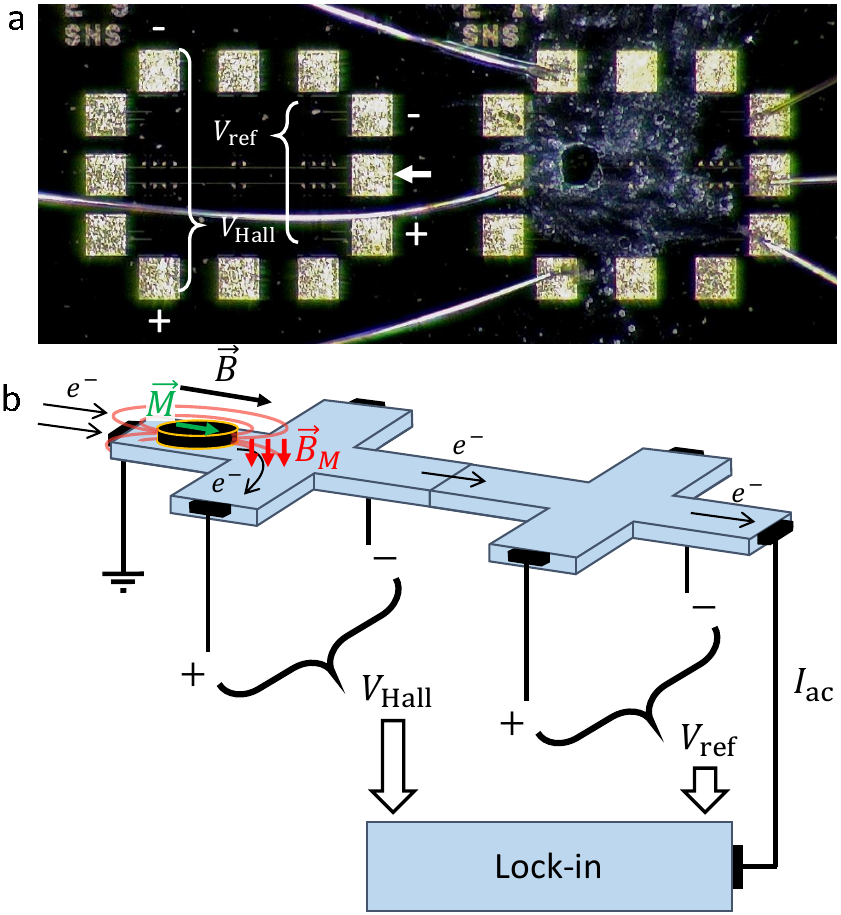}
\caption{\textbf{Micro-Hall magnetometry.} (a) Optical microscopy image showing a bare micro-Hall 
bar with two Hall crosses (left) and another one hosting a DPPH sample on top of one of the crosses (right). 
(b) Schematic image of the magnetometer, illustrating the differential detection method used 
to measure the Hall voltage associated with the sample's magnetization.}
\label{fig:micro_Hall}
\end{centering}
\end{figure}

\newpage

\section*{Supplementary Note 1: Magnetic characterization of DPPH}
\label{sec:magChar}

\subsection*{Magnetic susceptibility}
\begin{figure}[t!]
\begin{centering}
\includegraphics[width=0.55\columnwidth, angle=-0]{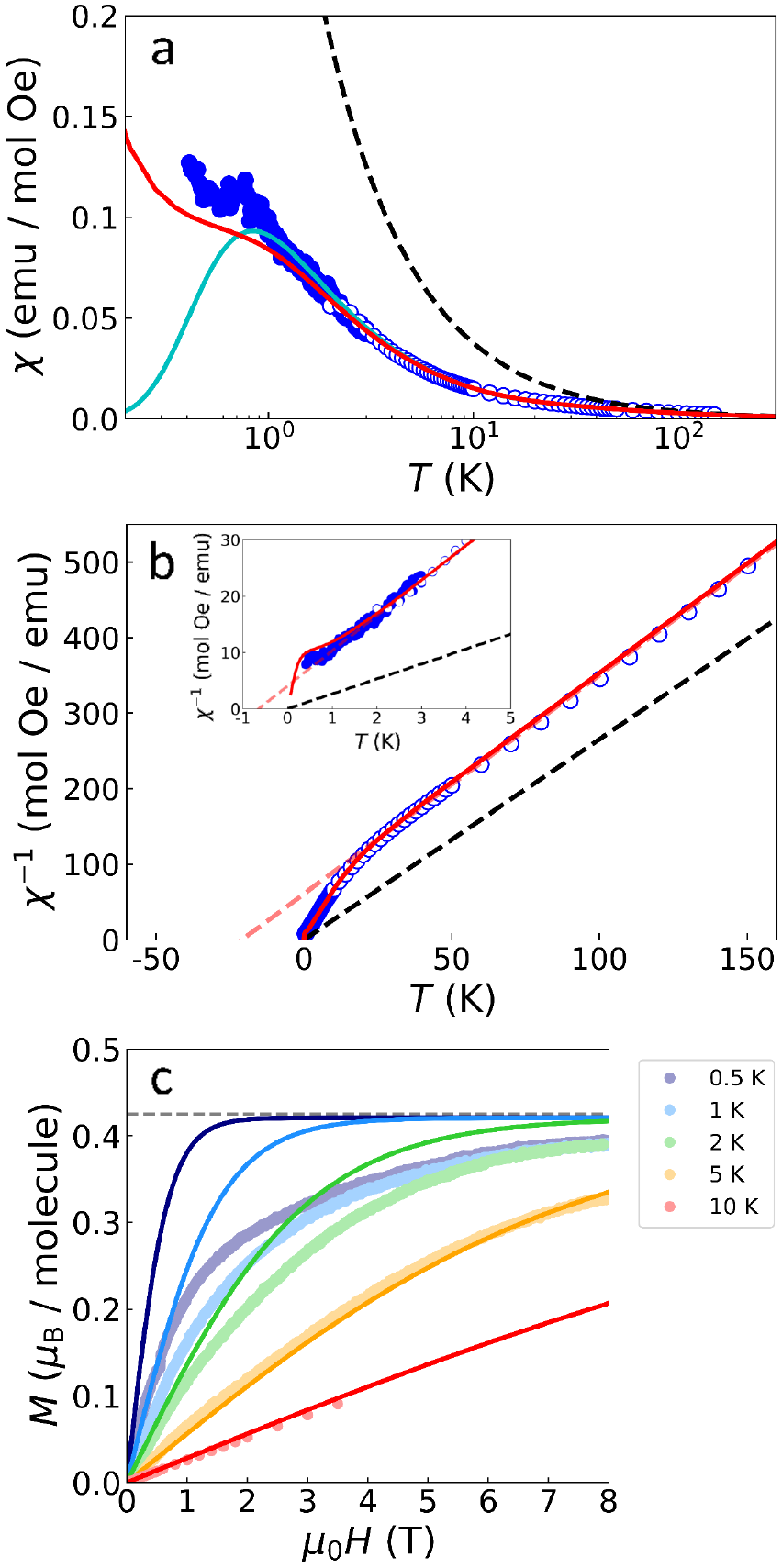}
\caption{\textbf{Magnetic characterization.} (a) dc-magnetic susceptibility $\chi$ of a DPPH powdered sample measured between $0.35$ and $300$ K with a magnetic field of $0.1$ T. Open dots were measured with a dc-SQUID magnetometer, solid dots with a micro-Hall magnetometer. The predictions of a 
model of infinite AF chains (red solid line) and a model of AF dimers (cyan solid line) are compared to the experimental data. The black dashed line is 
the Curie law for a fully paramagnetic DPPH sample. (b) Curie-Weiss plot of the reciprocal susceptibility $\chi^{-1}$ vs temperature. The light red dashed line is a least square linear fit of data measured above $50$ K. It gives a Weiss temperature $\theta = -21$ K associated with the formation of AF dimers in sublattice A. Inset: close up plot of the low temperature region 
data, which follow a Curie-Weiss law (light red dashed line) with $\theta = -0.65$ K associated to the formation of AF chains in sublattice B. (c) 
Magnetic isotherms, measured at $0.5$, $1$, $2$, $5$ and $10$ K for magnetic fields up to $8$ T. The lines show the dependence expected for non-interacting paramagnetic spins.}
\label{fig:magnetic_properties}
\end{centering}
\end{figure}

The dc-magnetic susceptibility $\chi$ measured between $0.35$ and $100$ K 
with a magnetic field $B = 0.1$ T can be found in Fig.~1 of the main text, as 
the product $\chi T$. Figure \ref{fig:magnetic_properties} shows two 
complementary representations of these data: $\chi$ and $1/\chi$ (Curie-Weiss 
plot) vs temperature. From room temperature down to about $35$ K, $\chi$ 
follows a Curie-Weiss law $\chi = C/(T-\theta)$, with $C$ the Curie constant 
and $\theta$ the Weiss temperature. We find that $C = 0.320 (1)$ emu K/mol Oe 
measured in this region is smaller, by a factor $x \equiv C / C_{S=1/2} \cong 0.85$, than $C_{S=1/2} = N_{\rm A} g_S^2/4 k_{\rm B} = 0.375$ emu K/mol Oe 
that would be expected for isolated $S = 1/2$ spins having the g-factor $g_S = 2.004$ reported for DPPH.\cite{Zilic2010} This suggests that about $15$ \% 
of DPPH molecules are in their oxidized form, which is diamagnetic, i.e. has $S = 0$. The presence 
of non-radical molecules is common in these materials.\cite{Zilic2010} The finite negative $\theta = -21$ K reflects the existence of 
relatively strong antiferromagnetic (AF) interactions. These interactions 
lead to the drop in the $\chi T$ product, proportional to the effective 
magnetic moment squared, that is observed below $100$ K. A quite similar 
behaviour has been observed in diverse DPPH 
derivatives.\cite{OhyaNishiguchi1979, Fujito1981, Zilic2010} It is associated 
with the coupling between nearest neighbour molecules, which tend to form AF 
dimers with an $S=0$ ground state. In the DPPH-III structure, such spin 
dimers are formed by DPPH molecules belonging to the A crystal sublattice, 
see Fig.~1a in the main text and Ref.~\citenum{Zilic2010}. 

\addcontentsline{toc}{subsubsection}{Supplementary Figure 4: Magnetic characterization.}

Below $10$ K, $\chi \, T$ shows a second drop. In this temperature region, 
$\chi$ follows also a Curie-Weiss law, but with a smaller $C \equiv C_{\rm B} = 0.154(1)$ emuK/mol Oe and a much smaller $\theta = -0.6$ K. This behaviour 
probably reflects the weaker interactions between spins in sublattice B of 
DPPH-III. The  response remains paramagnetic down to very low-$T$, with no 
clear indication of a phase transition. It is consistent with the formation 
of either AF dimers or low dimensional ($1D$) antiferromagnetic chains. 
However, specific heat data, discussed below, point 
to the formation of chains. We have calculated the dc susceptibility of 
finite chains of $N \leq 8$ $S=1/2$ spins coupled by a Heisenberg AF 
coupling. Details of the model and the calculations are given in Supplementary Note 5. The experimental $\chi T$ can be 
described by the combination of two contributions: one arising from AF dimers of A-type DPPH molecules and another one resulting from AF chains formed 
within the B sub-lattice. The reason why the low-$T$ $\chi$ can be accounted for with a model that considers small-$N$ chains is that the contribution per spin to the total susceptibility tends to vanish as $N \rightarrow \infty$. 
The uncompensated spins in odd-$N$ chains are responsible for the incomplete cancellation of $\chi T$ as $T \rightarrow 0$. The origin of finite size 
chains can be associated with the presence of about $15$\% of diamagnetic 
DPPH molecules, which are randomly distributed over the crystal lattice.

\subsection*{Magnetization isotherms}

Magnetic isotherms measured at $0.5$, $1$, $2$, $5$ and $10$ K for magnetic 
fields up to $8$ T are shown in Supplementary Fig.~\ref{fig:magnetic_properties}c. 
Below 5 K, the theory for non-interacting paramagnetic spins (solid lines) 
overestimates the observed magnetization (light dots), which is partially 
suppressed at low fields by AF interactions.

\subsection*{Specific heat}

\addcontentsline{toc}{subsubsection}{Supplementary Figure 5: Heat capacity of polycrystalline DPPH for different magnetic fields.}
\begin{figure}[b!]
\begin{centering}
\includegraphics[width=0.7\columnwidth, angle=-0]{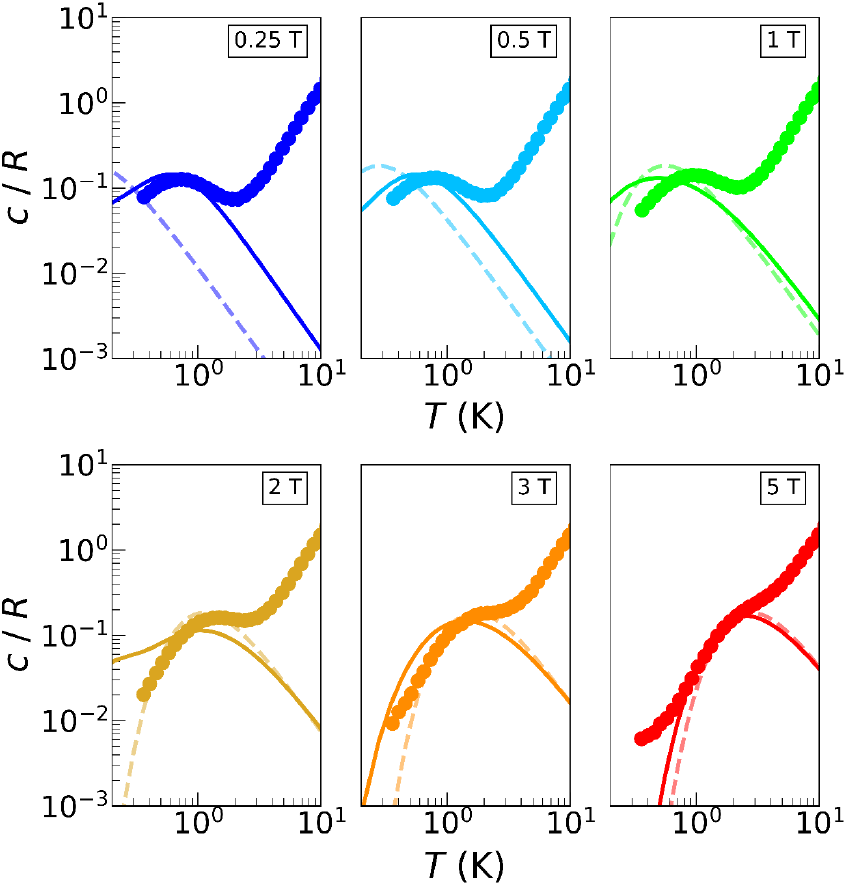}
\caption{\textbf{Heat capacity of polycrystalline DPPH for different magnetic fields.} The experimental results (dots) are compared with a theory of small AF chains (solid lines) and with the paramagnetic theory for non-interacting $S = 1/2$ spins (light dashed lines).
}
\label{fig:CPvsH}
\end{centering}
\end{figure}

The zero-field specific heat of DPPH is shown in Fig. 1c of the main text. 
Above $2$ K, it is dominated by a large contribution arising from lattice and 
molecular vibrations. At lower temperatures, an additional contribution shows 
up, which depends on magnetic field, as shown in Supplementary Fig.~\ref{fig:CPvsH}. 
Therefore, it reflects spin excitations. The broad shape of this anomaly 
confirms that no phase transition to long-range magnetic order takes place in 
this temperature region.

The zero-field data agree well with predictions for the specific heat of 
infinite AF chains,\cite{Blote1975} and seem to rule out the formation of 
dimers within the B sublattice. Lacking an analytical model for the 
$B \neq 0$ specific heat of infinitely long AF spin chains, we had to content 
ourselves with computing the heat capacity of finite chains of up to eight $S=1/2$ spins. The model is discussed in 
Supplementary Note 5. This simplified model 
accounts reasonably well for data measured at low ($B \leq 0.5$ T and high 
($B > 2$ T) magnetic fields. In the latter case, it tends towards the 
predictions for non-interacting spins, as expected. Deviations are observed 
at intermediate fields ($1-2$ T).

\addcontentsline{toc}{subsubsection}{Supplementary Figure 6: cw-EPR experiments.}
\begin{figure}[t!]
\begin{centering}
\includegraphics[width=0.7\columnwidth, angle=-0]{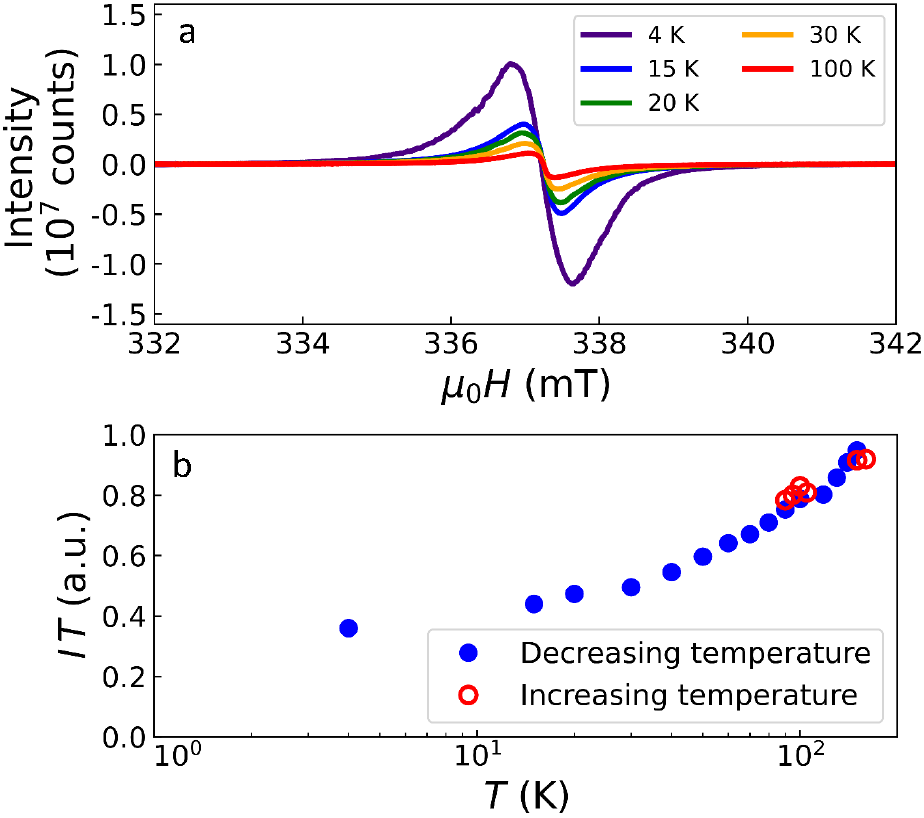}
\caption{\textbf{cw-EPR experiments.} (a) cw-EPR spectra of a powder DPPH 
sample measured at five different temperatures. (b) Evolution of the product 
of the cw-EPR spectrum intensity times temperature, measured first for 
decreasing temperatures (blue closed dots) and then for increasing 
temperatures (red open dots).}
\label{fig:cw_EPR}
\end{centering}
\end{figure}

\subsection*{Electron Paramagnetic Resonance}
The single narrow line shape measured with this technique, shown in Supplementary Fig.~\ref{fig:cw_EPR}a, agrees with a spin $S = 1/2$ and $g_S = 2.004$. The 
line intensity $I$ shows a paramagnetic behaviour, with $I T$ (Supplementary 
Fig.~\ref{fig:cw_EPR}) behaving very much as $\chi T$ (Fig. 1 in the main 
text). The line shape suggests that the line width is dominated by the 
homogeneous broadening down to $4$ K. It is known that, in the paramagnetic 
phase of DPPH, exchange interactions tend to suppress the inhomogenous 
broadening associated with dipole-dipole interactions \cite{VanVleck1948} via the exchange narrowing mechanism.\cite{Anderson1953} This property makes DPPH 
a good reference material for calibrating commercial EPR systems.\cite{Yordanov1996}
\addcontentsline{toc}{subsubsection}{Supplementary Figure 7: Pulsed-EPR experiments.}
\begin{figure}[h!]
\begin{centering}
\includegraphics[width=0.7\columnwidth, angle=-0]{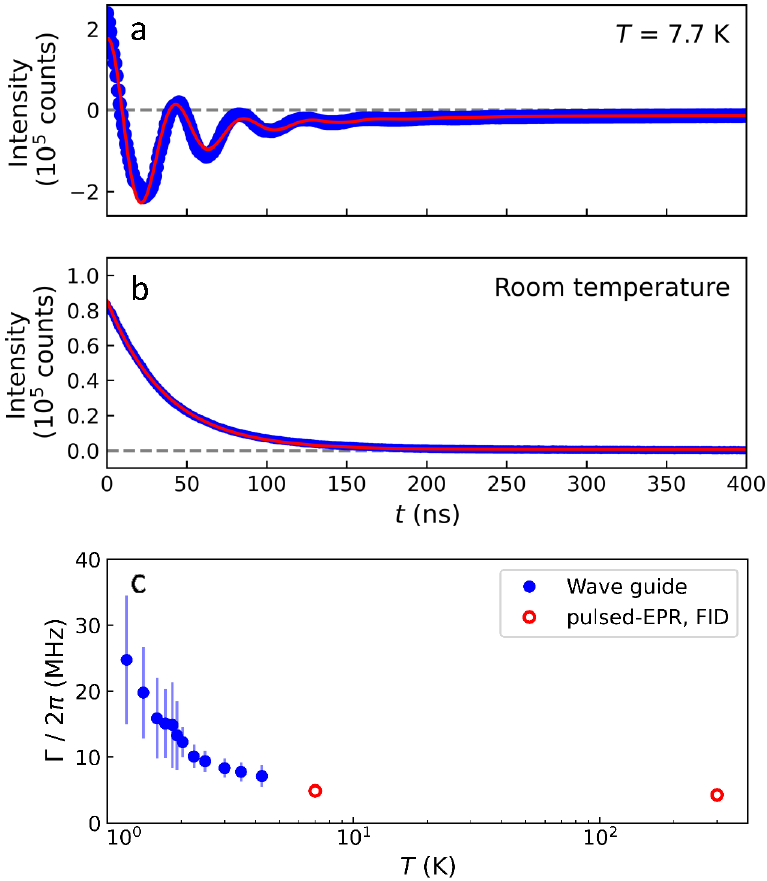}
\caption{\textbf{Pulsed-EPR experiments.} (a) Free Induction Decay (FID) of a DPPH powered sample measured at $7.7$ K. (b) FID of the same sample measured at room temperature. (c) Temperature dependence of the DPPH absorption 
spectrum width. The values extracted from FID data (red open dots) are 
compared to those derived from on-chip microwave transmission experiments (blue closed dots).}
\label{fig:pulsed_EPR}
\end{centering}
\end{figure}

Free induction decay measurements were carried out at $7.7$ K and at room temperature on DPPH powder samples. The results are shown in Supplementary
Fig.~\ref{fig:pulsed_EPR}. The fit of the two signals is compatible with a 
homogeneous broadening $\gamma_{\phi} / 2 \gpi = 1 / 2 \gpi T_{\rm{m}} = 4.8$ 
MHz and $4.3$ MHz at $7.7$ K and room temperature, respectively. These values 
correspond to $T_{\rm{m}} = 33$ ns and $37$ ns, respectively. Supplementary Fig. \ref{fig:pulsed_EPR}c shows that these values are also compatible with those 
obtained from microwave transmission experiments at lower temperatures.

\clearpage

\section*{Supplementary Note 2: Normalization of microwave transmission and reflection data}
\label{sec:transmission_normalization}

\addcontentsline{toc}{subsubsection}{Supplementary Figure 8: Normalization of transmission data.}
\begin{figure}[h!]
\begin{centering}
\includegraphics[width=0.7\columnwidth, angle=-0]{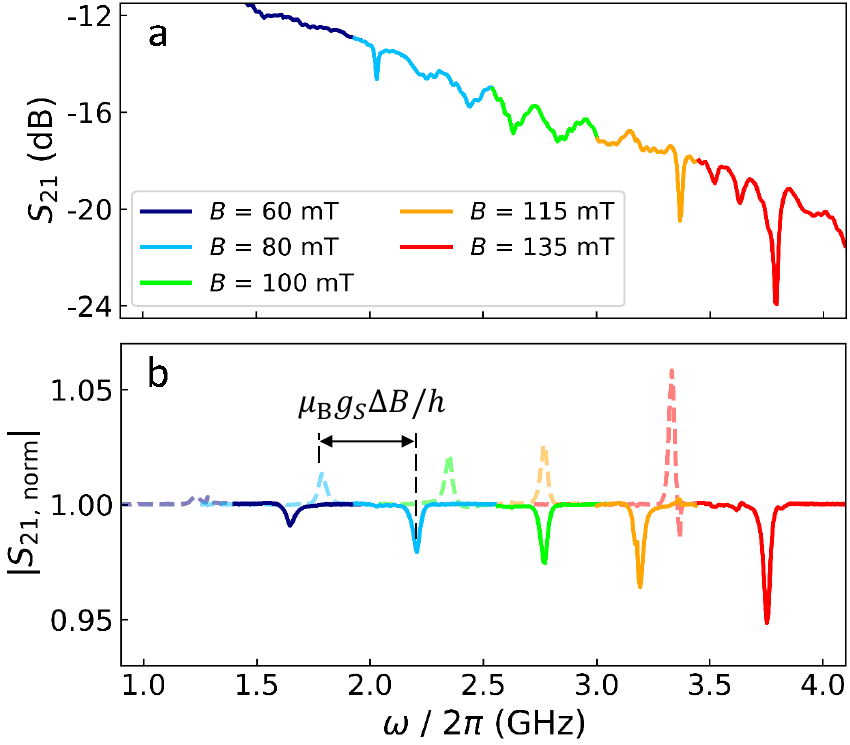}
\caption{\textbf{Normalization of transmission data.} (a) Raw 
microwave transmission data $\vert S_{21} \vert$ measured on 
a waveguide coupled to 
a powdered DPPH sample at $T=2$ K and for five different 
magnetic fields, as indicated. They are chosen in such a way 
that spin resonances occurring at 
$\omega = \Omega \equiv \mu_{\rm B} g_S B / \hbar$ do 
not overlap with each other. Some of these resonances are 
masked by spurious modes in 
${\cal F} (\omega)$ (see Eq. (\ref{eq:raw-transmission})). 
(b) Normalized transmission $\vert S_{21, {\rm norm}} \vert$ 
showing clear features associated with the spin resonances. 
It is obtained from the raw data via the application of 
the normalization procedure defined by  
Eq.~(\ref{eq: transmission_norm}). Only $\vert S_{21} \vert$ 
data measured at frequencies sufficiently close to the spin 
resonance frequency $\Omega$ (solid lines) are used to 
extract the spin-photon coupling $G$ and the dissipation rate 
$\Gamma$. The rest (dashed lines), 
which includes an upwards peak associated with the 
normalization with $|S_{21}(B+\Delta B)|$ (in this example, 
$\Delta B = -15$ mT), is discarded from 
the analysis. This peak is not relevant as long as it does not overlap 
with the 'true' absorption peak at $\omega = \Omega$. This sets the condition 
$\mu_{\rm B} g_S |\Delta B| / h \gg \Gamma / 2 \gpi$.}
\label{fig:transmission_normalization}
\end{centering}
\end{figure}

In the raw transmission data, the absorption by the DPPH 
spins is often masked by spurious modes, see Supplementary
Fig.~\ref{fig:transmission_normalization}a, which arise from 
unavoidable imperfections of the microwave circuit, e.g. 
near the chip boundaries or at the bonds to the PCB. Besides, 
the transmission shows an attenuation that increases with 
increasing frequency. The measured transmission can be understood as the 
convolution:
\begin{equation}
S_{21} \simeq S_{21,{\rm norm}} {\cal F} \; ,
\label{eq:raw-transmission}
\end{equation}
\noindent of the normalized $S_{21,{\rm norm}}$ for a 
perfect transmission waveguide coupled to the spins with 
a frequency-dependent function ${\cal F}$ that includes the 
contribution of the modes and the attenuation.
\addcontentsline{toc}{subsubsection}{Supplementary Figure 9: Normalized transmission and reflection measured at $T = 2$ K for magnetic fields between $100$ and $150$ mT.}
\begin{figure*}[t!]
    \centering
    \includegraphics[width = 0.7\textwidth]{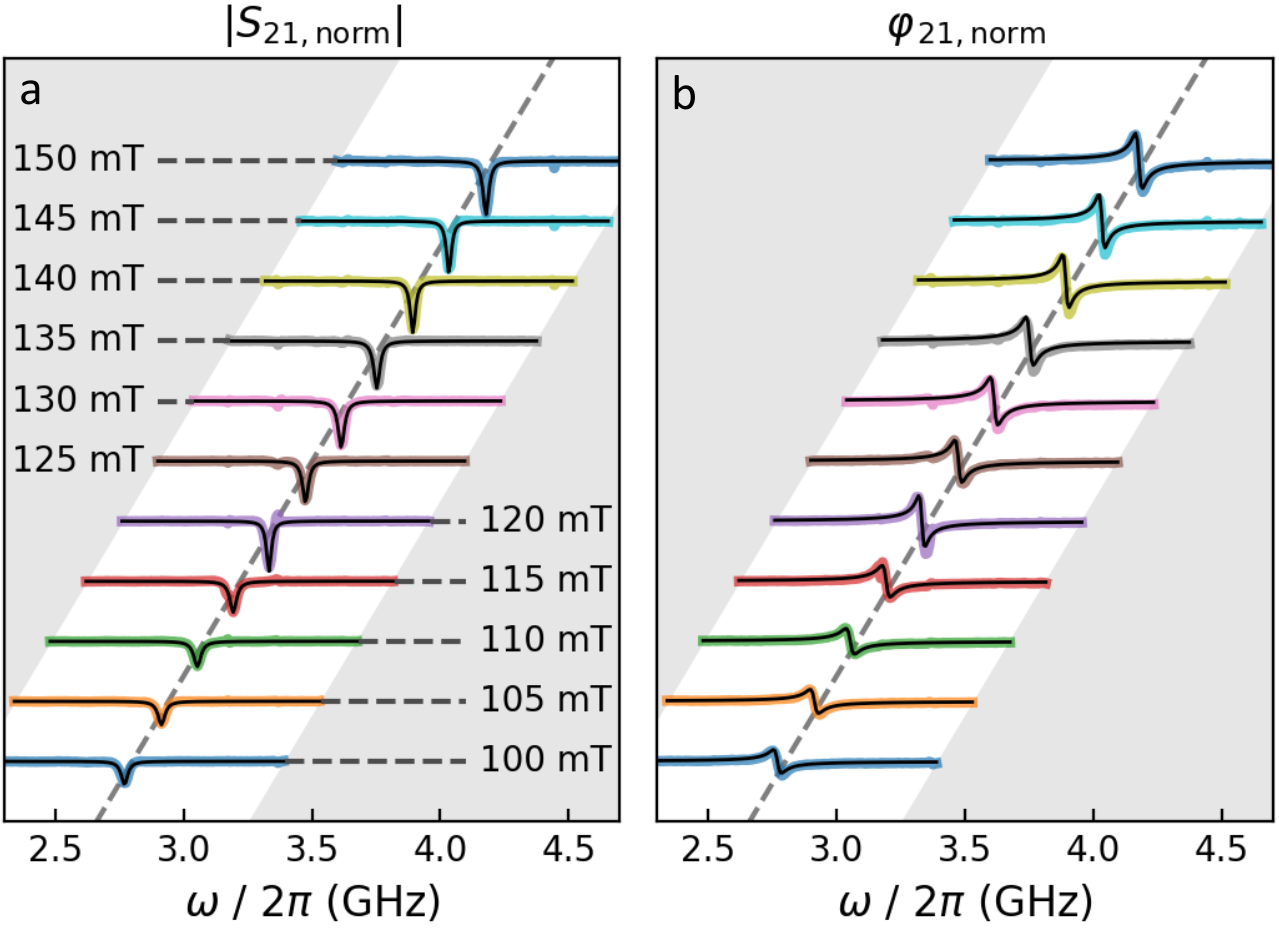}
    \caption{\textbf{Normalized transmission and reflection measured at $T = 2$ K for magnetic fields between $100$ and $150$ mT.} (a) and (b) Normalized transmission amplitude, $|S_{21,{\rm norm}}|$, and phase, $\varphi_{21,{\rm norm}}$. Black solid lines show the joint fit of the two quantities to the input-output theory (Eq. (1) in the main text), with the same parameters. In both plots, the position of the resonance at different fields follows the relation $\omega = \Omega = \mu_{\rm B} g_S B / \hbar$ (grey dashed line).}
    \label{fig:S21_amp_and_phase}
\end{figure*}
\addcontentsline{toc}{subsubsection}{Supplementary Figure 10: Temperature dependence of the normalized transmission measured at three different magnetic fields.}
\begin{figure*}[t!]
    \centering
    \includegraphics[width = 0.7\textwidth]{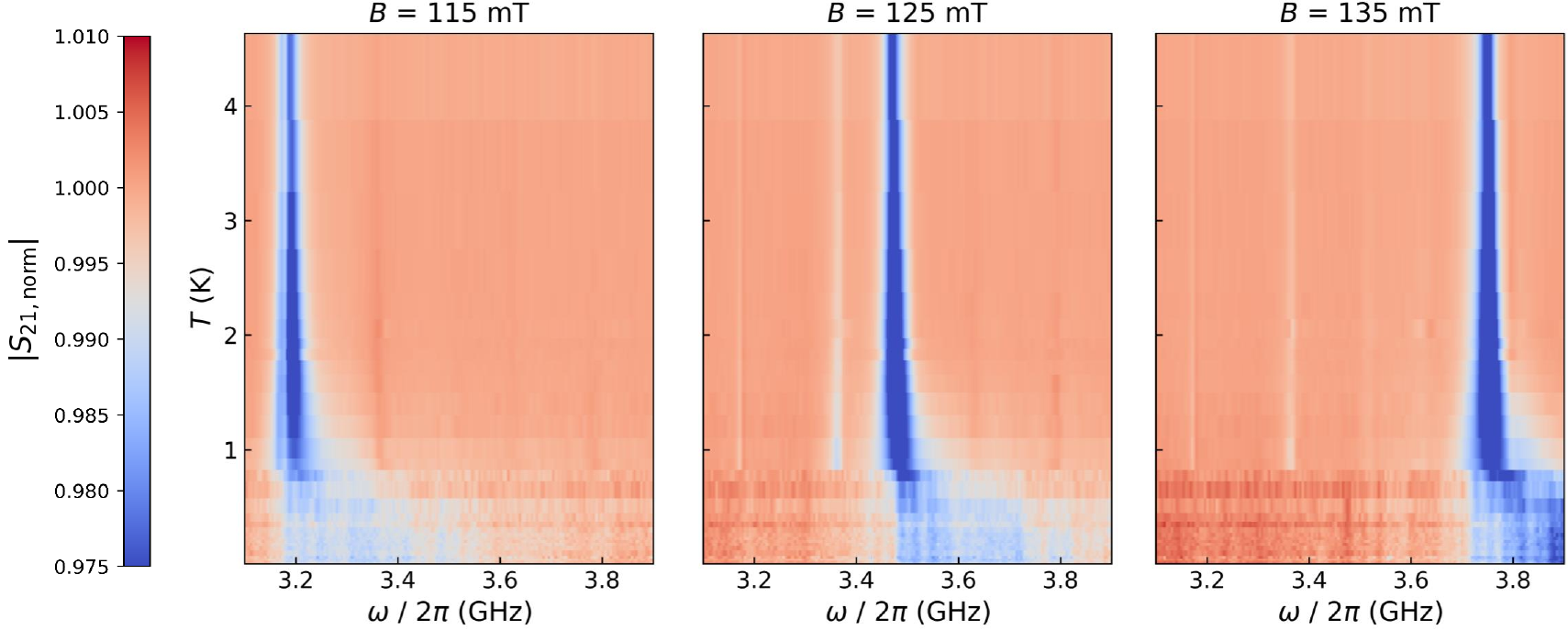}
    \caption{\textbf{Temperature dependence of the normalized transmission measured at three different magnetic fields.} The resonance frequency of the absorption signal associated to the spin ensemble increases as the magnetic field is increased. In comparison, spurious modes of the transmission line that are not completely removed by the normalization (see the two modes at $\sim 3.17$ and $3.37$ GHz) do not move.}
    \label{fig:S21_field_comparison}
\end{figure*}
\addcontentsline{toc}{subsubsection}{Supplementary Figure 11: Normalized transmission and reflection.}
\begin{figure*}[t!]
    \centering
    \includegraphics[width = \textwidth]{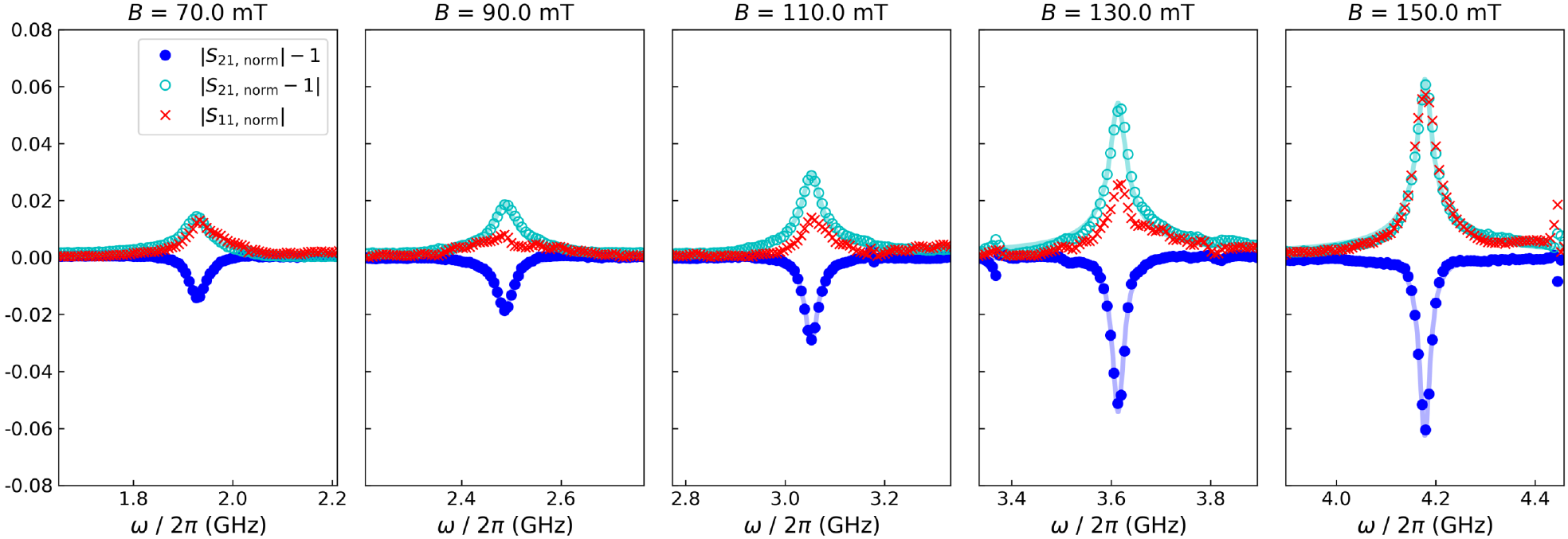}
    \caption{\textbf{Normalized transmission and reflection.} Normalized $\vert S_{21} \vert$ and $\vert S_{11} \vert$ measured at $2$ K and for five magnetic fields, in frequency ranges around the corresponding spin resonance condition $\omega = \Omega = \mu_{\rm B} g_S B / \hbar$. Lines show the fit of the transmission data to the paramagnetic theory (Eq. (1) in the main text). The normalized reflection shows a resonance consistent with the transmission results.}
    \label{fig:S21_S11_comparison}
\end{figure*}

In order to obtain $S_{21,{\rm norm}}$ at any magnetic field, 
the raw $S_{21}$ data need to be normalized. A simple 
normalization procedure exploits the fact that 
$S_{21,{\rm norm}}$ must be equal to 
$1$ unless there is a resonant absorption of photons by the 
spin system. Therefore, ${\cal F}$ can be approximated from  
data measured at a magnetic field $B+\Delta B$ that gives no 
spin resonance in the frequency region of interest. The 
normalized transmission is then estimated as follows 
\cite{Clauss2013,Gimeno2023}
\begin{equation}
\label{eq: transmission_norm}
|S_{21,{\rm norm}}(B)| = \dfrac{|S_{21}(B)|}{|S_{21}(B+\Delta B)|}
\; .
\end{equation}
\noindent As shown in 
Supplementary Fig.~\ref{fig:transmission_normalization}b, this 
method effectively removes the contribution of ${\cal F}$, 
provided that it is independent of $B$ or that it varies 
sufficiently slowly with it. The normalization step 
$\Delta B$ must be as small as possible while avoiding the 
overlap of the spin resonances in 
$S_{21}(B)$ and $S_{21}(B+\Delta B)$. This requires that 
$|\Delta B| \gg \hbar \Gamma / \mu_{\rm B} g_S$. Similarly, 
the phase $\varphi_{21,{\rm norm}}$ of $S_{21,{\rm norm}}$ 
can be retrieved as:\cite{Clauss2013}
\begin{equation}
\label{eq: transmission_phase}
\varphi_{21,{\rm norm}}(B) = \varphi_{21}(B) - \varphi_{21}(B+\Delta B) \; ,
\end{equation}
Supplementary Fig. \ref{fig:S21_amp_and_phase} shows 
normalized data for the amplitude and phase of the 
transmission measured at $T=2$ K and 
for different magnetic fields. The field-dependent features 
are solely related to the interaction of propagating photons 
with the DPPH spins and 
can, therefore, be used to estimate the collective coupling 
$G$ and the spin decay rate $\Gamma$ as described in the 
main text where. All transmission data shown in the main 
text correspond to data normalized as described above 
although, to simplify the notation, we refer to them as 
$S_{21}$. 

In practice, there always remain some weak spurious features 
in $S_{21,{\rm norm}}$, which are approximately independent 
of $\omega$ and of $B$ (see Figs. 2a and 3 in the main 
text). The latter property allows to 
unequivocally distinguish them from the changes associated 
to the resonant spin-photon coupling, as shown in Supplementary
Fig. \ref{fig:S21_field_comparison}. 

The reflection $S_{11}$ could only be retrieved from data measured in the dilution refrigerator that did not have attenuators in the input line, nor a 
cryogenic amplifier in the output line. Even in this case, reflections at different parts of the circuit introduce a large contribution $\mathcal{R}$ that must be removed in the normalization process:
\begin{equation}
S_{11} \simeq S_{11,{\rm norm}} {\cal F} + \mathcal{R} \; ,
\label{eq:raw-reflection}
\end{equation}
with the amplitude of the normalized reflection estimated as:
\begin{equation}
\label{eq: reflection_norm}
|S_{11,{\rm norm}}(B)| = \dfrac{|S_{11}(B) - S_{11}(B+\Delta B)|}{|S_{21}(B+\Delta B)|}
\; .
\end{equation}

Supplementary Fig. \ref{fig:S21_S11_comparison} compares the normalized reflection $S_{\rm 11,\,norm}$ obtained from Eq. (\ref{eq: reflection_norm}) with the normalized transmission $S_{\rm 21,\,norm}$, both at $T = 2$ K. Both signals 
show the expected paramagnetic spin resonance at a frequency $\omega = \Omega = \mu_{\rm B} g_S B / \hbar$, and with a similar visibility. 
At those fields for which the reflection signal is clean enough, the  condition $|S_{\rm 11,\,norm}| = |1 - S_{\rm 21,\,norm}|$ is observed.

\clearpage

\section*{Supplementary Note 3: Microwave transmission as a function of temperature: from the paramagnetic regime to the onset of spin-spin correlations}
\label{sec:transmission_paramagnetic}

\begin{figure}[b!]
\centering
\includegraphics[width=0.7\columnwidth, angle=-0]{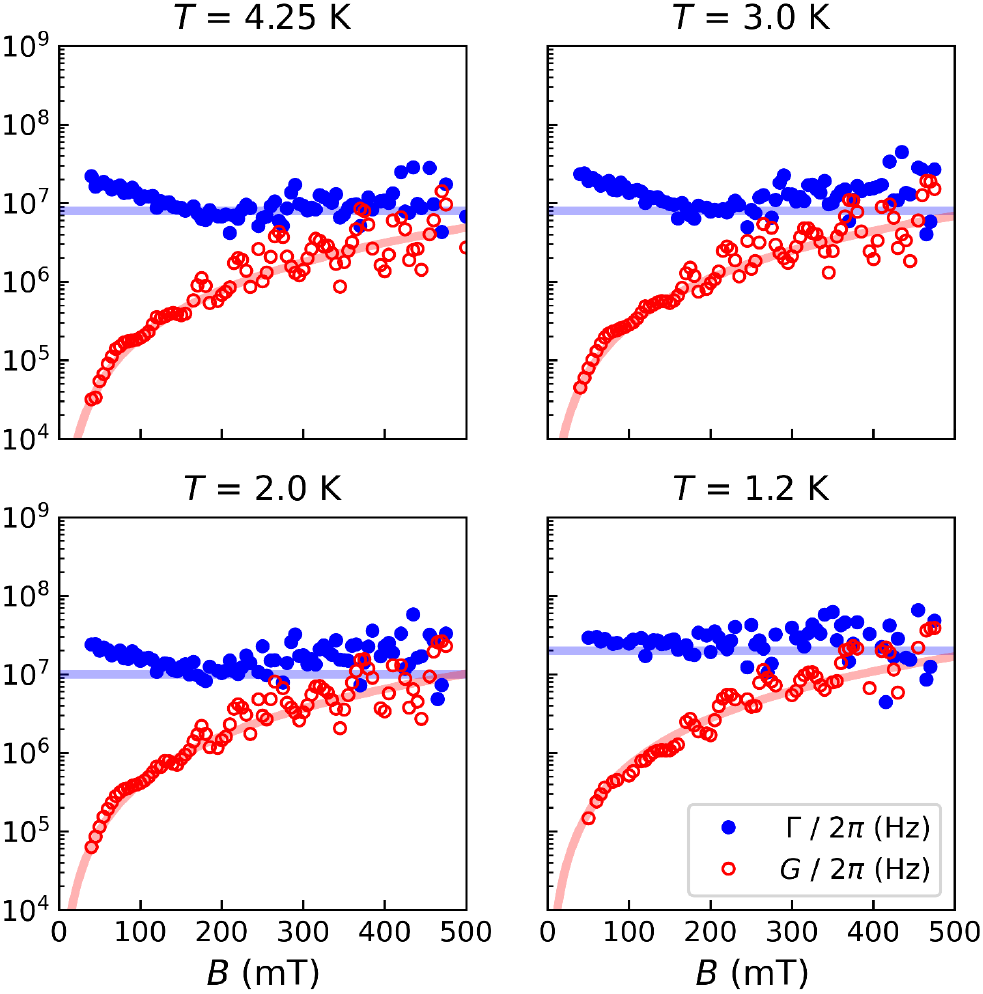}
\caption{\textbf{Magnetic field dependence of $G$ and $\Gamma$ measured 
at four selected temperatures within the paramagnetic phase of DPPH.} $G$ (red open dots) and $\Gamma$ (blue closed dots) values were obtained by 
fitting the normalized transmission data using Eq. (2) in the main text. Light solid lines show the average $\Gamma$ (discarding the low magnetic field region) and the result of a least-squares fit of $G$ with Eq. (\ref{eq:pm_phase_trans_params}).}
\label{fig:G_and_Gamma}

\end{figure}

Supplementary Fig. \ref{fig:G_and_Gamma} shows the field and temperature dependence of 
$G$ and $\Gamma$ in the paramagnetic phase of the B-type DPPH sublattice ($T > $ 0.7 K). In a one-dimensional waveguide, the coupling $G$ is proportional 
to $\omega$. Close to a spin resonance, 
$\omega$ approximately equals $\Omega = \mu_{\rm B} g_S B / \hbar$, and $G$ 
can then be written as (see Eq.~(3) in the main text):
\begin{equation}
\label{eq:pm_phase_trans_params}
G = 2 \gpi \alpha \Omega N
\tanh{\left(\dfrac{\hbar \Omega}{2 k_{\rm B} T}\right)} \; ,
\end{equation}
\noindent where $N$ is the total number of spins coupled to the microwave 
magnetic field and $\alpha$ is a temperature-independent and field-
independent dimensionless scaling factor. The only free parameter in the 
fit of $G$ for all temperatures and magnetic fields within the 
paramagnetic phase is $2 \gpi \alpha N = 0.00441(6)$. Therefore, the 
enhancement of $G$ by a factor 
$N_{\rm{eff}} = N \tanh{\left(\hbar \Omega / 2 k_{\rm B} T\right)}$ that 
is observed experimentally is a direct manifestation of the increase in 
the radiative rate decay from superradiant collective states.

\addcontentsline{toc}{subsubsection}{Supplementary Figure 12: Magnetic field dependence of $G$ and $\Gamma$ measured at four selected temperatures within the paramagnetic phase of DPPH.}

For $T > 1$ K, the dissipation rate $\Gamma$ is found to 
depend weakly on both magnetic field and temperature. We 
attribute the increase in $\Gamma$ observed below $100$ mT 
to the change in the inhomogeneous broadening associated 
with the magnetic depolarization of the sample. The values 
derived from these experiments are compatible with those 
obtained from EPR, see Supplementary Fig.~\ref{fig:pulsed_EPR}c. Besides, 
$\Gamma / 2 \gpi \cong 8$ MHz measured at $T = 4.2$ K is also 
compatible with that obtained for DPPH samples coupled to 
coplanar resonators at the same 
temperature.\cite{Gimeno2020} 

Remarkably, $G$ becomes 
comparable to $\Gamma$ for magnetic fields $B >$ 400 mT. 
This yields a spin resonance visibility 
$\eta \simeq G/(G + \Gamma)$ that becomes close to $0.6$ 
for $B = 0.5$ T. The spin ensemble and the waveguide then 
enter a regime in which $G$ is close to $\Gamma$. This 
condition resembles the threshold of the strong coupling 
limit ($G \geq \Gamma$) that is commonly reached when 
coupling spins to a resonator. This regime has been achieved 
with atomic ensembles and superconducting qubits coupled to 
waveguides \cite{sheremet2023waveguide}. Notice, however, 
than these systems couple to the photon electric field, 
often in a very high frequency range, whereas here we are 
dealing with the magnetic field of microwave photons. 

In order to fully appreciate the magnitude of $G$ attained in these 
experiments, it is worth estimating what it would lead to in the case of a 
coplanar resonator. Input-output theory\cite{Hummer2013} provides a relation 
between $G$ and the coupling $G_{\rm res}$ to a cavity resonating at a given 
frequency $\omega$
\begin{equation}
G_{\rm res} = \sqrt{\frac{G \omega}{\gpi}} \; .
\label{eq:Gres}
\end{equation}
\noindent Using this relation, the maximum coupling 
$G/2 \gpi \simeq 12$ MHz to the line that we measure at 
$B=500$ mT and at $\omega/2 \gpi = 14$ 
GHz would correspond to $G_{\rm res}/2 \gpi \sim 0.25$ GHz. 
This value is more than one order of magnitude larger than 
those previously reported for DPPH samples at coplanar 
superconducting resonators.\cite{Ghirri2016,Mergenthaler2017}

\begin{figure}[b!]
\begin{centering}
\includegraphics[width=0.7\columnwidth, angle=-0]{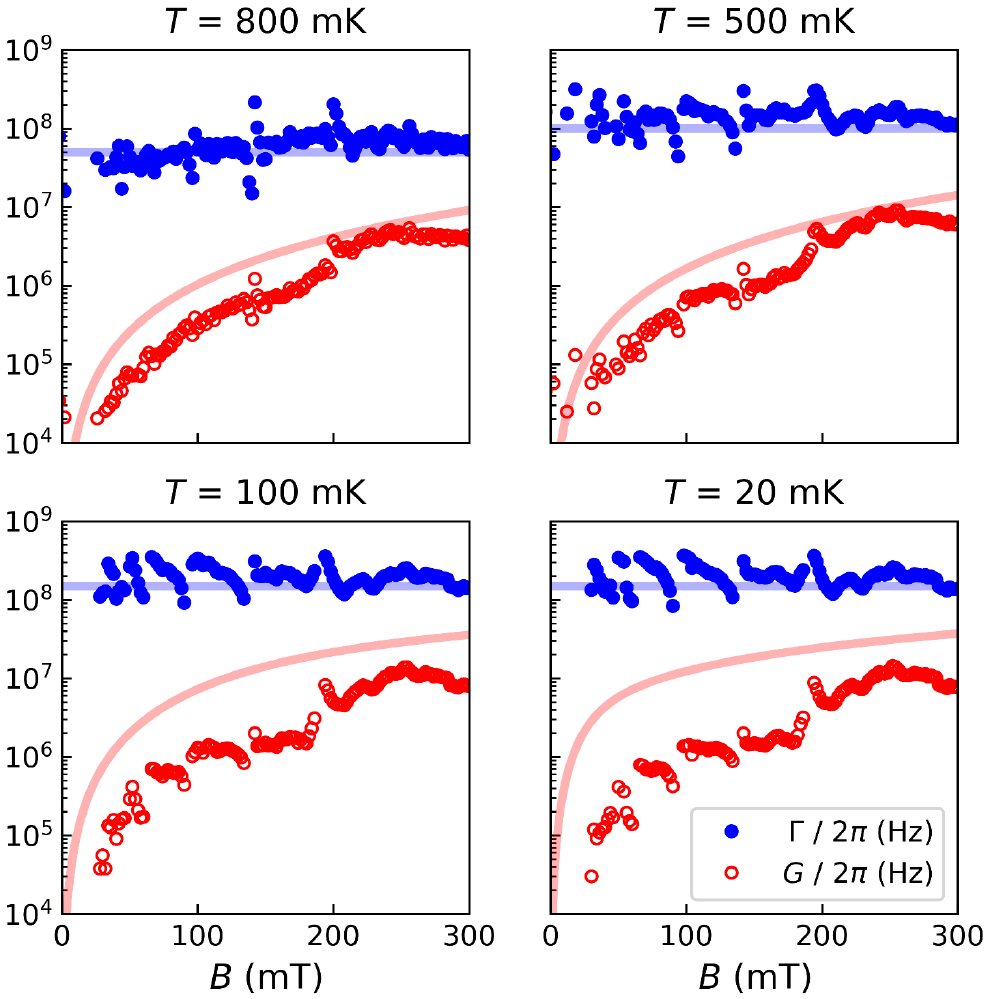}
\caption{\textbf{Magnetic field dependence of $G$ and $\Gamma$ measured 
at four selected temperatures below $1$ K, where the paramagnetic theory fails.} $G$ (red open dots) and $\Gamma$ (blue closed dots) values were obtained by 
fitting the normalized transmission data using Eq. (2) in the main text. Light solid lines show the average $\Gamma$ (discarding the low magnetic field region) and the prediction for the spin photon coupling that follows from the paramagnetic Eq. (\ref{eq:pm_phase_trans_params}), extended here to lower temperatures with the same $2 \gpi \alpha N = 0.00441(6)$.}
\label{fig:G_and_Gamma_lowT}
\end{centering}
\end{figure}

As a comparison, Supplementary Fig.~\ref{fig:G_and_Gamma_lowT} shows the 
dependence of $G$ and $\Gamma$ with temperature and magnetic field for 
$T < 1$ K, where the effective spin-photon coupling gradually falls below the 
predictions for the paramagnetic regime 
(Eq.~\eqref{eq:pm_phase_trans_params}), represented by the solid red lines, 
and the decay rate increases.

\addcontentsline{toc}{subsubsection}{Supplementary Figure 13: Magnetic field dependence of $G$ and $\Gamma$ measured at four selected temperatures below $1$ K, where the paramagnetic theory fails.}

\clearpage 

\section*{Supplementary Note 4: Microwave transmission as a function of magnetic field: Zeeman versus exchange interaction}

Very low-$T$ transmission experiments have been performed under magnetic 
fields $B \leq 0.3$ T. Illustrative examples are shown in Figs. \ref{fig:S21_field_comparison} and \ref{fig:S21_zeeman}. The main effect of 
$B$ is shifting the center frequency $\Omega$ of the spin resonance. Besides, 
increasing $B$ leads to a higher spin polarization, thus to a stronger spin-
photon coupling and to a higher visibility at any temperature (see Fig. 2 in 
the main text and Figs. \ref{fig:G_and_Gamma} and \ref{fig:G_and_Gamma_lowT}). 

A closer look at the data also reveals a competition between 
spin-spin interactions, the same that broaden and shift the spin resonance as 
$T$ decreases, and the coupling of each spin to $B$. This effect can be 
seen in Supplementary Fig. \ref{fig:zeeman_observables}, which shows
how the resonance width $\Gamma$, the spin-photon coupling $G$ and the shift 
$\delta \Omega$ of the mean spin resonance frequency depnd on $T$ for three 
different values of $B$. Increasing magnetic field tends to reduce those 
effects that we have associated with the onset of spin-spin interactions: the 
changes in $\Gamma$, $\delta \Omega$ and visibility become smaller, and take 
place less abruptly. However, fields much higher than $0.3$ T, corresponding 
to frequencies beyond our experimentally accessible window, would still be 
required to completely erase spin correlations and recover the paramagnetic 
regime, as can be seen from heat capacity data shown in Fig. \ref{fig:CPvsH}. 

\addcontentsline{toc}{subsubsection}{Supplementary Figure 14: Temperature dependence of the normalized transmission measured at three different magnetic fields.}
\begin{figure*}[t!]
    \centering
    \includegraphics[width = \textwidth]{SuppFigs/S21_2d_Zeeman_comparison.png}
    \caption{\textbf{Temperature dependence of the normalized transmission measured at three different magnetic fields.} As $B$ increases, the spin resonance shift $\delta \Omega$ and its broadening, which are related to the onset of spin-spin correlations along randomly oriented chains of DPPH molecules, are reduced and take place more smoothly with decreasing $T$.}
    \label{fig:S21_zeeman}
\end{figure*}

\addcontentsline{toc}{subsubsection}{Supplementary Figure 15: Dependence of the parameters describing the spin-photon coupling on temperature and magnetic field.}
\begin{figure}[h!]
\begin{centering}
\includegraphics[width=0.85\textwidth, angle=-0]{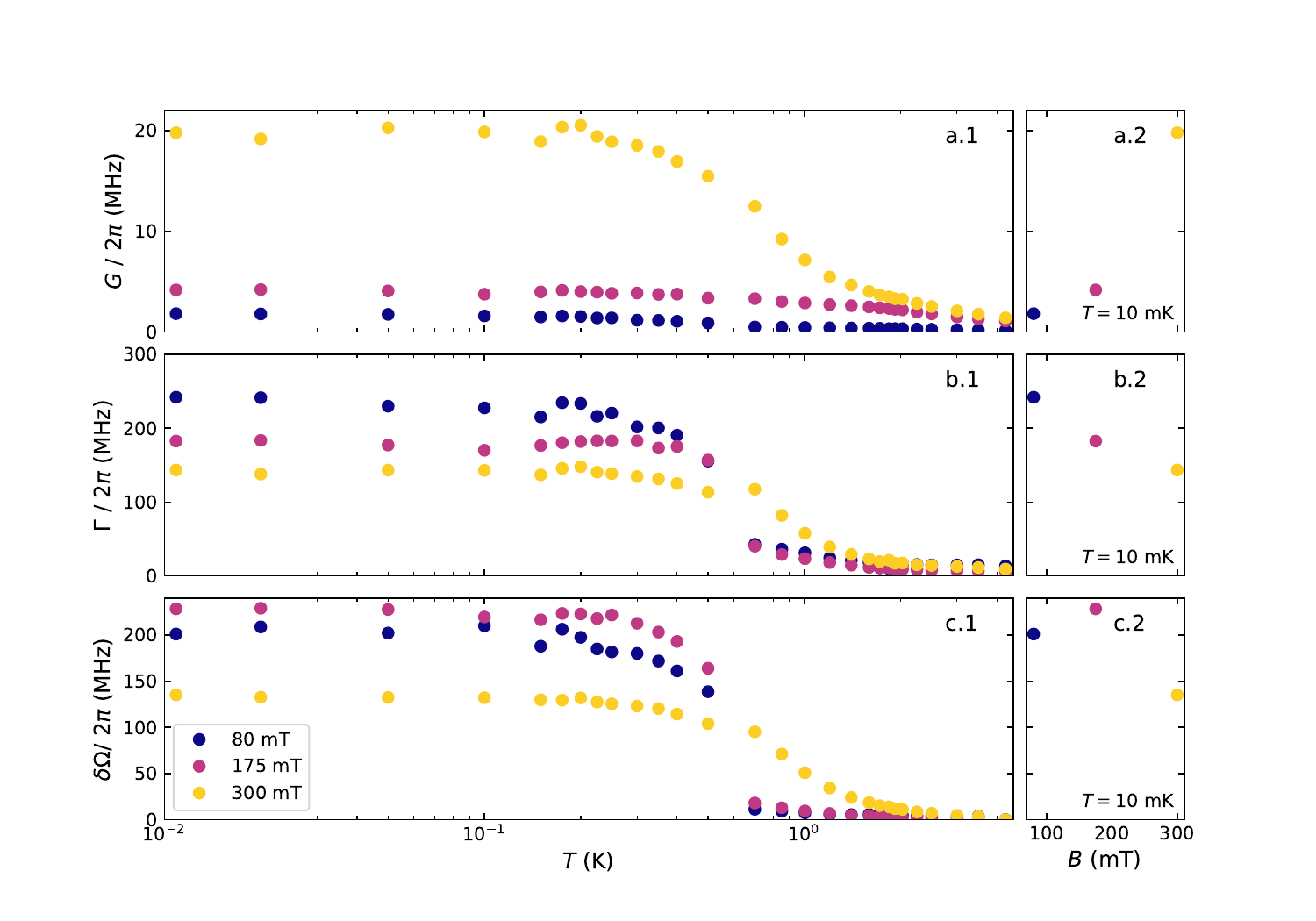}
\caption{\textbf{Dependence of the parameters describing the spin-photon coupling on temperature and magnetic field.}
Panels a.1, b.1 and c.1 show, respectively, the collective spin-photon 
coupling $G$, the spin resonance width $\Gamma$ and the shift $\delta\Omega$ 
of the spin resonance frequency as a function of temperature for three 
different magnetic fields. The right hand panels a.2, b.2 and c.2 show these 
quantities at $T=10$ mK for increasing magnetic fields.}
\label{fig:zeeman_observables}
\end{centering}
\end{figure}

\clearpage 

\section*{Supplementary Note 5: Effect of input microwave power}
\label{sec:microwave_power}

The resonance of the DPPH ensemble was measured at $T$ = 0.3 K and $B$ = 0.22 T with different input microwave powers $P_{\rm in}$ between -25 and -10 dBm. No saturation effect was observed in the resonance due to the change in $P_{\rm in}$, as shown in Supplementary Fig.~\ref{fig:power_sweep}.

\addcontentsline{toc}{subsubsection}{Supplementary Figure 16: Spin resonance versus microwave power.}
\begin{figure}[h!]
\begin{centering}
\includegraphics[width=0.7\columnwidth, angle=-0]{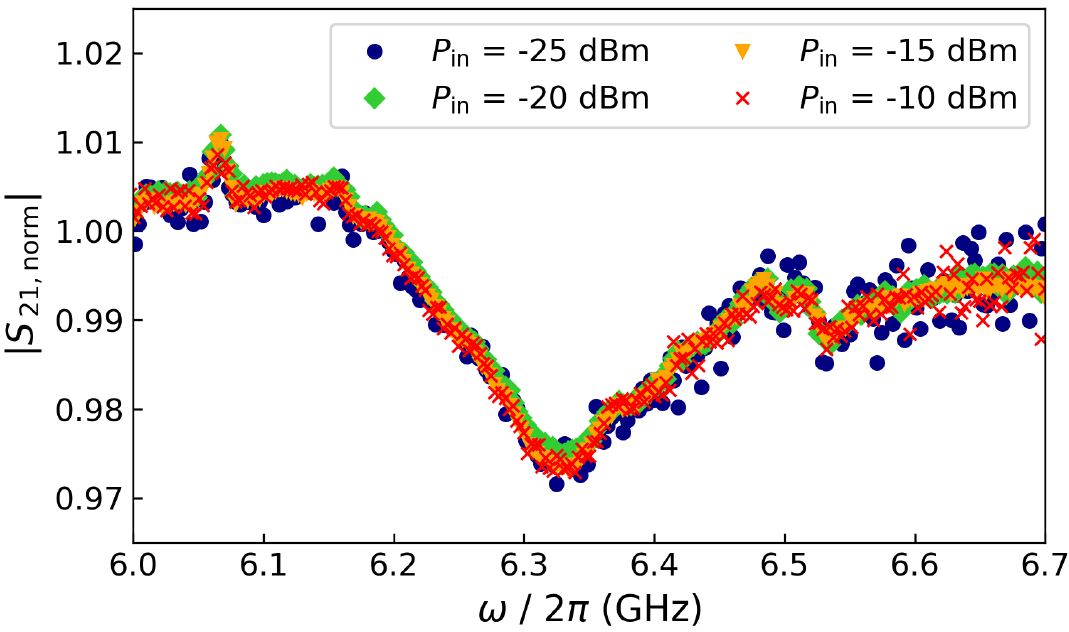}
\caption{\textbf{Spin resonance versus microwave power.} Normalized transmission of a waveguide coupled to a powdered DPPH sample measured at $T=0.3$ K and $B=0.225$ T.}
\label{fig:power_sweep}
\end{centering}
\end{figure}

\clearpage

\section*{Supplementary Note 6: Theory of antiferromagnetic spin chains}
\label{sec:AFtheory}

\subsection*{Exact diagonalization calculation of the magnetic and thermal responses of small AF spin chains}
\label{ssec:exact_diagonalization}

The 1D antiferromagnetic chains that form in the B-sublattice can described by ($\hbar = 1$):
\begin{equation}
\label{Hm-sup}
     \mathcal{H}_{m} = \sum_{i,j}\vect{S}_i\cdot\bar{J}\cdot\vect{S}_j
     - g_S \mu_B B \sum_i S_i^y \; .
\end{equation}
Here, $\vect{S_j} =(\sigma_j^x,\sigma_j^y,\sigma_j^z)$ are the Pauli matrices. The anisotropy of the spin-spin interactions is encoded in the coupling tensor $\bar{J}$:
\begin{equation}
\label{J-sup}
        \bar{J} = {\rm diag} ( J, J(1+\epsilon\sin\psi), J(1+\epsilon\cos\psi) )
        \; .
\end{equation}
Here, $\psi$ is the angle that the anisotropy axis $\vect{\epsilon}$ forms with $z$. Moreover, w.l.o.g we assume this axis lies in the $yz$-plane, see Supplementary Fig. \ref{fig:angles}. The last term in $\mathcal{H}_{m}$ is the Zeeman coupling to the externally applied DC magnetic field $B_0$. We define the $y$-axis as the orientation of this field, matching the laboratory frame defined in Fig.~1a in the main text.

\addcontentsline{toc}{subsubsection}{Supplementary Figure 17: Convention of angles employed for both the exact diagonalization and the mean field treatment.}
\begin{figure}[b!]
\begin{centering}
\includegraphics[width=0.85\textwidth, angle=0]{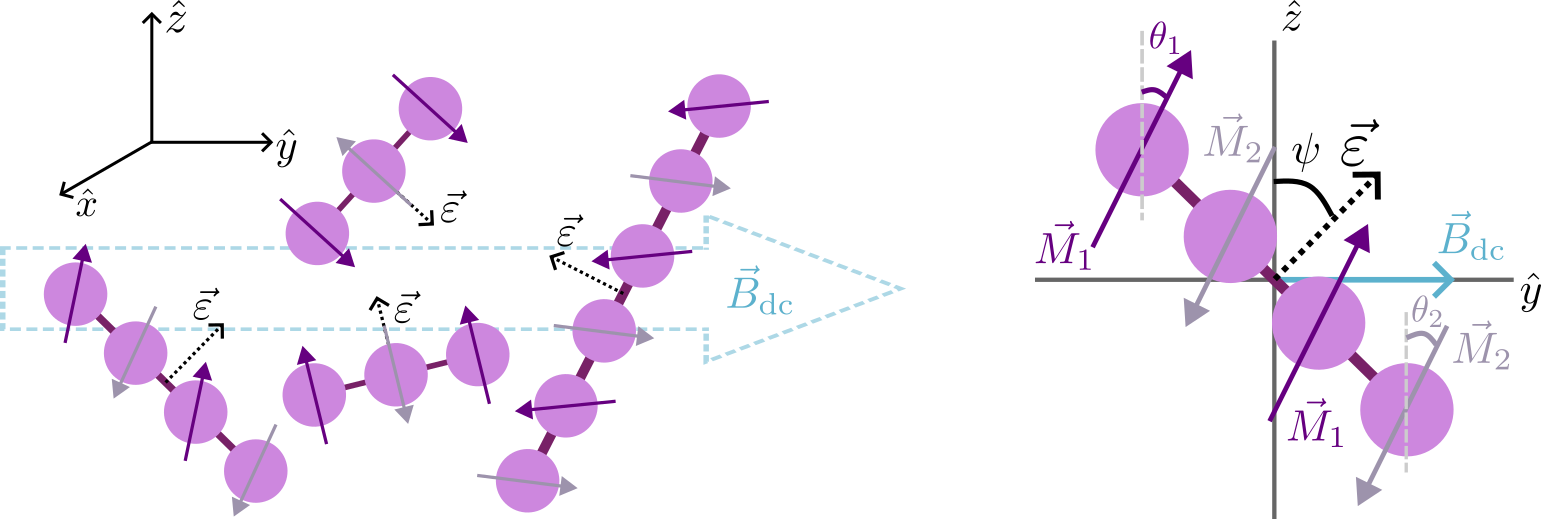}
\caption{\textbf{Convention of angles employed for both the exact diagonalization and the mean field treatment.} Each chain in the powder sample has a different anisotropy axis orientation with respect the magnetic field. }
\label{fig:angles}
\end{centering}
\end{figure}

Using exact diagonalization we can obtain all the eigenstates of the system up to sizes not much larger than 10 spins. With them, we can calculate observables such as the specific heat or the magnetic susceptibility. Starting with the specific heat, $c$, we calculate it from the eigenvalues of the Hamiltonian given in Eq.~\eqref{Hm-sup} at a certain temperature, $T$, with the following expression
\begin{equation}\label{eq:specheat}
    c = \beta\left[\frac{Z^{\prime\prime}}{Z} - \left(\frac{Z^\prime}{Z}\right)^2\right]\ ,
\end{equation}
where $\beta = 1/T$, $()^{\prime}=\partial_\beta$ and $Z = Tr\left({\rm e}^{-\beta\mathcal{H}}\right)$.\cite{Garcia2009}
The terms explicitly read $Z = \sum_m {\rm e}^{-\beta \epsilon_m}$, $Z^\prime = -\sum_m {\rm e}^{-\beta \epsilon_m}\epsilon_m$ and $Z^{\prime\prime} = \sum_m {\rm e}^{-\beta \epsilon_m}\epsilon^2_m$, with $\epsilon_m$ being the eigenstates of Eq.~\eqref{Hm-sup}.
We compute these energy fluctuations for the different angles $\psi$ and sizes up to $N = 8$ spins and then we take the average. The 
results of the simulations are presented in Fig. \ref{fig:CPvsH} and 1c of the main text and in Supplementary Fig.~\ref{fig:CPvsH}. We confine our analysis to small-sized lattices, as the complete spectrum is necessary to compute the specific heat at finite temperatures, as discussed in Supplementary Fig.~\ref{fig:CP-Lanczos}. This limits the quantitative approach to actual experimental values.

\addcontentsline{toc}{subsubsection}{Supplementary Figure 18: Specific heat obtained through exact diagonalization.}
\begin{figure}[b!]
\begin{centering}
\includegraphics[width=0.7\columnwidth, angle=-0]{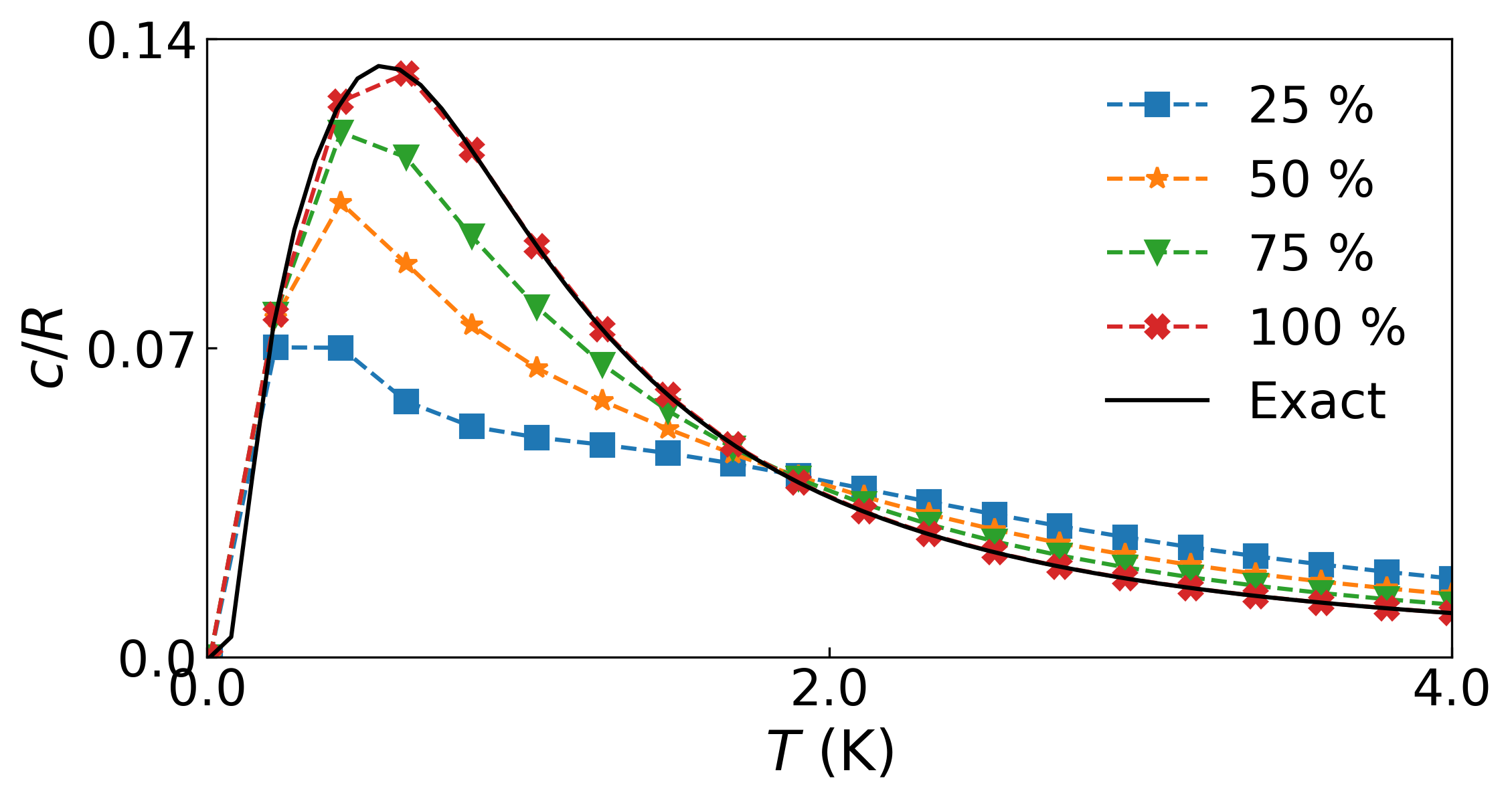}
\caption{\textbf{Specific heat obtained through exact diagonalization.} The results shown are for a 7-spin chain, as a function of temperature fixing the external magnetic field at $0.5$ T. As can be seen, we only recover the analytical curve (black curve in both plots) when the totality of the system's eigenstates is considered.}
\label{fig:CP-Lanczos}
\end{centering}
\end{figure}

On the other hand, for the magnetic susceptibility, we can treat perturbatively the partition function $Z$. In general, for a system described by some hamiltonian $\mathcal{H}_0$ that is under a perturbation $Q$ with intensity $f$, i.e. $\mathcal{H} = \mathcal{H}_0 - fQ$, the susceptibility, $\chi$, is defined as
\begin{equation}\label{gen-susc}
    \chi = \beta\left[\frac{Z_2}{Z_0} - \left(\frac{Z_1}{Z_0}\right)^2\right],
\end{equation}
where $Z \simeq Z_0 + \xi Z_1 + \xi^2 Z_2$ and $\xi = \beta f$.\cite{Garcia2009}
In our case, $Z_0 = \sum_m {\rm e}^{-\beta \epsilon_m}$, $Z_1=\sum_m {\rm e}^{-\beta \epsilon_m}Q_{mm}$, and $Z_2 = \sum_{mn}{\rm e}^{-\beta\epsilon_{mn}}|Q_{mn}|^2K_{mn}$, with $Q = \tfrac{1}{|\vect B|} \vect{B}\vect{S}$, $K_{mn} = K\left[\beta(\epsilon_m - \epsilon_n)\right]$ and $K(X) = ({\rm e}^X-1)/X$.

In this case, we observe that small chains accurately reproduce the experimental curve (see Fig.~\ref{fig:magnetic_properties}). This phenomenon can be attributed to the fact that, with increasing chain length, the per-spin contribution to the total susceptibility diminishes. As demonstrated in Supplementary Fig.~\ref{fig:susceptibility}, the susceptibility scales as $1/N$, in relation to the susceptibility of a single spin, which predominantly influences the behavior in the limit  $T \rightarrow 0$. Furthermore, only chains with an odd number of spins contribute, due to their $S=1/2$ ground state. At sufficiently high temperatures, however, chains of all lengths contribute equally, enabling the recovery of the experimental results using solely small chains, as shown in Fig. 1b in the main text. 

\addcontentsline{toc}{subsubsection}{Supplementary Figure 19: Dependence of the magnetic susceptibility on the spin chain size.}
\begin{figure}[t!]
\begin{centering}
\includegraphics[width=0.7\columnwidth, angle=-0]{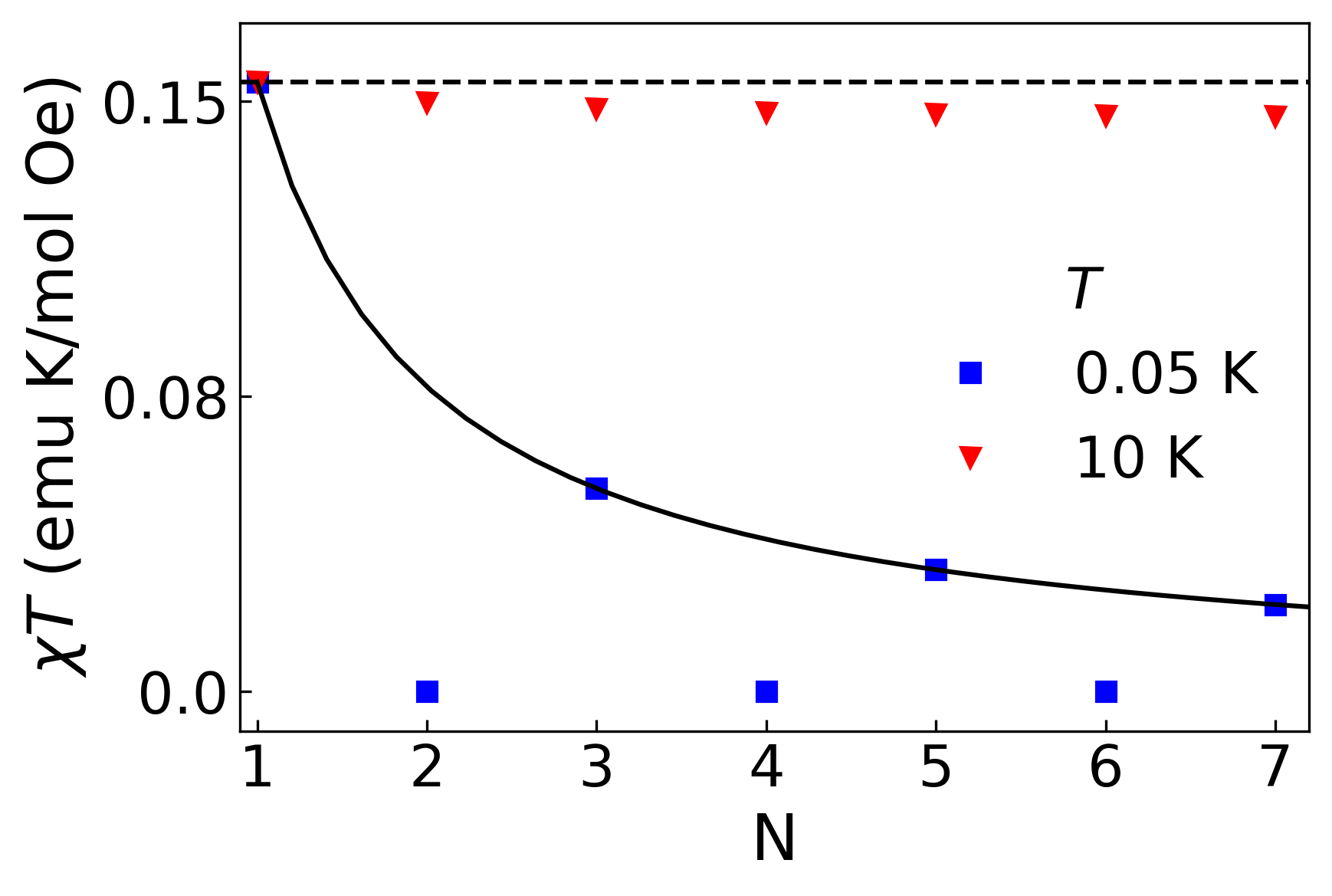}
\caption{\textbf{Dependence of the magnetic susceptibility on the spin chain size.} We show the $\chi T$ product calculated, at $B=0$, as a function of the number of spins in the chain in two regimes: low temperature (blue squares) and high temperature (red triangles). In the low-temperature regime only odd number spin chains contribute. This response (normalized by the spin number) decays with increasing chain length.}
\label{fig:susceptibility}
\end{centering}
\end{figure}

Comparing the specific heat and the magnetic susceptibility obtained from the exact diagonalization of our model of small chains with the experimental results we found that the anisotropy 
in the spin-spin interactions is very small and not relevant in 
these experiments. This is not the case in microwave transmission 
experiments, where the small anisotropy plays a crucial role and significantly changes the spin resonance features, as it is shown in the main text and in section VI below.

\subsection*{Mean field theory of antiferromagnetic spin chains}
\label{ssec:mag}

For computing the transmission at very low temperatures, it is necessary to consider large chains that are beyond the scope of exact diagonalization. Therefore, we employ a mean field theory, which not only explains the underlying physics but also accounts for the experimental data both qualitatively and quantitatively, as detailed in the main text. Here, we aim to describe our application of this mean field (MF) theory to DPPH. Additionally, within the MF theory, we can compute the linear response, enabling us to characterize the transmission experiments.

We use the Bogolyubov inequality, which serves as a method for approximating the equilibrium partition function of an interacting system. This inequality arises from convexity and requires the Hamiltonian to be divided into a part that is solvable and the remaining part.
\begin{equation}
\label{Bogo}
{\cal H} = {\cal H}_0 + {\cal H}_1 \;  \Longrightarrow \;  \frac{Z}{Z_0} \geq {\rm e}^{-\beta \langle
  H_1 \rangle_0} \; , 
\end{equation}
with $Z= {\rm Tr} [{\rm e}^{-\beta {\cal H}}]$ the actual partition function,
$Z_0= {\rm Tr}[{\rm e}^{-\beta {\cal H}_0}]$ and
$\langle ...\rangle_0 = {\rm Tr} [ ...{\rm e}^{-\beta {\cal H}_0} ] $.
Taking the
logarithm in Eq.~\eqref{Bogo}, an upper bound for the free energy is found:
\begin{equation}
\label{F-ine}
F \leq F_0  + \langle {\cal H}_1 \rangle_0
\; .
\end{equation}

In antiferromagnetic systems, spins are divided in two sub-lattices, namely $1$ and
$2$.  In each sub-lattice the spin averages are the same. 
Therefore, we choose for ${\cal H}_0$:
\begin{equation}
{\cal H}_0 = -\vect \lambda_1 \vect M_1 -  \vect \lambda_2 \vect M_2 \; ,
\end{equation}
where $\vect \lambda_\alpha$ are variational parameters that 
minimize the left hand side of Eq.~\eqref{F-ine} and we have used the following compact notation for the magnetization vectors:
\begin{align}
  \vect M_1 = & M_1 \vect u_1 \; ,
  \\
  \vect M_2 = & M_2 \vect u_2 \; .
\end{align}
It turns out that $Z_0$ only depends on the modulus $|\vect \lambda_\alpha|$. Thus, we can set $\lambda_\alpha \geq 0$. Considering nearest neighbor interactions, together with the angle convention used in Supplementary Fig.~\ref{fig:angles}, we obtain for 
the average:
\begin{align}
  \label{H0}
  \langle {\cal H}_1 \rangle_0
  = \; &
  \lambda_1 M_1 + \lambda_2 M_2
  \\ \nonumber
  & -
  g_S\mu_{\mathrm{B}}B (M_1 \sin \theta_1 + M_2 \sin \theta_2 )
  \\ \nonumber
  & +
  M_1 M_2 \vect u_1 \cdot \bar J \cdot \vect u_2
   \; ,
\end{align}
where $\bar J$ is given in Eq.~\eqref{J-sup}.  We assume that 
the magnetization vectors lie also in the $zy$-plane, the plane spanned by the anisotropy and magnetic field vectors as depicted in Supplementary Fig.~\ref{fig:angles}. Simulations performed considering a generic three-dimensional orientation for the magnetisation vectors gave the same results as in the 2D case. This is easy to see since the directions of interest are always going to be the anisotropy axis (when there is no field) or the field axis (when the field is very strong).\\

Inserting Eq.~\eqref{H0} in Eq.~\eqref{F-ine} and minimizing over the variables $\{\lambda_1, \theta_1, \lambda_2, \theta_2 \}$, we end with the consistency relations:
\begin{subequations}
\begin{align}
\label{minH}
  \frac{\partial \langle {\cal H} \rangle_0}{\partial \theta_1}&=0 \; ,
 \\
 \label{la}
\lambda_1 &= g_S\mu_{\mathrm{B}}B \sin \theta_1 - M_2 \vect u_1 \cdot \bar J \cdot \vect u_2 \; ,
\\
\label{ma}
M_1 &= \tanh ( \beta \lambda_1) \; ,
\end{align} 
\end{subequations}
and similar equations for sublattice 2.\\

In general, these equations are solved numerically. However, analytical progress can be made using symmetry arguments. We assume \(\lambda_1 = \lambda_2 = \lambda\) and \(M_1 = M_2 = M\). With our convention for angles (Supplementary Fig. \ref{fig:angles}), we find that in either the paramagnetic phase or the spin-flop transition (see below), \(\theta_1 + \theta_2 = \gpi\). This condition is also applicable, in principle, to the antiferromagnetic scenario. By imposing these conditions, we derive the following expression:
\begin{equation}\label{eq:mf_eq_angle}
    \sin\theta_2^{\mathrm{eq}} = \frac{g_S\mu_{\mathrm{B}}B}{M J[2+\epsilon(\sin\psi+\cos\psi)]} \; ,
\end{equation}
with:
\begin{align}
\label{m-mf}
    M = \tanh \bigg[ \beta \big( & g_S\mu_{\mathrm{B}}B
    + JM\cos2\theta_2^{\mathrm{eq}} \\ \nonumber
    & -
    JM\epsilon(\sin^2\theta_2^{\mathrm{eq}}\sin\psi-\cos^2\theta_2^{\mathrm{eq}}\cos\psi)
    \big) \bigg] \; .
\end{align}

Equations \eqref{eq:mf_eq_angle} and \eqref{m-mf} are used for determining the phase diagram, as illustrated in Supplementary Fig.~\ref{fig:mf_angles}. Here, we depict the numerical solutions for the equilibrium angles, $\Delta\theta_{\mathrm{eq}}$, yielding the distinct phases. A zero value of $\Delta\theta_{\mathrm{eq}}$ indicates that temperature and/or magnetic field are sufficient to align the magnetization vectors. A non-zero $\Delta\theta_{\mathrm{eq}}$ signals the emergence of magnetic correlations, revealing two distinct phases: the spin-flop phase for $0 < \Delta\theta_{\mathrm{eq}} < \gpi$, and the antiferromagnetic phase for antiparallel spins. Panel (a) of Supplementary Fig.~\ref{fig:mf_angles} displays the case with perpendicular anisotropy to the magnetic field, while panel (b) considers parallel orientation. Specifically, when $\vect{B} \perp \vect{\epsilon}$ and $\psi < \gpi/4$, a magnetic field weaker than $J\sqrt{2\epsilon(\sin\psi-\cos\psi)}$ results in antiparallel magnetization vectors orthogonal to the anisotropy axis, where $\theta_1 + \theta_2 = 2\gpi$. This condition persists until the magnetic field $B$ is strong enough to rotate both vectors towards the paramagnetic phase. This process is elaborated in subplots (a.1) and (b.1) of Supplementary Fig.~\ref{fig:mf_angles}. The critical field at $T=0$ is derived from Eq.~\eqref{eq:mf_eq_angle} as $g_S\mu_{\mathrm{B}}B^{\mathrm{c}} = J(2+\epsilon(\sin\psi+\cos\psi))$, while the Néel temperature is fixed at $T_{\mathrm{N}} = J/k_{\mathrm{B}}$.

\addcontentsline{toc}{subsubsection}{Supplementary Figure 20: Equilibrium magnetization angles.}
\begin{figure*}[t!]
    \centering
    \includegraphics[width = \textwidth]{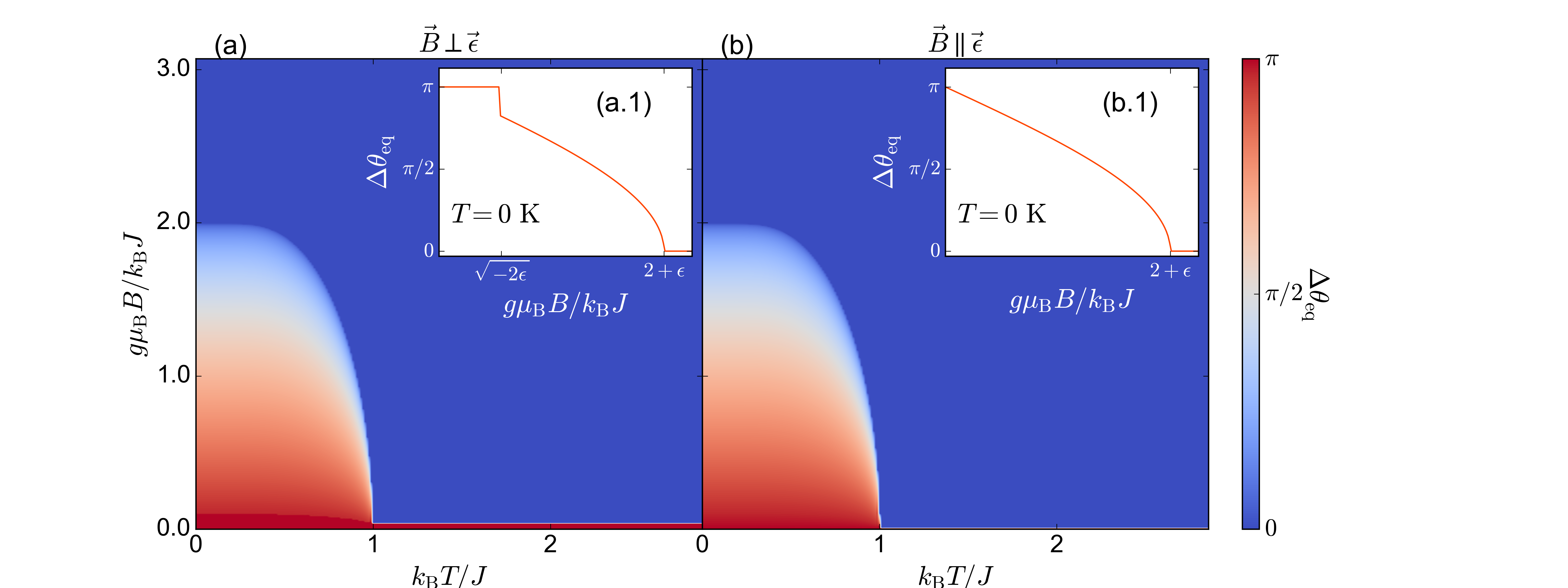}
    \caption{\textbf{Equilibrium magnetization angles.} Phase diagram for $\Delta \theta_{\rm eq}$ as a function of temperature and magnetic field. In (a) the anisotropy is perpendicular to the magnetic field while in (b) they are parallel.
    In subplots (a.1) and (b.1) we see the different behaviour at $T=0$ that the anisotropy shows up. Since $J(1+\epsilon)<J$, the magnetization
    vectors lie perpendicular to the anisotropy axis to minimize the energy. When $\psi < \gpi/4$, there is a magnetic field threshold for which,
    below that point, magnetization remains in the perpendicular axis to $\vect \epsilon$. When the field is strong enough, both vectors flip
    to start coming parallel between them symmetrically until the paramagnetic phase is reached. This point is found to be $g_S\mu_{\rm B} B = J\sqrt{2\epsilon(\sin\psi-\cos\psi)}$.
    The parameters used were $J/k_{\rm B} = 0.7\ \rm K$ and $J\epsilon/ k_{\rm B} = -0.06\ \rm K$ for panels (a) and (b). In (a.1) and (b.1)
    an arbitrarily bigger $\epsilon$ was used for better visualization.}
    \label{fig:mf_angles}
\end{figure*}

%

\subsection*{Spin dynamics: computing the resonance frequency of the chains}
\label{ssec:dyn}

Mean field treatment is consistent with using the Landau-Lifshitz-Gilbert (LLG) equation for the dynamics. The LLG equation reads
\begin{equation}
\label{LLG}
\frac{ d \vect M_\alpha}{d t}
= - g_S \mu_B 
\; 
\vect M_\alpha \times \vect B_{{\rm mol}, \alpha} - \gamma
\;
\vect M_\alpha \times ( \vect M_\alpha \times \vect B_{{\rm mol}, \alpha}
) \, 
\end{equation}
with $\alpha = 1,2$. The dissipation parameter $\gamma$ is related in our simulations to the inhomogenous broadening of the sample in the paramagnetic regime ($\Gamma$ in the main text) as $\gamma = \Gamma / g_S\mu_BB_0$. The molecular field 
is given by
\cite{Vonsovskii,Ross2013} 
\begin{equation}
\vect B_{{\rm mol}, \alpha}
=
- \frac{1}{g_S \mu_B} \vect \nabla  \,  F \, ,
\end{equation}
with $F$ the free energy computed by means of the MF, Cf. Eq.~\eqref{F-ine}.

As the LLG preserves $|\vect{M}_\alpha|$, it is convenient to work in spherical coordinates, which transform Eq.~\eqref{LLG} into the following:
\begin{equation}
\begin{split}
\sin \theta_\alpha \; \dot \theta_\alpha &=
- 
\frac{1}{ M_\alpha }  F _{\varphi_\alpha}
-
\gamma  \sin \theta_\alpha \; F_{\theta_\alpha} 
\; ; \\ 
\sin \theta_\alpha \;  \dot \varphi_\alpha &=
\;\;\; \frac{1 }{ M_\alpha} F_{\theta_\alpha} 
-
\gamma \frac{1 }{\sin \theta_\alpha} F_{\varphi_\alpha} 
\; .
\end{split}
\end{equation}
We have introduced the notation $F _{\theta_\alpha}\equiv \partial_{\theta_\alpha} F
$. 
At equilibrium, the derivatives of the free energy are zero. Therefore, in the linear response regime, we perform a Taylor expansion of these derivatives, resulting in the set of linear equations:
\begin{widetext}
\begin{align}
\label{LLGmatrix}
\begin{pmatrix}
\sin \theta_1 \; \dot  \theta_1
\\
\sin \theta_1 \; \dot  \varphi_1
\\
\sin \theta_2 \; \dot  \theta_2
\\
\sin \theta_2 \; \dot  \varphi_2
\\
\end{pmatrix}
= &
\begin{pmatrix}
- \frac{F_{\varphi_1, \theta_1}}{M_1} 
& - \frac{F_{\varphi_1,    \varphi_1}}{M_1} 
& - \frac{F_{\varphi_1, \theta_2}}{M_1} 
& - \frac{F_{\varphi_1,    \varphi_2}}{M_1} 
\\
 \frac{F_{\theta_1, \theta_1}}{M_1} 
&  \frac{F_{\theta_1,    \varphi_1}}{M_1} 
&  \frac{F_{\theta_1, \theta_2}}{M_1} 
&  \frac{F_{\theta_1,    \varphi_2}}{M_1} 
\\
- \frac{F_{\varphi_2, \theta_1}}{M_2} 
& - \frac{F_{\varphi_2,    \varphi_1}}{M_2} 
& - \frac{F_{\varphi_2, \theta_2}}{M_2} 
& - \frac{F_{\varphi_2,    \varphi_2}}{M_2} 
\\
 \frac{F_{\theta_2, \theta_1}}{M_2} 
&  \frac{F_{\theta_2,    \varphi_1}}{M_2} 
&  \frac{F_{\theta_2, \theta_2}}{M_2} 
&  \frac{F_{\theta_2,    \varphi_2}}{M_2} 
\end{pmatrix}
\\ \nonumber
& - \gamma
\begin{pmatrix}
- \sin \theta_1 \; F_{\theta_1, \theta_1}
& - \sin \theta_1  \; F_{\theta_1, \varphi_1}
& -  \sin \theta_1 \; F_{\theta_1, \theta_2}
& - \sin \theta_1  \; F_{\theta_1, \varphi_2}
\\
 \frac{F_{\varphi_1, \theta_1}}{\sin \theta_1} 
&  \frac{F_{\varphi_1,    \varphi_1}}{\sin \theta_1} 
&  \frac{F_{\varphi_1, \theta_2}}{\sin \theta_1} 
&  \frac{F_{\varphi_1,    \varphi_2}}{\sin \theta_1} 
\\
- \sin \theta_2 \; F_{\theta_1, \theta_2}
& - \sin \theta_2  \; F_{\theta_2, \varphi_1}
& -  \sin \theta_2 \; F_{\theta_2 \theta_2}
& - \sin \theta_2  \; F_{\theta_2, \varphi_2}
\\
 \frac{F_{\varphi_2,  \theta_1}}{\sin \theta_2} 
&  \frac{F_{\varphi_1,    \varphi_2}}{\sin \theta_2} 
&  \frac{F_{\varphi_2, \theta_2}}{\sin \theta_2} 
&  \frac{F_{\varphi_2,    \varphi_2}}{\sin \theta_2} 
\end{pmatrix}
\begin{pmatrix}
 \theta_1
\\
\varphi_1
\\
 \theta_2
\\
\varphi_2
\\
\end{pmatrix}
.
\end{align}
\end{widetext}

\noindent Here, both $\theta_\alpha$ and $\varphi_\alpha$ are the 
equilibrium
values (obtained as explained in the previous section). Besides, the 
derivatives
are evaluated in the corresponding equilibrium values. Note that 
here we use a 3D orientation for the magnetization vectors for convenience. The 2D choice that we made in the previous section allowed us to derive some analytical expression whereas here we will resort to numerical solutions.

The resonance frequency $\Omega$ is derived by seeking oscillatory solutions of the form
\begin{subequations}
\begin{align}
\theta_\alpha &= \delta \theta_\alpha \mathrm{e}^{\mathrm{i}\Omega t}, \\
\varphi_\alpha &= \delta \varphi_\alpha \mathrm{e}^{\mathrm{i}\Omega t}.
\end{align}
\end{subequations}
Substituting these expressions into Eq.~\eqref{LLGmatrix}, we obtain a system of equations for $\Omega$ that is amenable to numerical solution. Among the four solutions, two correspond to the normal modes of each sublattice, indicating that for each sublattice, two frequencies describe the motion of $\theta_\alpha$ and $\varphi_\alpha$. These solutions present both real and imaginary components, which turn to be crucial for understanding experimental outcomes. In Supplementary Fig.~\eqref{fig:wres}, we present several numerical examples to illustrate the dependence of the real component of $\Omega$ on the angle $\vect{\psi}$. Specifically, in panel \ref{fig:wres}a, we plot the angle dependence of the non-vanishing resonant frequency as a function of the field for different anisotropy orientations, $\psi$. The resonant frequencies are found to follow:
\begin{equation}\label{eq:omegares_angle}
\text{Re}(\Omega) = \sqrt{\bar B^2 - 2J^2\epsilon(\sin\psi-\cos\psi)},
\end{equation}
where $\bar B = g_S\mu_{\mathrm{B}}B$. For angles less than $\gpi/4$, grey dashed lines show the expected frequency curve absent a spin-flop transition. Orientations with complementary angles ($\psi_1 + \psi_2 = \gpi/2$) exhibit identical resonant frequencies at zero field. The black dashed line denotes the magnetic field value for which panel \ref{fig:wres}b is simulated, corresponding to the experimental and simulation data presented in Fig.~3 of the main text. Below the Néel temperature ($T_{\mathrm{N}}$), magnetic correlations lead to distinct resonant frequencies for each anisotropy orientation, resulting in a distribution where larger angles, i.e., orientations closer to the magnetic field direction, exhibit higher frequencies. Note that from the top panel we can see how as the magnetic field increases, the spread in resonant frequencies with the anisotropy angle decreases, as observed in the experimental measurements shown in the Supplementary Note 4.

The imaginary component of $\Omega$ also merits special attention. It has been recently proven \cite{Wang2024} that the line width in antiferromagnetic resonances is broader than what one whould expect from Eq.~\eqref{LLG}. The intrinsic dissipation of the system 
becomes larger as the antiferromagnetic spin-spin correlations grow by increasing the imaginary component of $\Omega$. Thus, the line width increases below $T_{\mathrm{N}}$ (see Fig. 4 in the main text). However, as explained in the main text, this does not suffice to account for the experimental broadening, which is still dominated 
by the distribution of resonance frequencies $\Omega(\psi)$ that the anisotropy in the spin-spin interactions introduces when spin waves emerge.

\addcontentsline{toc}{subsubsection}{Supplementary Figure 21: Real component of the resonant frequency of the chains according to its orientation.}
\begin{figure}[h!]
    \centering
    \includegraphics[width = 0.7\columnwidth]{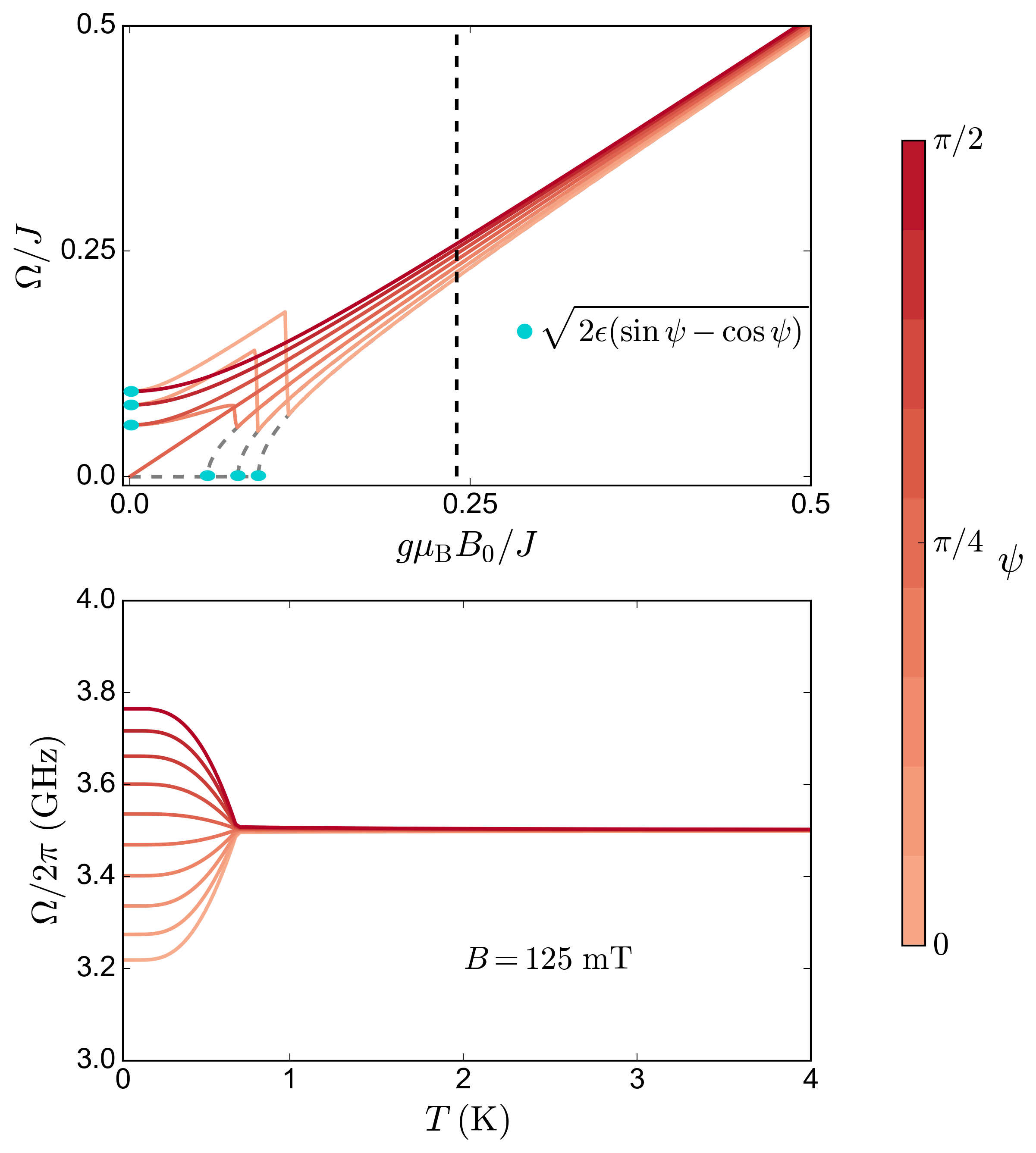}
    \caption{\textbf{Real component of the resonant frequency of the chains according to its orientation.} In a) we show the dependence of $\Omega$ at $T = 0$ with the anistropy orientation, $\psi$. 
    We find that $\Omega$ follows Eq.~\eqref{eq:omegares_angle} so angles between $0$ and $\gpi/4$ would follow the grey dashed line if they
    did not pass through the particular spin flop transition described. The black dashed line indicates the magnetic field at which the results in panel b) were obtained.
    There, the temperature behaviour of the frequencies for each orientation $\psi$ is shown. Over $T_{\mathrm{N}}$, the system is paramagnetic and
    anisotropy does not play any role but below that temperature, magnetic correlations show up its strong influence on the frequencies. The parameters used were
    $J/k_{\mathrm{B}} = 0.7\ \mathrm{K}$ and $J\epsilon/k_{\mathrm{B}} = -0.06\ \mathrm{K}$.}
    \label{fig:wres}
\end{figure}

\newpage
\section*{References}
\stoptoc

\bibliographystyle{apsrev4-1}
\bibliography{superra}